\documentclass[final,10pt,5p,times]{elsarticle}
\journal{Future Generation Computer Systems}
\usepackage{blkarray, bigstrut}
\usepackage{xparse}
\usepackage{multirow} 
\usepackage{algorithmicx}
\usepackage{algpseudocode}
\usepackage{algorithm}
\usepackage{amsmath}
\usepackage{rotating}
\usepackage{tabularx}

\usepackage{mathtools}
\usepackage{blkarray}
\usepackage{amsmath}
\usepackage{stfloats}
\DeclarePairedDelimiter{\abs}{\lvert}{\rvert}

\usepackage{kantlipsum} 
\usepackage{commath}
\allowdisplaybreaks
\usepackage{mathtools}
\usepackage{verbatim}
\usepackage{xr-hyper} 
\usepackage{enumitem}
\usepackage{multirow}
\usepackage{hhline}
\usepackage[table,xcdraw]{xcolor}
 \usepackage{graphicx,booktabs}
 \usepackage{longtable}
 \usepackage{subcaption}
\usepackage[hyphens]{url}
\usepackage{hyperref}
\hypersetup{breaklinks=true}
\usepackage{epstopdf}
\usepackage{wrapfig}
\usepackage{latexsym}
\usepackage{amssymb}
\usepackage{amsthm}
\usepackage{amsfonts}
\usepackage{amsmath} 
\usepackage{graphicx}
\usepackage{latexsym}
\usepackage{booktabs}
\usepackage[titletoc, toc, page]{appendix}

\usepackage[style=base]{caption}
\usepackage{subcaption} 
\usepackage{breqn}
\newtheorem{thm}{Definition}

\definecolor{LightCyan}{rgb}{0.88,1,1}
\algnewcommand\algorithmicinput{\textbf{INPUT:}}
\algnewcommand\INPUT{\item[\algorithmicinput]}
\algnewcommand\algorithmicoutput{\textbf{OUTPUT:}}
\algnewcommand\OUTPUT{\item[\algorithmicoutput]}
\algdef{SE}[DOWHILE]{Do}{doWhile}{\algorithmicdo}[1]{\algorithmicwhile\ #1}

\makeatletter
\newcounter{algorithmbis}
\renewcommand{\thealgorithmbis}{\arabic{algorithmbis}}
\def\algorithmbis{\@ifnextchar[{\@algorithmbisa}{\@algorithmbisb}}
\def\@algorithmbisa[#1]{%
  \refstepcounter{algorithmbis}
  \trivlist
  \leftmargin\z@
  \itemindent\z@
  \labelsep\z@
  \item[\parbox{0.49\textwidth}{%
    \hrule
    \noindent\strut\textbf{Algorithm \thealgorithmbis} #1
    \hrule
  }]\hfil\vskip+0em%
}
\def\@algorithmbisb{\@algorithmbisa[]}

\makeatother

\usepackage[table]{xcolor}

\usepackage{tikz}
\usetikzlibrary{fit} 

\usepackage{color}

\colorlet{BLUE}{blue}
\usepackage{colortbl}
\definecolor{LightCyan}{RGB}{155, 227, 247}

\newif\ifcommentson

\newif\ifextended
\newif\ifshortver

\shortvertrue

\newcommand{\optional}[1]{\ignorespaces}

\newif\ifrevisionactive
\newif\ifshowdeleted
\revisionactivetrue
\allowdisplaybreaks

\definecolor{maroon}{cmyk}{0,0.87,0.68,0.32}

\begin{document}

	\begin{frontmatter}
\title{\huge Similarity-based Android Malware Detection Using Hamming Distance of Static Binary Features}
\author[label1]{Rahim Taheri}
\address[label1]{Department of Computer Engineering and Information Technology, Shiraz University of Technology, Shiraz, Iran}
\ead{r.taheri@sutech.ac.ir}
\author[label1]{Meysam Ghahramani}
\ead{m.ghahramani@sutech.ac.ir}
\author[label1]{Reza Javidan}
\ead{javidan@sutech.ac.ir}
\author[label2]{Mohammad Shojafar\corref{cor1}}
\address[label2]{Department of Mathematics,University of Padua, Via Trieste 63, 35131, Padua, Italy}
\ead{mohammad.shojafar@unipd.it}
\cortext[cor1]{Corresponding author}
\author[label2]{Zahra Pooranian}
\ead{zahra@math.unipd.it}
\author[label2]{Mauro Conti}
\ead{conti@math.unipd.it}



\begin{abstract}

In this paper, we develop \textit{four} malware detection methods using Hamming distance to find similarity between samples which are first nearest neighbors (FNN), all nearest neighbors (ANN), weighted all nearest neighbors (WANN), and k-medoid based nearest neighbors (KMNN). In our proposed methods, we can trigger the alarm if we detect an Android app is malicious. Hence, our solutions help us to avoid the spread of detected malware on a broader scale. We provide a detailed description of the proposed detection methods and related algorithms. We include an extensive analysis to asses the suitability of our proposed similarity-based detection methods. In this way, we perform our experiments on \textit{three} datasets, including benign and malware Android apps like Drebin, Contagio, and Genome. Thus, to corroborate the actual effectiveness of our classifier, we carry out performance comparisons with some state-of-the-art classification and malware detection algorithms, namely Mixed and Separated solutions, the program dissimilarity measure based on entropy (PDME) and the FalDroid algorithms. We test our experiments in a different type of features: API, intent, and permission features on these three datasets. The results confirm that accuracy rates of proposed algorithms are more than 90\% and in some cases (i.e., considering API features) are more than 99\%, and are comparable with existing state-of-the-art solutions.

\end{abstract}
\end{frontmatter}

\section{Introduction}\label{sez:1}

Nowadays, the widespread use of mobile devices in comparison with personal computers has begun a new era of information exchange. 
Besides, the increased power of mobile devices, coupled with the portability of user attention has attracted. Smartphones and tablets are prevalent in recent years. By the end of 2014, the number of active mobile devices around the world was about 7 billion, and in developed countries, the proportion of mobile devices and people are estimated to be 120.8\%, respectively. Due to their widespread distribution and their abilities, mobile devices have become the main target of the attackers in recent years~\cite{Globalmobilestatistics}. Android is currently the most widely used mobile smartphone platform in the world, which occupies 85\% of the market share. Recent reports indicate an increase in the number of Android programs in recent years. As the number of Android applications on Google Play in December 2009 was 16,000, in July 2013 one million, in February 2016 it was about 2 million and in December 2017 it was five million.~\cite{Sophosmobilesecurity,kasperskysecurity}.
\subsection{General Definition}\label{sec:sec1.1}
Android app is in two categories: Benign and Malware. Samples that are safe and do not show malicious behaviors are called \textit{benign} samples. In contrast, examples of software that create a security threat are named \textit{malware} samples. In recent years, the variety of malware in Android mobile networks is continuously increasing and thus causes a risk to users' privacy. 
Furthermore, the popularity of Android with cyber-criminals is also high and creates a lot of malicious programs to steal sensitive information and compromise mobile systems, and these conditions represent the need for security in the mobile app. Unlike other smartphone platforms like iOS, Android users can install their apps from unverified sources, such as file sharing websites. In Android apps, the issue of malware infection is very serious, and recent reports show that 97\% of the attacks on mobile malware came from Android devices. In 2016 alone, more than 3.25 million Android malicious apps were detected. That means almost every 10 seconds a new malicious Android application is created~\cite{GooglePlay,Gdata,razaque2018naive}.
Malware term is created by combining the words ``malicious'' and ``software''. Malware is a serious threat to the computer world, and this threat is increasing and complicated. When malicious software finds its way into the system, it scans the OS's vulnerabilities, performs unwanted actions on the system, and ultimately reduces system performance ~\cite{Vinod2009}. Hence, an important problem with cyber-security is malware analysis~\cite{Wang2017,al2018live}. 

In addition to accurate precision and the precision recognition rates, a malware detection system could generalize to new malicious families. For Android malware detection, two types of solutions, namely \textit{Static} and \textit{Dynamic}, have been proposed. Features like APIs, permissions, intent, URLs are analyzed in static solutions. In another category of malware, malicious components are downloaded at run-time, which requires dynamic analysis to detect these malwares~\cite{cai2019droidcat}. For instance, the authors~\cite{MalDAE2019} have provided a method for detecting malware concerning the correlation between static and dynamic features. Also, the authors~\cite{Meng2019} have come up with a way to detect malware in Android applications, by combining static analysis and outlier detection.

Another important point is that the system does not need to compute too much to deploy on mobile devices. Hence, the system should adopt models (e.g., machine learning models) to estimate the malicious behavior in a short time~\cite{Feng2018}. Machine learning (ML) methods are part of the artificial intelligence-based system in which solutions are provided to improve the decision-making process~\cite{Zhuo2019}. An ML method is widely used for specific decision-making tasks such as detecting malware, network penetration detection, and general pattern recognition issues. This method is very effective in identifying well-known and unknown malware families with high accuracy. In various studies, they design ML-based classification methods to categorize different types of samples (for example, static-based, logic-based, perception-based and sample-based types samples) and detect traffic networks on Android mobile devices~\cite{Ignacio2019}.

An advantage of using the ML method is its ability to identify different types of malware~\cite{Ajit2018}. In ML methods, complex pattern recognition and optimization of parameters are well investigated~\cite{Babak2011}. The current study indicates that the damage caused by malware programs, hidden among millions of mobile applications, is increasing, and this has been a visible motivation for researchers to deal with more complex applications.

Some Android software analyzes the malware behaviors at the API level. For example, the authors~\cite{Shanmugam2013} give a precise analysis of an opcode-based Android software based on finding the similarity measurements inspired by simple substitution distance of the features. They indicate that their technique provides a useful means of classifying metamorphic malware.

Some ML solutions adopt several distance calculation mechanisms to find similar samples to a specific sample. For example, the authors in~\cite{Radkani2017} add new distance measure using entropy for two computer programs which are called \textit{program dissimilarity measure} or PDME. PDME introduces a measure for the degree of metamorphism for samples. Also, the authors in~\cite{wangwei12017} elicit several types of behavior static features from Android apps and apply Support Vector Machine (SVM), $K$ Nearest Neighbor (KNN), Naive Bayes (NB), Classification and Regression Tree (CART) and Random Forest (RF) classifiers to detect malware from benign apps. KNN algorithm is classified as a supervised ML algorithm that could solve the classification and regression problems. KNN is easy to implement, no need to build a model, tune several parameters, or make additional assumptions. However, it is a slow method for large datasets. KNN algorithm can find the $k$ nearest samples to a specific query which have distances between a query and all the samples in the dataset. Then, it votes for the most frequent label or averages the labels.

Among different methods to calculate the distance, the \textit{Hamming distance} applies between two vectors with the same length and indicates the number of entries where injected elements are different. In other words, the Hamming distance achieves the minimum number of errors while converting one vector to another one. Suppose $x \triangleq (x_1, x_2,\ldots,x_n)$ is a sample vector and $y \triangleq (y_1, y_2,\ldots,y_n)$ is a corresponding label of vector $x$ on $n$ dimensional space, the Manhattan distance is the sum of the peer to peer distances between same indexes (see equation~\eqref{eq:eq1}).                                          
   
\begin{equation}
\label{eq:eq1}\small
d_{1}(X,Y)=\sum_{i=1}^{n}{|x_{i}-y_{i}}|,  
\end{equation}

And Minkowski distance presents by equation~\eqref{eq:eq2}:
\begin{equation}
\label{eq:eq2}\small
d_{2}(X,Y)=\sqrt[p]{\sum_{i=1}^{n}{|x_{i}-y_{i}|^p}},
\end{equation}
where $p\geq 1$.

In this paper, using replacement method we prove that with the binary representation of the data, we calculate the Hamming distance, and the distance calculated by this method is the same as the distance used by other methods like Euclidean distance and Manhattan distance. 

\subsection{Motivation and open issues} \label{sec:sec1.2}
As we described earlier, ML has been widely used in the classification of various types of Android OS like API, permission, intent and Android malware detection. For example, the paper~\cite{FanMing2018} applies API system call and shapes the API graph, the reference~\cite{KumarAjit2018} utilizes a score function to the extracted permission feature set, and finally, the paper~\cite{Varsha2017} adopts weighted mutual information to select prominent features. All of these research papers used the KNN algorithm to detect malware; however, due to the lack of binary representation of data, they need several calculations to extract malware vectors from benign samples.

Finding a threshold for $k$ in the KNN algorithm has been considered in many studies which are important in the malware detection methods~\cite{Xiong2018}. Another category of studies has suggested methods using ensemble learning that employ other algorithms such as decision tree, SVM and RF for malware detection. However, due to the simultaneous using of multiple algorithms, these methods have a high time complexity~\cite{Feng2018}. In some studies, a framework for detecting malware has been presented, which different classification methods such as SVM are applied in them~\cite{Saracino2016}. In~\cite{Saracino2016}, the authors propose a structure that uses the KNN algorithm based on Hamming distance for malware detection system. It used a fixed $k$ value for KNN which limits their structure.

The purpose of this paper is to investigate the effect of the distance between samples to classify into malware and benign. Due to the sparse feature vectors, the Hamming distance is an appropriate measure for the discrimination of samples. We propose a modified supervised KNN Algorithm using the Hamming distance to classify the samples. Then, we combine it with an unsupervised K-Medoids algorithm to detect malware based on static features. In the proposed framework of this paper, we use the Hamming distance to apply proposed classification methods which are the modified form of the KNN method.

\subsection{Problem Definition}\label{sec:sec1.3}
     Due to the widespread use of Android apps, finding a way of identifying malicious files is a critical problem that needs to be solved instantly. This paper use static analysis technology and propose \textit{four} detection methods based on similarity for Android malware by calculating distance of samples using a Hamming distance measure. The proposed methods are flexible solutions for the problem. It means, the generated model by each scenario learns the patterns in the features and can be used to classify the samples into malware and benign. Our proposed methods well generalize the patterns even for new samples. To do so, first, we find the related set of features  from the {manifest} part of {apk} file. Then, we use the RF regressor as a feature selection algorithm and rank the features. The main reason behind selecting the RF as a feature selection algorithm is that we could have better control over the results using RF when we consider different random subsamples of the original dataset~\cite{Robin2010}. Finally, we use the proposed methods based on the nearest neighbors of each sample and classify them.

\subsection{Contribution} 
In this research, we deploy several methods that applied on APIs, Permissions, and Intents used by Android applications to identify malware samples or apps. We carry out extensive experiments to compare proposed solutions with existing solutions and examine the validity of the proposed detection model. To sum up, we make the following contributions:

        \begin{itemize}[leftmargin=*]
        \item We prove that the result of using the Hamming distance with other methods is the same for the binary vectors and apply the Hamming distance in the distance-based malware detection methods.
     
        \item We propose \textit{four} scenarios for malware detection based on the nearest neighbor approach in which we use Hamming distance to find neighbors.
        
        \item We obtain the maximum achievable accuracy with the Hamming distance method as a threshold. We present the accuracy threshold calculation strategy in Section \ref{Testmetrics}.
        
        \item We evaluate the proposed malware detection methods using \textit{three} standard datasets: Drebin, Contagio, and Genome. Besides, by analyzing the time and space complexity, we performed a theoretical analysis to realize the scalability of our approach.
        
        \item We compare the proposed malware detection methods against the state-of-the-art methods applied for malware detection. At first, the proposed methods are compared to~\cite{Xiong2018}, which is Android malware detection based on a combination of clustering and classification. The next comparison solution in literature uses an entropy-based distance measure to detect malware~\cite{Radkani2017}. In the third comparison method~\cite{FanMing2018}, malware samples classify into different families, making it possible for each family to share the features of the samples in a better way. The main reason behind selecting such schemes for comparison is that our proposed methods and these cutting-edge solutions using similarity-based metrics for detecting malware. Moreover, the papers~\cite{FanMing2018} and ~\cite{Xiong2018} carry out their numerical validations in Drebin dataset in which we adapt our results on the same dataset.
        \end{itemize}

 \subsection{Roadmap}
    The remainder of the paper organizes as follows: We discuss related work in Section~\ref{relatedWork}. In Section \ref{Preliminaries} we study the preliminary essential malware analysis. Section~\ref{DetectionِDefinition} describes the distance calculation measures in binary representation, explore the detection strategies, our defined scenarios, designs our proposed architecture for malware detection systems and provides a toy scenario and delineates the proposed algorithms, while Section~\ref{resultAnalysis} presents the experimental results of our proposed scenarios. Section Section~\ref{discussion} reports the achievement of the experiment and provide some discussions regarding our method. Finally, in Section~\ref{conclusion} we summarize our research and provides future directions.

\section{Related Work}\label{relatedWork} 
Machine learning techniques use static, dynamic, and hybrid analysis methods to classify Android applications. In the following subsections, we introduce them. Also, we study some important researches in malware analysis and malware detection.

\subsection{Static analysis}
\par Some techniques using static permission features, such as Drebin ~\cite{arp2014drebin}, StormDroid~\cite{Chen2016}, and DroidSIFT~\cite{Zhang2014} which are applied on Android apps ~\cite{Distributiondashboard}.

\par The authors in~\cite{Fereidooni2016} propose a new detection system called \textit{ANASTASIA} to identify malicious samples using intents, permissions, system commands, and API calls features. ANASTASIA uses several classifiers by applying deep learning method and can extract several feature types from Android applications using the conditions of the app. Additionally, The authors in~\cite{Canfora2016} investigate Android apps to describe their resource usage and leverage the profiles to detect Android malware.

The authors in~\cite{Faruki2013} present an automatic signature generation approach called \textit{AndroSimilar} in which to detect malware for the static syntactic features in Android apps. Also, AndroSimilar can detect unintelligible malware with techniques such as junk method insertion, renaming method, string encryption, and changing control flow that can be used to evade fixed signatures working against malware. Besides, AndroSimilar can detect unknown types of existing malware. Also, the authors build an AndroSimilar generation approach based on digital forensics Similarity Digest Hash (SDHash) to distinguish similar documents. In SDHash, unrelated apps receive a lower probability of having standard features. Also, it helps to control false positive rates for two separate apps that share some features. Another method~\cite{Roussev2009} applies the same strategy to extract fixed-size byte-sequence features using their entropy values and searches for popular features and selects some of them using KNN strategy. 

\subsection{Dynamic analysis}
Dynamic solutions could run an Android app in a protected environment and provide all the required emulated resources to identify malicious activities. In literature, we find some implemented dynamic analysis methods -- however, they suffer from resource constraints of a smartphone. 
In another group, some papers concentrate on the behavioral class of the malware detection solutions. For example, in~\cite{Saracino2016}, the authors define the malware types based on their behavioral class. They propose a new scheme which identifies the misbehavior classes modified by each malware type by correlating the features extracted at \textit{four} different levels: kernel level, application level, user level, and package level. At the kernel level, their solution could monitor the system calls and hijacks them if any app triggers them. At the application level, it controls the critical APIs to detect the malicious behaviors posed by the apps such as the installation of new applications, requests for administrative privileges, generating too many processes, constant app monitoring on the active application. At the user level, they monitor user activities and detect malicious events when the user is idle or not interacting with the device. At the package level, they propose a new system to identify the risky applications under observation based on permissions requested by the app and market information. 

The fatal limitation of dynamic approaches is if they trigger with some non-trivial events, then they can miss some malicious execution path. For example, anti-emulation techniques such as Sandbox~\cite{Enck2009} and reference~\cite{Vidas2014} detection mechanism are unable to timely analyze the environment and lead to delaying the identifying malware and raise the evasion of the dynamic analysis methods.

\subsection{Hybrid analysis}
We can generate hybrid solutions when we apply static and dynamic approaches in the same time. Hybrid solutions can borrow the characteristics of static and dynamic solutions to improve malware detection strategies like DroidDetector~ \cite{Yuan2016}. DroidDetector could apply static and dynamic analysis usign deep learning to distinguish malicious software from normal applications. It uses permissions and sensitive API for static analysis. These static behaviors extract the features using TinyXml \cite{TinyXML}, 7-zip \cite{7-Zip}, and Baksmali tools \cite{Blasing2010}. After that DroidDetector dynamic features analysis using DroidBox tool.

Furthermore, Shanmugam~\cite{Shanmugam2013} propose an alternative distance for metamorphic malware. Their distance measurement solution is based on the opcode-based similarity method and simple encryption reported in ~\cite{Jakobsen1995}.  They use this distance measurement to classify malicious programs. The application, which is sufficiently similar to the metamorphic malware is classified as malicious. Some malware detection methods use Euclidean histogram distance metrics to compare two program files -- for example, Rad et al.~\cite{BashariRad2011} suggest that a histogram of opcodes can be used to detect metamorphic viruses. Some studies apply statistical methods to detect malware. For example, Toderici ~\cite{Toderici2013} use an analytical approach based on a chi-squared test to improve the hidden markov models Based malware detection. In another work, Ambra Demontis et al. ~\cite{Demontis2017} elaborate a solution to mitigate evasion attacks like malware data manipulation. In that paper, their method utilizes a secure SVM algorithm that can enforce its features to have evenly-distributed weight. 

\section{Preliminaries}\label{Preliminaries}

\begin{figure}[!htb]
\centering 
\includegraphics[width=0.95\columnwidth]{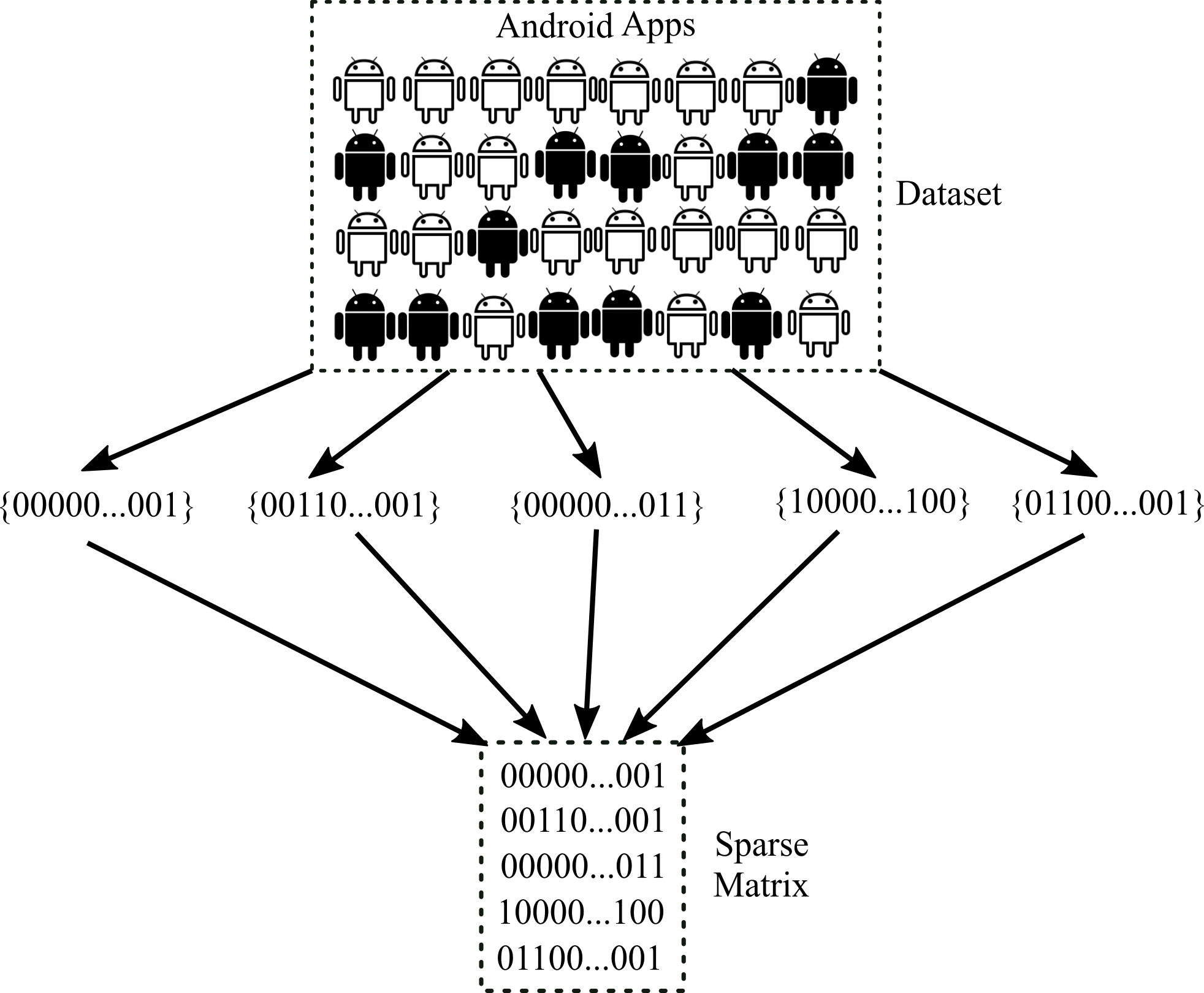}
\caption{\small State transition diagram for converting samples to sparse matrix.}
\label{fig:fig1}
\end{figure}

In this section, we review some of the essentials for malware analysis and how to model malware. In applied mathematics and computer science, a sparse matrix is a matrix in which most of the elements are zero. In Fig.~\ref{fig:fig1}, we use sparse matrix representation which contains important information related to Android app features such as APIs, permissions and intents.\\

In this study, we follow the general setting for designing a malware detection system that contains the benign $B$ and malware samples $M$. To do so, we select the performance evaluation settings and store a dataset that includes the labeled examples (i.e., with $n$ samples) and the $m$ elements for each sample. Hence, in equation~\eqref{eq:eq3} we have
\begin{equation}
\label{eq:eq3}\small
D=\{(x_i,y_i)\mid \forall i=1,\ldots,n\},
\end{equation}
where $x_i$ is the $i$-th malware sample vector of each component presents the selected feature; $y_i \in \{0, 1\}$ is the corresponding label of the features; $x_{ij}$ is the binary value of the $j$-th feature in $i$-th sample where $\{\forall j=1,\ldots,m\}$. Also, we can set $x_{ij} = 1$ if $x_i$ has the $j$-th component and $x_{ij} = 0$  otherwise;   $n$ is the total number of samples, and $X\subseteq \{0,1\}^m$ is an $m$-dimensional feature space.

\section{Proposed Approaches for Malware Detection System}\label{DetectionِDefinition}In this section, we first apply replacement method and prove that in the binary representation, Manhattan distance, Minkowski distance, and Hamming distance are the same. Then, due to the simplicity of computation, we use the Hamming distance method in the proposed detection algorithms. After that, we present our proposed architecture.  The main notations and symbols used in this paper are listed in Table~\ref{NotationTab}.

\begin{table}[!htpb]
\centering
\caption{\small   {Notation and symbols used in this paper.}\vspace{-5px}}
\label{NotationTab}
\scriptsize{
\setlength\tabcolsep{1.5pt} 
\begin{tabular}{|c|
>{\columncolor[HTML]{fcfcf4}}l|}
\hline
\rowcolor{LightCyan}
\textbf{Notations}  
&\multicolumn{1}{|c|}{\textbf{Description}}\\ \hline
\cellcolor[HTML]{E8E8AB}\textbf{$n$}& Number of Samples in Input Dataset  \\\hline
\cellcolor[HTML]{E8E8AB}\textbf{$m$}& Number of Features  in Each Sample\\\hline
\cellcolor[HTML]{E8E8AB}\textbf{$X$}& Input data ,  $\mid X \mid=n$ \\\hline
\cellcolor[HTML]{E8E8AB}\textbf{$x$}& A Sample from Input data , $x\in \{0,1\}^m$ \\\hline
\cellcolor[HTML]{E8E8AB}\textbf{$Y$}& Label of class in the classification problem, $y\in \{0,1\}$ \\\hline
\cellcolor[HTML]{E8E8AB}\textbf{$f$}& ML model, $f:X\rightarrow Y$ \\\hline
\end{tabular}}
\end{table}

\subsection{Equivalence of distance calculation measures in binary representation}
    In our paper, we introduce methods to identify malware samples from benign samples using the distance measure. Given the fact that the samples are binary vectors, the existence of a feature means a value of 1 and the absence of a feature means zero. The proposed method for computing the distance between the samples is to use the Hamming distance of the two vectors. On the other hand, it can easily be shown that in the binary mode of vectors, the result of using different criteria is to calculate the same distance.
    Suppose the binary vector $Y = [y_1, ..., y_n]$ is the most similar vector to $X = [x_1, ..., x_n]$. It means, $d(X,Y) \le d(X,Z)$, $\forall Z$. Several distance formulas apply to find the most similar vector to vector $X$. We list some of them as follows. 
    \begin{itemize}[leftmargin=*]
        \item  \textit{Taxicab distance} which is also called the \textit{Manhattan distance} presents in the equation~\eqref{eq:eq1}:
        
    
    \item \textit{Minkowski distance} presents as equation~\eqref{eq:eq2}.\\
     Since we need the most similar vector so we have equation~\eqref{eq:eq6}:\\
    \begin{equation}
    \label{eq:eq6}\small
    d_{1}(X,Y) \le d_{1}(X,Z), \quad \forall Z   
    \end{equation}
which is equivalent to:
    \begin{equation}
    \label{eq:eq7}\small
    \sum_{i=1}^{n}{|x_{i}-y_{i}}|) \le  \sum_{i=1}^{n}{|x_{i}-z_{i}}|).
    \end{equation}
    On the other hand, since our vectors are binary, so we have:
    \begin{equation}
    \label{eq:eq8}\small
     |x_{i}-z_{i}|=(|x_{i}-z_{i}|){^p}.
    \end{equation}
       {Different values of $p$ determine the application of this equation. For $p\geq 1$, the Minkowski distance is a metric as a result of the Minkowski inequality. Minkowski distance is typically used with p being 1 or 2, which corresponds to the Manhattan distance and the Euclidean distance, respectively. }\\
 
We can rewrite equation~\eqref{eq:eq6} using equation~\eqref{eq:eq8} as 
    \begin{equation}
    \label{eq:eq9}\small
    \sum_{i=1}^{p}\abs{x_{i}-y_{i}} \le \sum_{i=1}^{p}\abs{x_{i}-y_{i}}^{p}.
    \end{equation}

    Then, we have:
    \begin{equation}
    \label{eq:eq10}\small
    \sqrt[p]{\sum_{i=1}^{n}(\abs{x_{i}-y_{i}})^{p}} \le \sqrt[p]{\sum_{i=1}^{n}(\abs{x_{i}-z_{i}})^{p}}.
    \end{equation}

    and we can conclude
    \begin{equation}
    \label{eq:eq11}\small
    d_2 (X,Y) \le d_2 (X,Z),
    \end{equation}
The last equation imposes that the $X$ and $Y$ vectors result from each other. Formally speaking, we show that using the $d_{2}$ measure, vector $Y$ is the most similar vector to the binary vector $X$. 
  \end{itemize}

\subsection{Proposed architecture}

\begin{figure}[!htbp]
\centering 
\includegraphics[width=0.98\columnwidth]{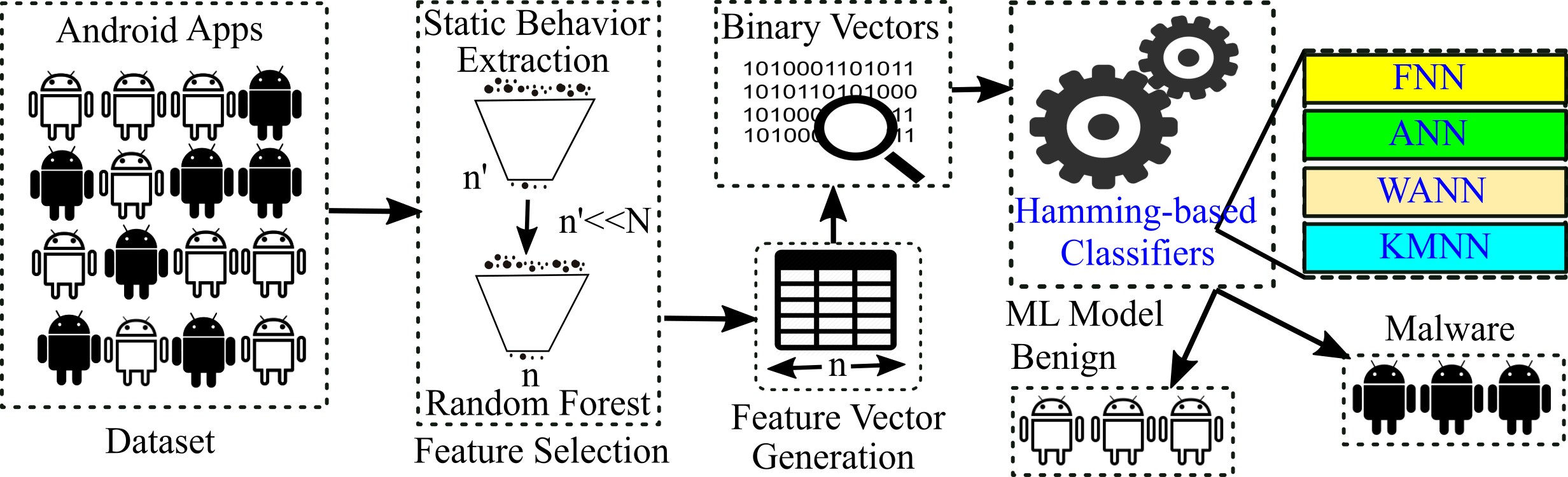}
\caption{\small Architecture overview of proposed method. ML= machine learning.   {$n$= number of features in each sample}}
\label{fig:fig2}
\end{figure}

In Fig.~\ref{fig:fig2}, we introduce our proposed architecture. In this figure, we select the static features of data samples (out of $N$ samples) in the dataset (see the rectangular feature selection component in Fig.~\ref{fig:fig2}). Then, using the Random Forest feature selection algorithm, we select the $\alpha$ percentage of features-- $n/n'=\alpha$. The $\alpha$ value will be in the following set.
\begin{equation}
\label{eq:eq125}\small
\alpha \in \{10,20,30,40,50,60,70,80,90,100\}
\end{equation}
For example, if $\alpha=10$, it means we select 10\% of features from feature selection component. After that, we convert the selected features of the samples to a vector. Then, we generate a binary vector for each sample by placing the value of 1 for each feature that exists in the sample and the value of 0 for each non-existent feature (see the Binary Vectors component in Fig.~\ref{fig:fig2}). Then, we generate our ML model using each of our proposed detection algorithms as classification algorithms based on Hamming distance similarities between the samples and use the ML model to detect malware among benign samples.

\subsection{Detection strategy and scenarios}

Measuring the similarity between samples is a significant operation in the classification algorithms.  Classifiers which use similarity strategy can estimate the label of a sample in test set based on the similarities between that sample and label of samples in a training set, and the pairwise similarities between the training samples. In the following, there are several ways to detect malware, which, despite the simplicity, represents a good result. Suppose that we want to find the most similar members (i.e., find the most similar vectors) of the train set to the vector $x$ which belongs to the test set. From the mathematical point of view, the element $y$ is the most similar to $x$ if we have equation~\eqref{eq:eq12}:
\begin{equation}
 \label{eq:eq12}\small
  d(x,y) \le d(x,z) \quad \forall z \in trainset
  \end{equation}
 In which, $d (x, y)$ represents the difference between $x$ and $y$, which is also called \textit{distance}. There are several methods to calculate the distance such as the Hamming distance, Minkowski distance and so on, which discussed earlier. Due to the binary nature of the elements (samples), we will show that the distance results of all these methods are similar, and therefore there is no ambiguity in selecting the specific method. For malware detection, we introduce several scenarios and present the results. To this end, we summarize each proposed malware detection method as follows:\\

 \noindent\textbf{FNN : First Nearest Neighbors--} In FNN, the first member of the training dataset is considered as the most similar member of the input data, and if a member is found to be more similar, the new member is considered as the most similar. The pseudocode of this method is shown Alg.~\ref{alg:algorithm}.\\ 
 
\begin{algorithm}[t]
\caption{Methods 1-4:  FNN, ANN, WANN, KMNN.}
\label{alg:algorithm}
\textbf{Input:} $test set$\quad  $X$\\
\textbf{Output:} 
 $Y$
\begin{algorithmic}[1]
\color{black}
\footnotesize
\For{$\forall x\in X$}
\Statex{\texttt{FNN:}}
\State{$MinX\leftarrow$ The first nearest sample from $train set$ to $X$}
\State{$y\leftarrow$ Label of $MinX$}
\Statex{\texttt{ANN:}}
\State{$NN\leftarrow$ All nearest samples from $train set$ to $X$}
\State{$y\leftarrow$ The most repeated label of samples in $NN$}
\Statex{\texttt{WANN:}}
 \State{$W\leftarrow$ Weight of each feature in $trainset$  $X$}
\State{$WX\leftarrow$ Weight of $trainset$ samples considering $W$}
\State{$WNN\leftarrow$ All weighted nearest samples from $trainset$ considering $WX$}
\State{$y\leftarrow$ The most repeated label of samples in $WNN$}
\Statex{\texttt{KMNN:}}
\State{$NN\leftarrow$ All nearest samples from $trainset$ to $x$}
\State{$C\leftarrow$ Run clustering on $NN$ and find two CHs}
\State{$d\leftarrow$ Compute sum of distance between each samples in $NN$ and CH}
\State{$D\leftarrow$ Sorted $d$}
\State{$D'\leftarrow$ Remove $10\%$ of the last samples in $ِِD$}
\State{$y\leftarrow$ Votes on the samples which are in $D'$}
\Statex{\texttt{Add $y$ to the $Y$}}
\EndFor
\State{\textbf{return} $Y$}
\end{algorithmic}
\end{algorithm}

\noindent\textbf{ANN: All Nearest Neighbors --}
     ANN is similar to the FNN, with the difference that at each stage all similar neighbors are stored and all of these elements are involved in the conclusion process. The pseudocode of this method is as Alg.~\ref{alg:algorithm}. Note that in this method, the voting process is based on the population of the labels, and the features of the malware will have no effect on the decision-making process. This issue is discussed further.\\ 
    

\noindent\textbf{WANN: Weighted All Nearest Neighbors --}
    In WANN, the importance of features is  examined. For this purpose, first, a variable $W$ is defined, each element of which holds the percentage of the frequency of its corresponding feature. Similar to the previous method, in this method, all the neighbors are firstly calculated with the input element $x$ and among them, we store elements whose features are close to $x$ according to the frequency of the features. In this case, if we find several similar elements, we will take the voting process. In this method, the probability of the features is also effective in the voting process.\\



    



    
\noindent\textbf{KMNN: K-Medoid based Nearest Neighbors --} K-Medoid clustering method is a type of K-means that can be more robust to noises and outliers. Medoid is the central point of the cluster, which is an actual point of the cluster and has the minimum distance to other points  s~\cite{Jin12011}. This scenario is a combination of KNN and K-medoids. It is similar to previous scenarios, with the difference that the label recognition process is based on the closest nearest neighbor. First, the most similar neighbors are calculated using the second scenario (i.e., the scenario is based on finding all the same neighbors with the same similarity). Then, the neighboring set is divided into two clusters. In each cluster, one of the samples, which has the smallest distance with other samples, is selected as the cluster head. Then, the distance to each of the samples is calculated from the cluster heads and sort in terms of distance. In the last step, ten percent of the samples, which have more distance than the clusters, are ignored and voted between the rest of the remaining samples to obtain the label for the sample $x$. The reason for this clustering is that one of the clusters is likely to represent malware and another to represent benign software. After that, a cluster head is calculated for each of these clusters and used to determine the probable unrelated neighbors (outlier data). To do this, we consider $k$ percent of the neighbors with the most distance to these clusters as the outlier data, and remove them from the list of neighbors. In this paper, we consider $k = 10$. Finally, similar to the previous scenarios, the voting process is performed, and the test data label is determined. The process is presented as Alg.~\ref{alg:algorithm}.\\

  {To better understand the proposed methods, we use a toy example presented in appendix~\ref{appendix1} that clearly outlines the algorithms.}
    

\subsection{Time complexity of the detection algorithms}\label{ComplexityofDetection}

In following section, we conduct time complexity analysis on the presented detection methods. Hence, we detail the time complexity of our proposed methods as follows:

\begin{itemize}[leftmargin=*]
  \item \textbf{FNN.} In FNN, we first obtain the distance between each sample and other samples in the training dataset. Then, we aim to select the first nearest sample. Assuming that in the training dataset we have $n$ samples and each sample has $m$ features, the time complexity of finding the first similar sample in the worst case will be $\mathcal{O}(n\times m)$.
  
 \item \textbf{ANN.} In ANN, similar to FNN, we first obtain the distance between each sample and other samples in the training dataset. The time complexity of finding the most similar samples in the worst case is $\mathcal{O}(n\times  m)$. The next step in the ANN algorithm is voting on all similar samples. Suppose that we have $k$ similar samples. The time complexity of this step also is $\mathcal{O}(n\times  k)$, where $k<n$. As a result, the time complexity of ANN algorithms is equivalent $\mathcal{O}(n\times  m)+\mathcal{O}(n\times  k)$=$\mathcal{O}(n\times m)$.\\

\item \textbf{WANN.} The first step in the WANN algorithm is finding the vector $W$, which indicates the weight of the features. To compute the weight of the features, we calculate the abundance of features in the training dataset and divide it to the number of samples. Assuming that there are $n$ samples in the training dataset, and each sample contains $m$ features. The duration takes to find the vector $W$ in training dataset is in the order of $\mathcal{O}(n\times m)$. The next step in this method is similar to the previous methods and includes finding all similar instances and voting between them. In this way, the time complexity of the WANN method is $\mathcal{O}(n\times  m)+\mathcal{O}(n\times  m)+\mathcal{O}(n\times k)=\mathcal{O}(n\times m)$.

\item \textbf{KMNN.} In KMNN method, K-Medoid is used with the nearest neighbor method. Assuming that $k$ is the number of clusters which anyone has $c_i$ elements, the time complexity for this algorithm is about $\mathcal{O}(n\times m)+\mathcal{O}(i\times k \times  {c_i}^2\times m)$, that $i$ is the number of algorithmic repetitions to achieve the optimal answer. Given that only two clusters are chosen, $k$ is equal to 2, and we set  $i=10$. Therefore, the time complexity of this part is $\mathcal{O}(n^2\times m)$. In the second part of this algorithm, the distance of the selected samples with the CHs are calculated and these distances are sorted. The time complexity of this part is $\mathcal{O}(n\times log(n))$. Therefore, the time complexity of the KMNN algorithm is $ \mathcal {O} (n ^ 2\times m) $. 
\end{itemize}
In calculating the time complexities of proposed methods, we estimate duration takes for finding the distance between two arbitrary vectors $X$ and $Y$ with $m$ features. In worse case, we consider the samples $X$ and $Y$ as binary vectors with length $m$ and compare the elements of them by computing Hamming distance between the similar entry of vectors. In implementation, we present examples in the form of sparse collections. In this case, we can apply the following mathematical equation to calculate the distance between two sets of $X$ and $Y$ vectors as follows:
\begin{equation}
    \label{eq:eq100}\small
       D(X,Y)=\#((X\cup Y)-(X\cap Y))
    \end{equation}
In this regard,the symbol (\#) presents the number of members for each set. We confirm that this mathematical equation provide more accurate results. For example, In the tested dataset in this paper, in the worst case, we have $m = 21,492$ features per sample, while using the above equation, the distance will be at most the order of 925 simple calculations.

\section{Experimental Evaluation}\label{resultAnalysis}
In this section, we report an experimental evaluation of the proposed clustering algorithms by testing them under different scenarios. 

\subsection{Simulation setup}\label{sec:secsetup}
In the following, we describe the datasets, mobile application static features, test metrics, and comparison solutions' tuning.

\subsubsection{Datasets}
We conduct our experiments on \textit{three} datasets which are explained below:
\begin{itemize}[leftmargin=*]
\item \textit{Drebin dataset:} The Drebin dataset is a Android example collection that we can apply directly. The Drebin dataset includes 118,505 applications/samples from various Android sources~\cite{arp2014drebin}. 

\item \textit{Genome dataset:} The genome project is supported by the National Science Foundation (NSF) of the United States. From August 2010 to October 2011, the authors collected about 1,200 samples of Android malware from different categories as a \textit{genome} dataset~\cite{jiang2012dissecting}.  

\item \textit{Contagio dataset:} it consists of 11,960 mobile malware samples and 16,800 benign samples~\cite{Contagio}.
\end{itemize}

\subsubsection{Static features}
In this paper, we consider various malicious sample features like permissions, APIs and intents. We summarize them as follows:
\begin{itemize}[leftmargin=*]
    \item \textit{Permission:} permission is a essential profile of an Android  application (apk) file that includes information about the application. The Android operating system processes these permission files before installation.  
    
    \item \textit{API:} API feature monitors various calls to APIs on an Android OS, such as sending SMS or accessing a user's location or device ID.
    
    \item \textit{Intent:} Intent feature applies to represent the communication between different components which is known as a \textit{medium}.
\end{itemize}
\subsubsection{Parameter setting}
  {Due to a large number of features, we first ranked the features using the \textit{RandomForestRegressor} algorithm. Then, we repeat our experiments for \{10\%, 20\%, 30\%, 40\%, 50\%, 60\%, 70\%, 80\%, 90\%, 100\%\} of the manifest features with higher ranks to determine the optimal number of features for modification in each method. The evaluation of a model skill on the training dataset would result in a biased score. Therefore the model is evaluated on the held-out sample to give an unbiased estimate of model skill. This is typically called a train-test split approach to algorithm evaluation. In this paper,  at each test, we randomly consider 60\% of the dataset as training samples, 20\% as validation samples and 20\% as testing samples. We repeated this operation ten times for each algorithm and averaged the results.Each of these 10 sets of train, validation and test were generated using non-duplicate seed. Table~\ref{trainvalidationtest} shows the accuracy of the proposed methods in this paper on train, validation and test data. We run our experiments on an eight-core Intel Core i7 with speed 4 GHz, 16 GB RAM, OS Win10 64-bit.}

  \begin{table}[!htb]
  \centering
  \caption{\small   {Comparing accuracy of Algorithms without feature selection on train-validation and test data.}\vspace{-5px}}
\label{trainvalidationtest}
  \scriptsize{
\setlength\tabcolsep{0.5pt} 
  \begin{tabular}{|c|c|c|c|c||c|c|c||c|c|c|}
  \hline
  \rowcolor{LightCyan} 
  \multicolumn{11}{|c|}{\textbf{train-validation-test}}\\\hline\hline
\rowcolor{yellow}
   &&\multicolumn{3}{c|}{\textbf{Drebin Dataset}}&\multicolumn{3}{c|}{\textbf{Contagio Dataset}}&\multicolumn{3}{c|}{\textbf{Genome Dataset}}\\\hline
   \rowcolor{maroon!10}
    &\textbf{Algor..}&\textbf{train}&\textbf{valid.}&\textbf{test}&\textbf{train}&\textbf{valid.}&\textbf{test}&\textbf{train}&\textbf{valid.}&\textbf{test}\\
 \hline
 \cellcolor[HTML]{FAEC0D}&\cellcolor[HTML]{E8E8AB}FNN&99.35&99.29&99.36&99.06&99.05&99.06&	99.87&	99.84&	99.88\\\cline{2-11}
 \cellcolor[HTML]{FAEC0D}&\cellcolor[HTML]{E8E8AB}ANN&99.43&	99.47&	99.48&	99.26&99.27&99.27&99.88&	99.86&99.88\\\cline{2-11}
\cellcolor[HTML]{FAEC0D}&\cellcolor[HTML]{E8E8AB}WANN&99.33&99.35&99.33&99.08&	99.08&	99.06&99.83&	99.81&	99.82\\\cline{2-11}
\cellcolor[HTML]{FAEC0D}&\cellcolor[HTML]{E8E8AB}KMNN&99.26&99.25&99.26&99.17&99.11&99.18&	99.69&	99.68&99.70\\\cline{2-11}
\cellcolor[HTML]{FAEC0D}&\cellcolor[HTML]{E8E8AB}PDME&99.31&99.28&99.28&98.95&	98.92&98.92&99.55&	99.54&99.55\\\cline{2-11}
\cellcolor[HTML]{FAEC0D}&\cellcolor[HTML]{E8E8AB}FalDroid&94.54&94.49&94.54&93.58&93.58&93.58&	98.38&	98.37&98.38\\\cline{2-11}
\cellcolor[HTML]{FAEC0D}&\cellcolor[HTML]{E8E8AB}MAA&99.75&99.72&99.74&	99.42&	99.42&	99.41&	99.88&	99.87&	99.88\\\cline{2-11}
\cellcolor[HTML]{FAEC0D}\multirow{-5}{*}{\begin{sideways}{\textbf{API}}\end{sideways}}&\cellcolor[HTML]{E8E8AB}RF&99.23&	99.03&	99.17&	98.93&	98.92&98.82&	99.67&	99.66&	99.67\\\cline{2-11}
\cellcolor[HTML]{FAEC0D}&\cellcolor[HTML]{E8E8AB}SVM&98.89&	99.01&	98.95&	98.31&98.3&98.3&	99.6&	99.6&	99.61\\\cline{2-11}
\cellcolor[HTML]{FAEC0D}&\cellcolor[HTML]{E8E8AB}DT&98.16&	98.11&	98.14&	98.38&	98.39&	98.39&	99.26&	99.25&	99.25\\\cline{2-11}
\cellcolor[HTML]{FAEC0D}&\cellcolor[HTML]{E8E8AB}NN&91.29&	91.19&	91.20&	98.36&98.35&	98.36&	99.78&	99.76&	99.79\\
\hline\hline

\cellcolor[HTML]{FAEC0D}&\cellcolor[HTML]{E8E8AB}FNN&96.91&	96.72&	96.83&	98.94&	98.68&	98.77&	99.69&	99.46&	99.43\\\cline{2-11}
\cellcolor[HTML]{FAEC0D}&\cellcolor[HTML]{E8E8AB}ANN&98.12&	98.03&	98.02&	99.28&	99.01&	99&	99.54&	99.48&	99.49\\\cline{2-11}
\cellcolor[HTML]{FAEC0D}&\cellcolor[HTML]{E8E8AB}WANN&98.74&98.51&	98.45&	98.97&	98.76&	98.80&	99.56&99.39&99.37\\\cline{2-11}
\cellcolor[HTML]{FAEC0D}&\cellcolor[HTML]{E8E8AB}KMNN&97.99&97.86&97.88&99.03&98.83&98.83&	99.61&	99.43&99.43\\\cline{2-11}
\cellcolor[HTML]{FAEC0D}&\cellcolor[HTML]{E8E8AB}PDME&98.05&97.91&	97.92&98.97&98.78&	98.77&98.98&	98.88&	98.89\\\cline{2-11}
\cellcolor[HTML]{FAEC0D}&\cellcolor[HTML]{E8E8AB}FalDroid&70.23&70.09&	70.03&	88.81&88.43&	88.44&	92.18&	91.82&91.81\\\cline{2-11}
\cellcolor[HTML]{FAEC0D}&\cellcolor[HTML]{E8E8AB}MAA&99.64&99.45&	99.48&	99.64&	99.51&	99.53&	99.91&	99.85&	99.85\\\cline{2-11}
\cellcolor[HTML]{FAEC0D}\multirow{-5}{*}{\begin{sideways}{\textbf{Permissions}}\end{sideways}}&\cellcolor[HTML]{E8E8AB}RF\%&85.91&	85.76&	85.78&	98.39&	98.16&	98.07&	95.41&	95.09&	95.10\\\cline{2-11}
\cellcolor[HTML]{FAEC0D}&\cellcolor[HTML]{E8E8AB}SVM&87.59&	87.46&	87.45&	97.14&	97.02&97.05&94.65&	94.56&94.57\\\cline{2-11}
\cellcolor[HTML]{FAEC0D}&\cellcolor[HTML]{E8E8AB}DT&86.48&	86.31&86.33&	99.75&	97.67&	97.66&	95.13&	94.95&	94.96\\\cline{2-11}
\cellcolor[HTML]{FAEC0D}&\cellcolor[HTML]{E8E8AB}NN&82.51&	82.32&	82.38&	97.43&	97.16&	97.16&	94.97&	94.71&	94.72\\\cline{2-11}
\hline\hline
\cellcolor[HTML]{FAEC0D}&\cellcolor[HTML]{E8E8AB}FNN&90.41&	90.29&	90.29&	98.93&	98.79&	98.77&	99.17&	89.99&99.01\\\cline{2-11}
\cellcolor[HTML]{FAEC0D}&\cellcolor[HTML]{E8E8AB}ANN&99.91&	91.78&	91.79&	99.29&99.11&	99.09&	99.41&	99.23&	99.22\\\cline{2-11}
\cellcolor[HTML]{FAEC0D}&\cellcolor[HTML]{E8E8AB}WANN&99.39&92.21&	92.19&	99.18&99.06&	99.06&	99.32&	99.16&	99.16\\\cline{2-11}
\cellcolor[HTML]{FAEC0D}&\cellcolor[HTML]{E8E8AB}KMNN&92.09&91.92&91.91&	99.11&	98.9&98.89&	99.21&	99.13&99.13\\\cline{2-11}
\cellcolor[HTML]{FAEC0D}&\cellcolor[HTML]{E8E8AB}PDME&91.97&91.79&91.82&	99.04&98.82&	98.83&	99.04&	98.86&	98.89\\\cline{2-11}
\cellcolor[HTML]{FAEC0D}&\cellcolor[HTML]{E8E8AB}FalDroid&91.24&91.15&	91.15&88.91&88.61&	88.62&	85.22&85.09&	85.11\\\cline{2-11}
\cellcolor[HTML]{FAEC0D}&\cellcolor[HTML]{E8E8AB}MAA&99.69&99.52&	99.52&	99.82&	99.64&	99.62&	99.91&	99.93&	99.91\\\cline{2-11}
\cellcolor[HTML]{FAEC0D}\multirow{-5}{*}{\begin{sideways}{\textbf{intents}}\end{sideways}}&\cellcolor[HTML]{E8E8AB}RF\%&99.19&	92.04&	92.03&	98.75&	98.56&	98.57&	96.97&	96.82&	99.82\\\cline{2-11}
\cellcolor[HTML]{FAEC0D}&\cellcolor[HTML]{E8E8AB}SVM&91.21&	91.07&	91.08&99.01&	98.91&	98.89&	97.71&	97.44&	99.45\\\cline{2-11}
\cellcolor[HTML]{FAEC0D}&\cellcolor[HTML]{E8E8AB}DT&91.23&	91.16&	91.17&	98.94&	98.83&	98.80&	97.31&	97.18&	97.12\\\cline{2-11}
\cellcolor[HTML]{FAEC0D}&\cellcolor[HTML]{E8E8AB}NN&88.89&88.81&88.81&	88.89&98.75&98.74&	95.82&	95.68&	95.68\\\cline{2-11}
\hline
  \end{tabular}}
  \end{table}

\subsubsection{Comparison of solutions}
We compare our proposed algorithms against the corresponding ones of some state-of-the-art classification and malware detection algorithms, namely joint solution~\cite{Xiong2018}, the program dissimilarity measure based on entropy (PDME)~\cite{Radkani2017} and the FalDroid~\cite{FanMing2018} algorithms. In detail, two detection approaches are proposed in~\cite{Xiong2018}. In the first method, we consider a random member called $k$ from train set as a CH and divide the train set into $k$ clusters. Then, we calculate the distance between all the elements of each cluster and consider the element that minimizes the total distance from all members as a new CH. All members are re-clustered using a new cluster. We repeat this process and change the elements of the clusters in each step. Once the end cluster has been identified in the last replication, these clusters are used to identify the label. The label of the element (test element) is considered to be the label of the cluster label that has the smallest distance with the desired element. In the event that several clusters are found with this feature, they will be voted on. The second method presented in~\cite{Xiong2018} is very similar to the first method. The only difference between the first and second methods is that we divide the train set into \textit{two} parts: malware and benign. We now run the first method for both malware and benign separately. Therefore, the entire train set in the second method is divided into $2k$ clusters. In the following, similar to the first method, the test element spacing is calculated with all the clusters of the malware and benign sets. Finally, the labels are select by voting between the nearest clusters of malware and benign sets.

\par In~\cite{Radkani2017}, the proposed method is based on entropy. First, all neighbors with the least distance from the sample are calculated and voted between them. The only difference in this method is calculating distance. In this method, using the concept of entropy, which is one of the most famous concepts of information theory, the distance between the two samples $S_1$ and $S_2$ is calculated as equation~\eqref{eq:eq13}:
\begin{equation}
 \label{eq:eq13}\small
  d_{En}(S_1, S_2)=1-Sim(S_1, S_2)
  \end{equation}
  In which, the $Sim (S_1, S_2)$ is a similarity measure for two samples and is computed by \eqref{eq:eq14}:
\begin{equation}
 \label{eq:eq14}\small
 Sim(S_1,S_2)=\frac{En(S_1)+En(S_2)-En(S_1\cup S_2)}{\max\left(En(S_1),En(S_2),En(S_1\cup S_2)\right)}
  \end{equation}
In this definition, $En(S_1)$ represents an entropy that indicates the probability of the random variable $S_1$. In this paper, $S_1$ is considered as an $j$-{th} binary vector with a maximum of $m$ features. The entropy is calculated as equation \eqref{eq:eq15}, in which $p(S_{i})$ is the probability of $i$-th feature occurrence of the vector $S_1$:
 \begin{equation}
 \label{eq:eq15}\small
  En(S_1)=-\sum_{i=1}^{m}p(S_i)\log(p(S_i)))
  \end{equation}
 
In our paper, we use FalDroid as a method for classifying the samples and compare their results against our proposed methods. Obviously, other innovations expressed in~\cite{FanMing2018} are not recovered.

\subsection{Test metrics}\label{Testmetrics}
In the ML-based malware detection methods the confusion matrix is compute and then performance metrics are calculated using the confusion matrix. The confusion matrix contains the following parts:
\begin{itemize}[leftmargin=*]
\item\textit{True Positive ($\tau$):}{ the number of correctly classified samples that belongs to the class.}
\item\textit{True Negative ($\delta$):}{ the number of correctly classified samples that do not belong to the class.}
\item\textit{False Positive ($\rho$):} {the number of samples which were incorrectly classified as belonging to the class.}
\item \textit{False Negative ($\mu$):} {the number of samples which were not classified as class samples.}

\item\textit{Accuracy:} This metric is described as equation \eqref{eq:eq16}:
\begin{equation}
\label{eq:eq16}\small
Accuracy=\frac{\tau+\delta}{\tau+\delta+\rho+\mu}
\end{equation}

\item\textit{Precision:} This is the fraction of relevant samples between the retrieved samples. Equation \eqref{eq:eq17} shows how to calculate this metric:
\begin{equation}
\label{eq:eq17}\small
Precision=\frac{\tau}{\tau+\rho}
\end{equation}

\item\textit{Recall:} This Recall is defined as equation \eqref{eq:eq18}:
\begin{equation}
\label{eq:eq18}\small
Recall=\frac{\tau}{\tau+\mu}
\end{equation}
\item \textit{f1-Score:} The harmonic mean of precision and recall defines as F1-score which we describe it in equation \eqref{eq:eq182}:
\begin{equation}
\label{eq:eq182}\small
f1-Score=\frac{1}{\frac{1}{Recall}+\frac{1}{Precision}}=2*\frac{Precision*Recall}{Precision+Recall}
\end{equation}

\item\textit{False Positive Rate (FPR):} The FPR is computed as the ratio between the number of negative events incorrectly classified as positive (false positives) and the total number of actual negative events. This metric is described as equation \eqref{eq:eq19}:
\begin{equation}
\label{eq:eq19}\small
FPR=\frac{\rho}{\tau+\delta}
\end{equation}
\item\textit{Area Under Curve (AUC):} This metric is a method for determining the best model for predicting the class of samples using all thresholds. That is, AUC measures the trade off between misclassification rate and FPR. This metrics can be calculated as \eqref{eq:eq20}:
 \begin{equation}
\label{eq:eq20}\small
AUC=\frac{1}{2}\left(\frac{\tau}{\tau+\rho}+\frac{\delta}{\delta+\rho}\right)
\end{equation}
\item\textit{Receiver Operating characteristic Curve (ROC):} ROC a graphical plot that illustrates the detection ability of a binary classifier system as its discrimination threshold is varied.
\end{itemize}
In the similarity-based methods which are proposed in this paper, for each sample in the test set, we opt the samples that are the most similar to the test sample. Afterward, we determine the label of each sample using the labels of samples in the most relevant neighbors. We determine it when the main label of the test sample is not the same as the neighboring set. As an example, let us assume that the test element has a label of 1, and all the elements in the nearest neighbors are labeled 0. If we select the first element that is located in the nearest neighbor or select it based on polling which is conducted between neighbors, then, in both cases the 0 labels will suggest for the sample incorrectly. We can use this strategy to calculate the upper-bound and the lower-bound for accuracy in a similarity-based method. Therefore, we consider this ranges in calculating the maximum accuracy of similarity methods and add them to the Tables~\ref{tab5}-\ref{tab7} as the \textit{last three columns}. Formally speaking, suppose that we want to calculate the maximum value of $F\triangleq\frac{x}{x+y}$, hence, we have
\begin{equation}
\label{eq:eq195}\small
F=\frac{x}{x+y}=\frac{x}{x(1+\frac{y}{x})}=\frac{1}{1+(\frac{y}{x})}
\end{equation}
To maximize the expression $\frac{x}{x+y}$, we just need to minimize $1+\frac{y}{x}$. Therefore, by calculating the minimum $y$ and the maximum $x$ we can obtain the maximum $\frac{x}{x+y}$ expression. Regarding the method of calculating the accuracy presented in equation~\eqref{eq:eq16}, the following substitution defines as below:
\begin{equation}
\label{eq:eq196}\small
\tau+\delta   \longrightarrow  x  
\end{equation}
\begin{equation}
\label{eq:eq197}\small
\rho+\mu    \longrightarrow y
\end{equation}
Where maximizing the value of accuracy depends on minimizing the summation of $\mu$ and $\rho$ and maximizing the summation of the value of $\tau$ and $\delta$. Therefore, in the case of the maximum value for $\tau + \delta$ and minimum value for $\rho + \mu$, the accuracy maximizes. To obtain the maximum value of $\tau + \delta$ and the minimum $\rho + \mu$ value, we can write
 \begin{equation}
\label{eq:eq183}\small
MAX(\tau+\delta)=\frac{\# X \in testset}{\# testset}\ \textbf{where} \  \text{Real Label} (x) \in \text{ANN($X$)}
\end{equation}
\begin{equation}
\label{eq:eq184}\small
MIN(\rho+\mu)=\frac{\# X \in testset}{\# testset}\ \textbf{where} \ \text{Real Label} (x) \notin \text{ANN($X$)}
\end{equation} 
Similarly, the minimum FPR and the maximum AUC can be calculated. As recall it, we report these cases in Tables~\ref{tab5}-\ref{tab7} as MAA.

\subsection{Experimental results\label{results}}

In this section, we apply the proposed methods for detecting malware on three datasets, \textit{Drebin}, \textit{Contagio} and \textit{Genome}, and compare the results with three of the state-of-art researches.
	\begin{figure*}[!htb]
    \centering
	\begin{subfigure}{0.32\textwidth}
 		\centering 
 		\includegraphics[width=1.05\linewidth]{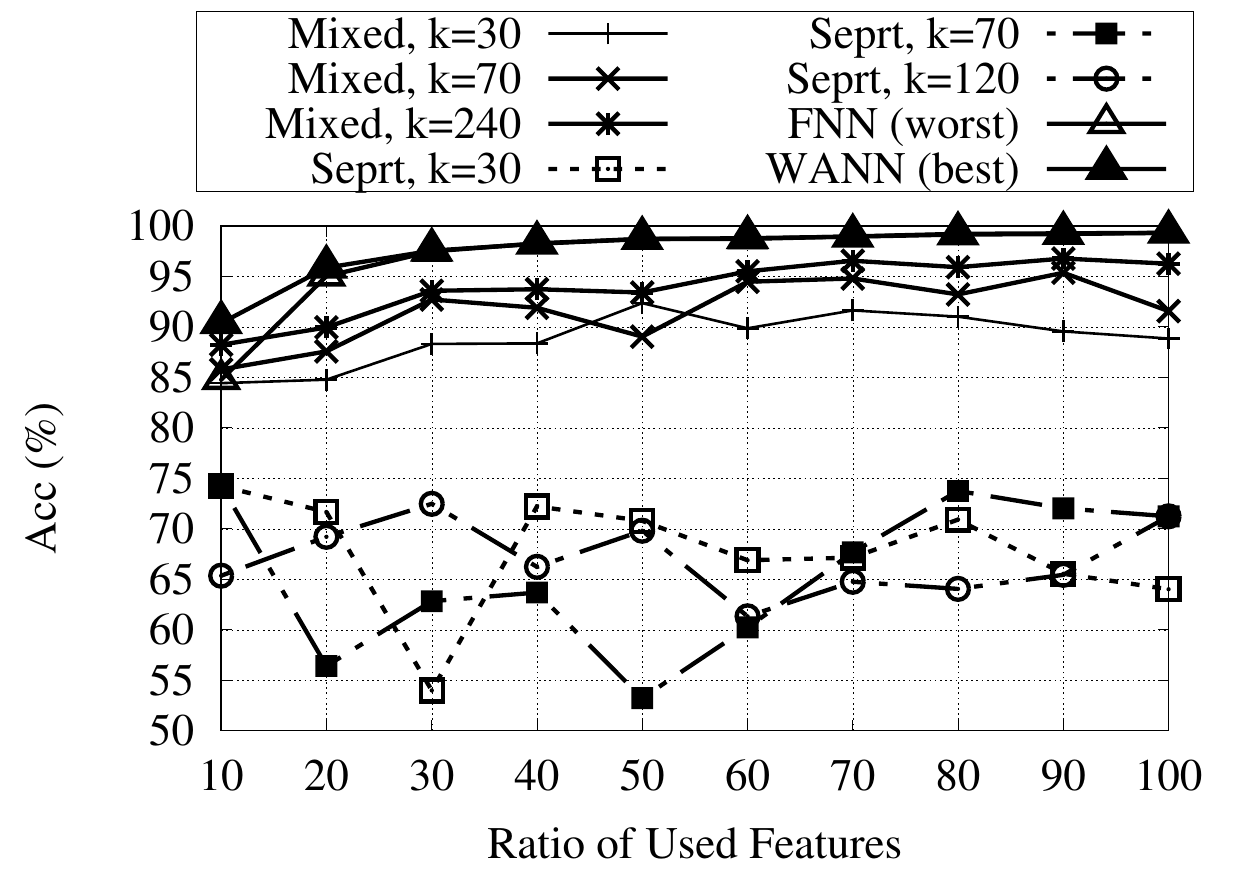}
			\caption{\small Drebin-API}
			\label{fig:fig4a}
 	\end{subfigure} 
   \begin{subfigure}{0.32\textwidth}
 		\centering
 		\includegraphics[width=1.05\linewidth]{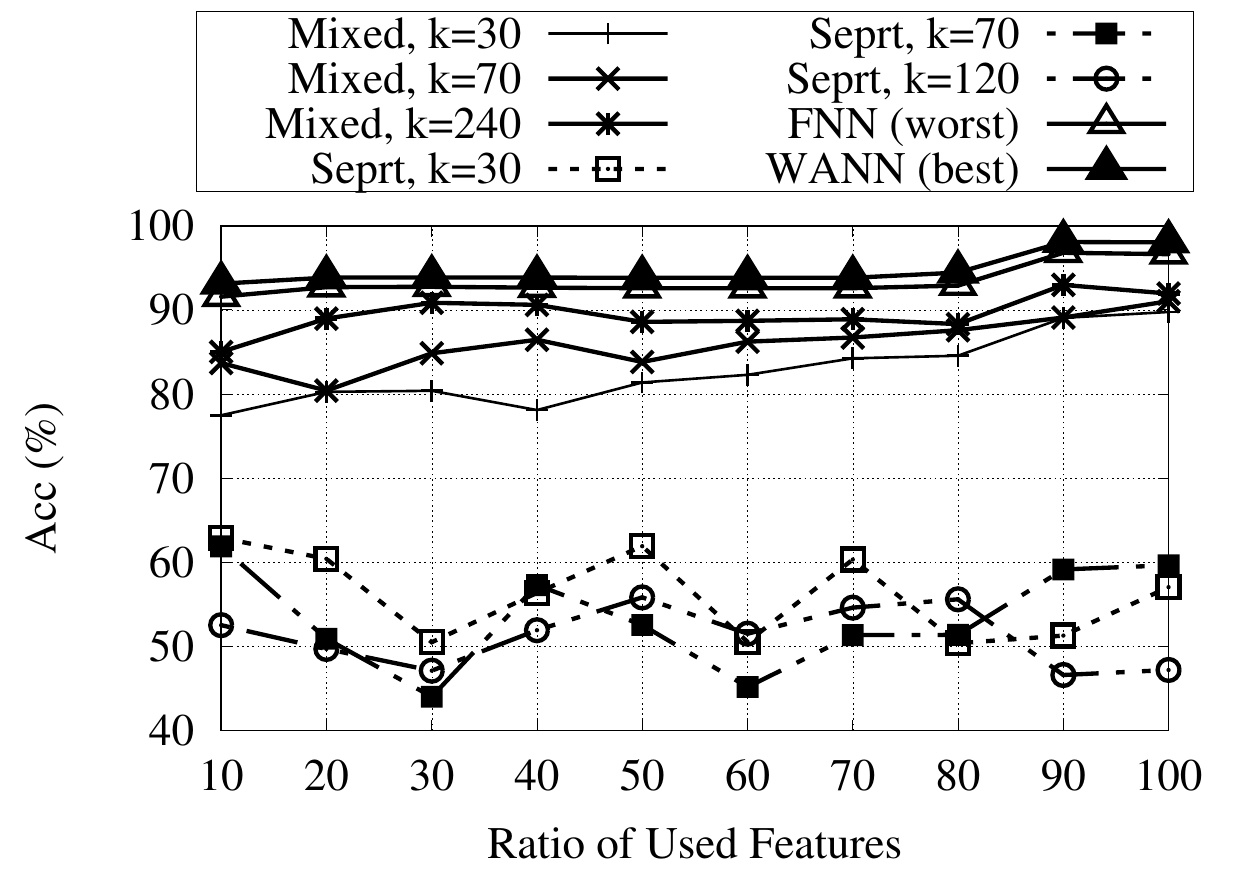}
			\caption{\small Drebin-permission}
			\label{fig:fig4b}
 	\end{subfigure} 
    \begin{subfigure}{0.32\textwidth}
 		\centering 
 		\includegraphics[width=1.05\linewidth]{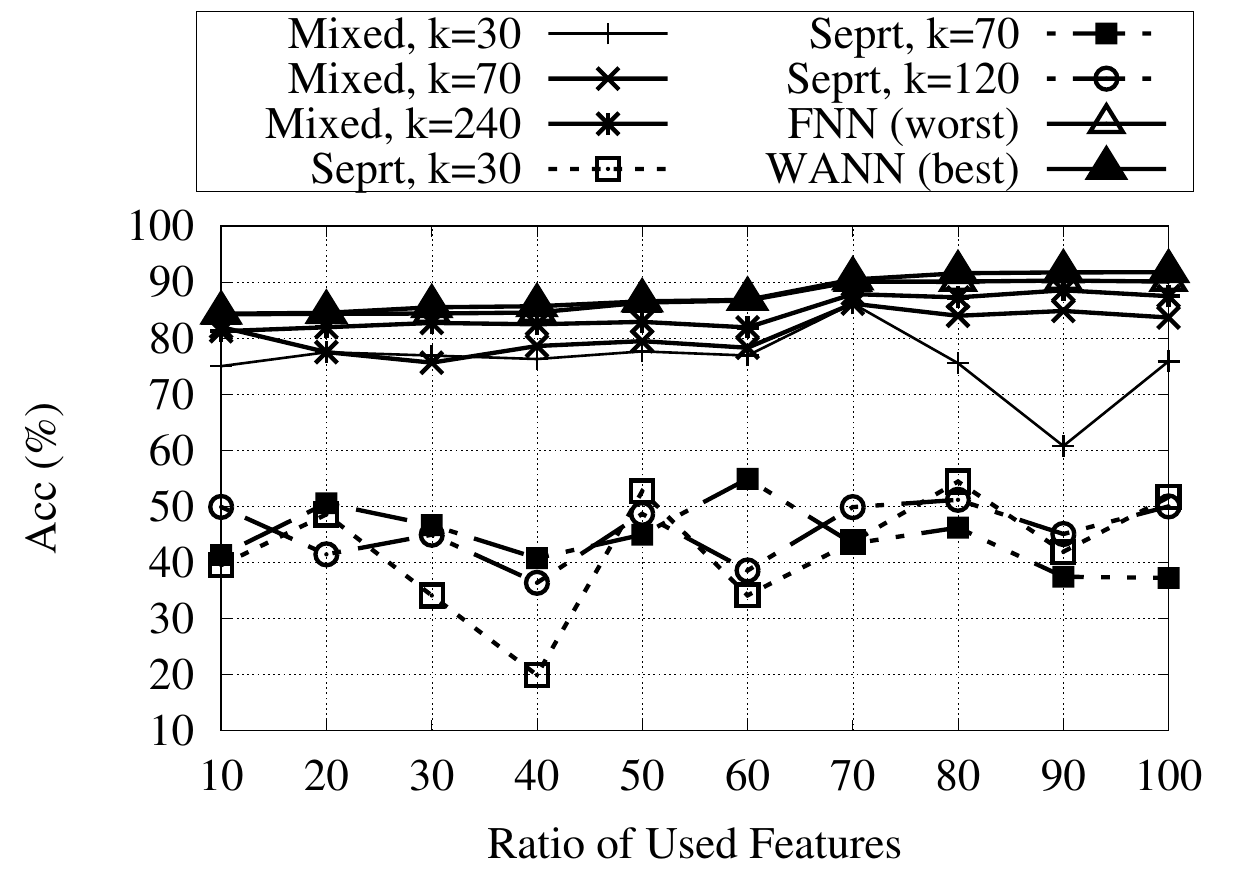}
		\caption{\small Drebin-intent}
			\label{fig:fig4c}
 	\end{subfigure}
 		\hfill
 	\begin{subfigure}{0.32\textwidth}
 		\centering
 		\includegraphics[width=1.05\linewidth]{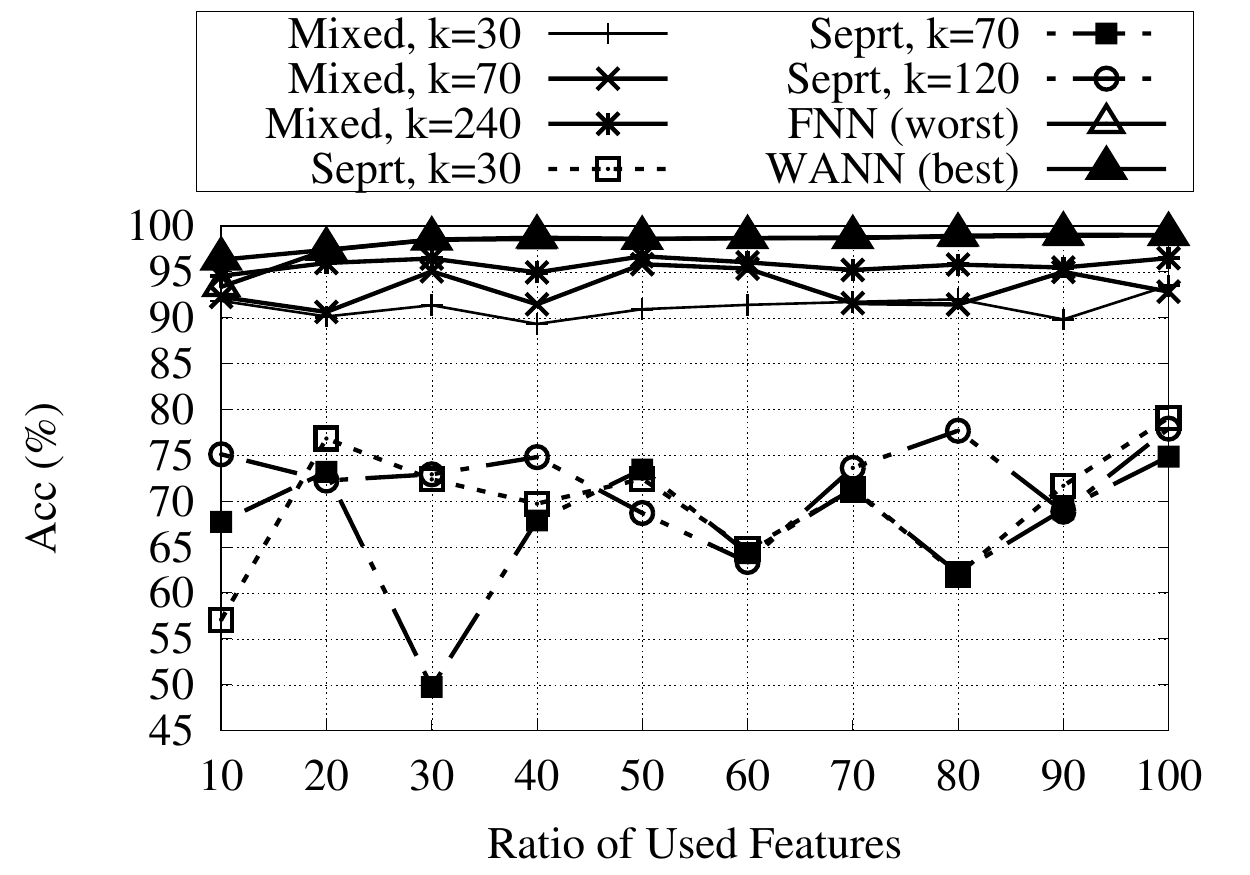}
			\caption{\small Contagio-API}
			\label{fig:fig4d}
 	\end{subfigure}  
    \begin{subfigure}{0.32\textwidth}
 		\centering 
 		\includegraphics[width=1.05\linewidth]{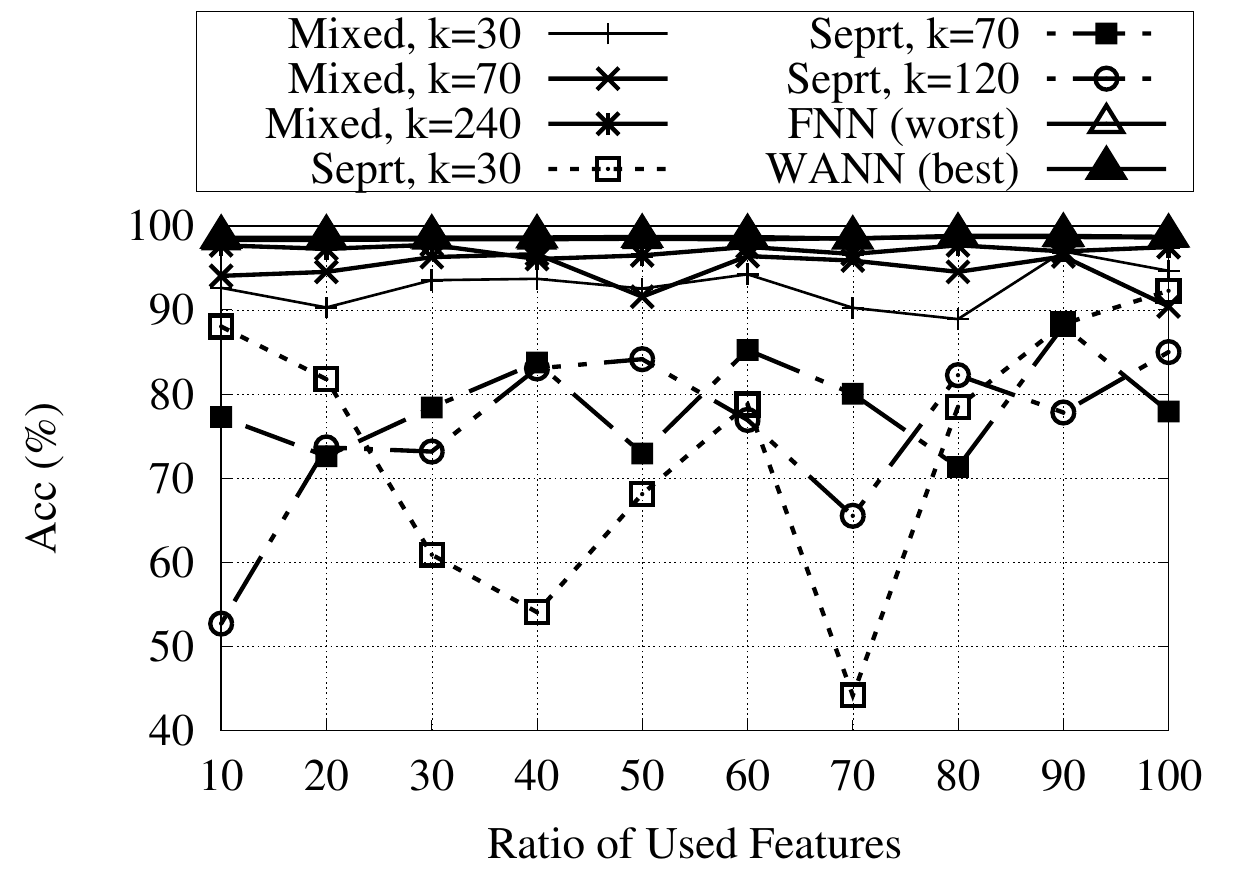}
			\caption{\small Contagio-permission}
			\label{fig:fig4e}
 	\end{subfigure}
   \begin{subfigure}{0.32\textwidth}
 		\centering
 		\includegraphics[width=1.05\linewidth]{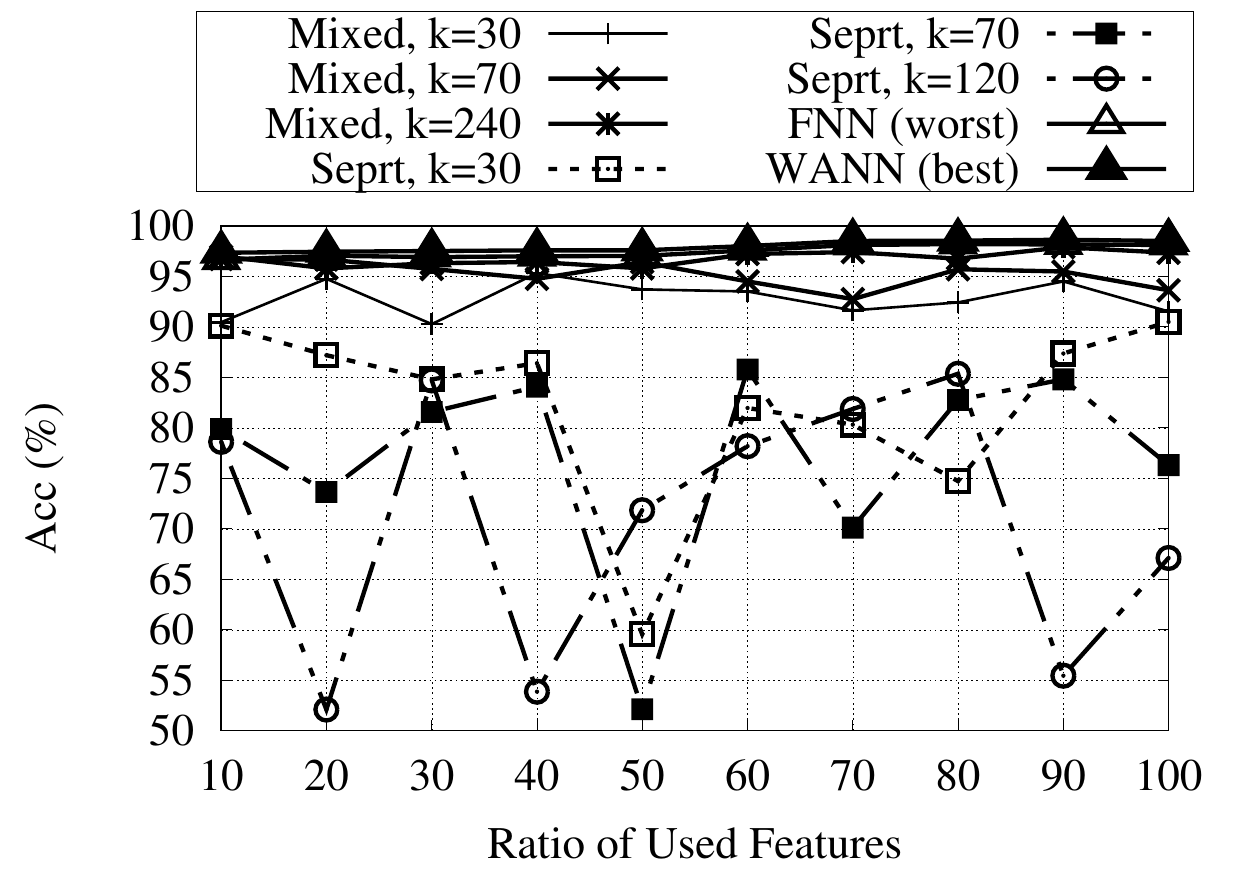}
		\caption{\small Contagio-intent}
			\label{fig:fig4f}
 	\end{subfigure} 
 	\begin{subfigure}{0.32\textwidth}
 		\centering
 		\includegraphics[width=1.05\linewidth]{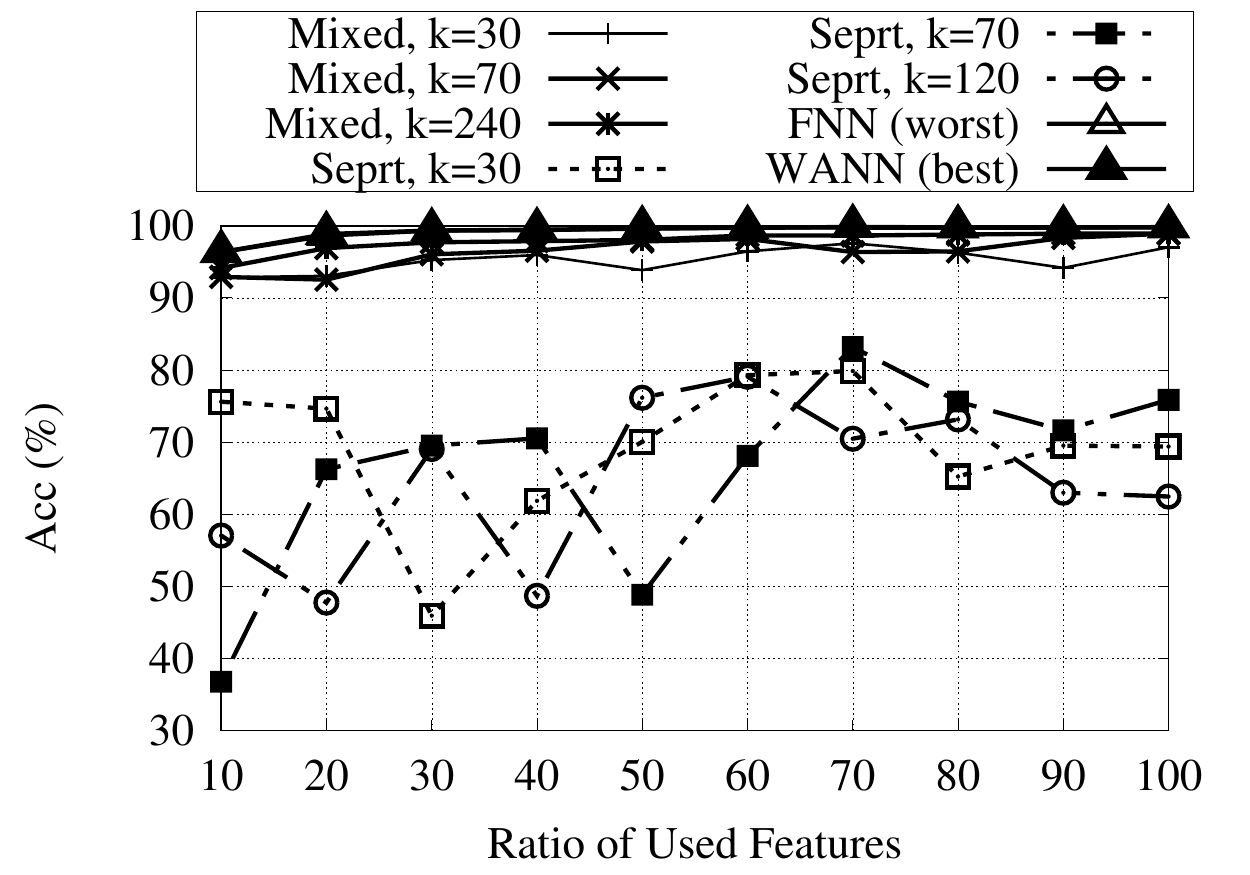}
			\caption{\small Genome-API}
			\label{fig:fig4g}
 	\end{subfigure}  
    \begin{subfigure}{0.32\textwidth}
 		\centering 
 		\includegraphics[width=1.05\linewidth]{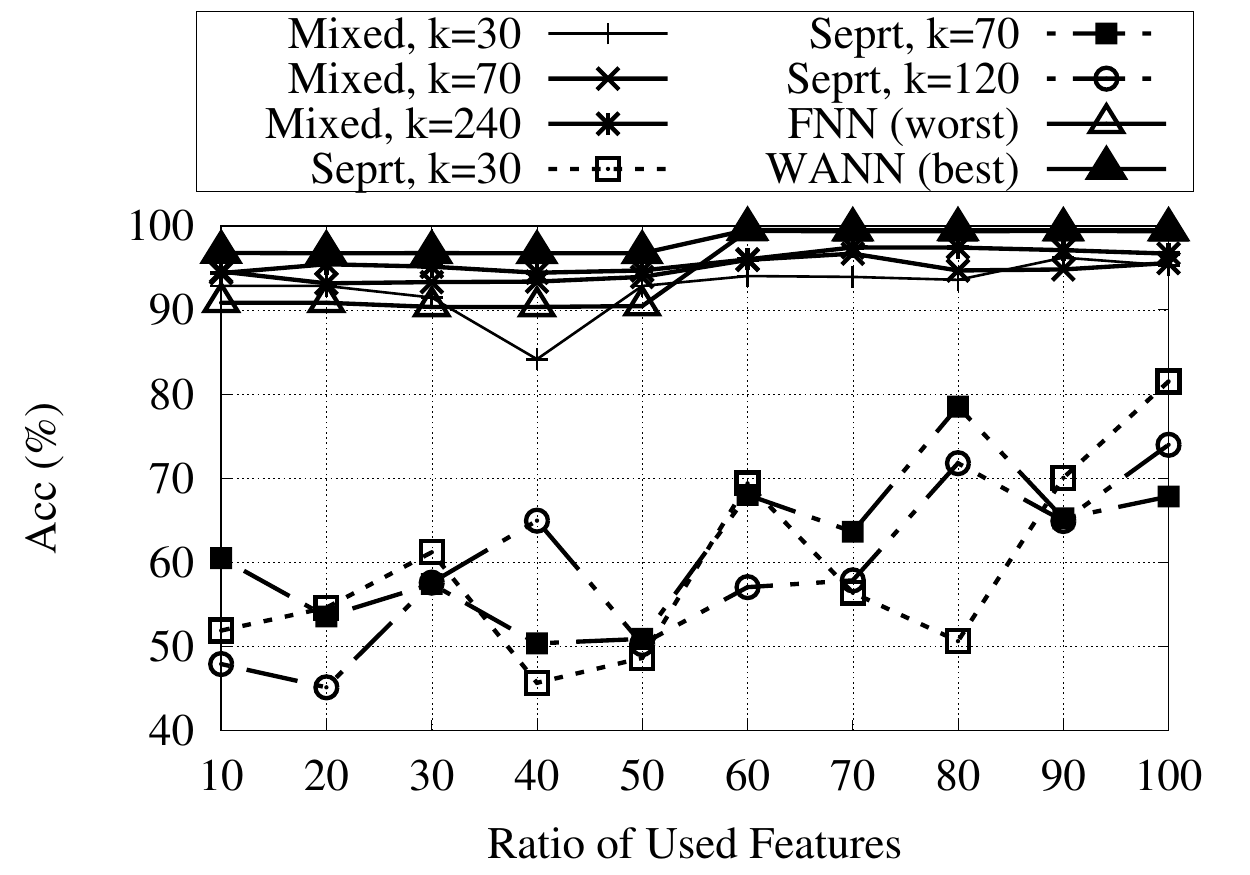}
			\caption{\small Genome-permission}
			\label{fig:fig4h}
 	\end{subfigure}
   \begin{subfigure}{0.32\textwidth}
 		\centering
 		\includegraphics[width=1.05\linewidth]{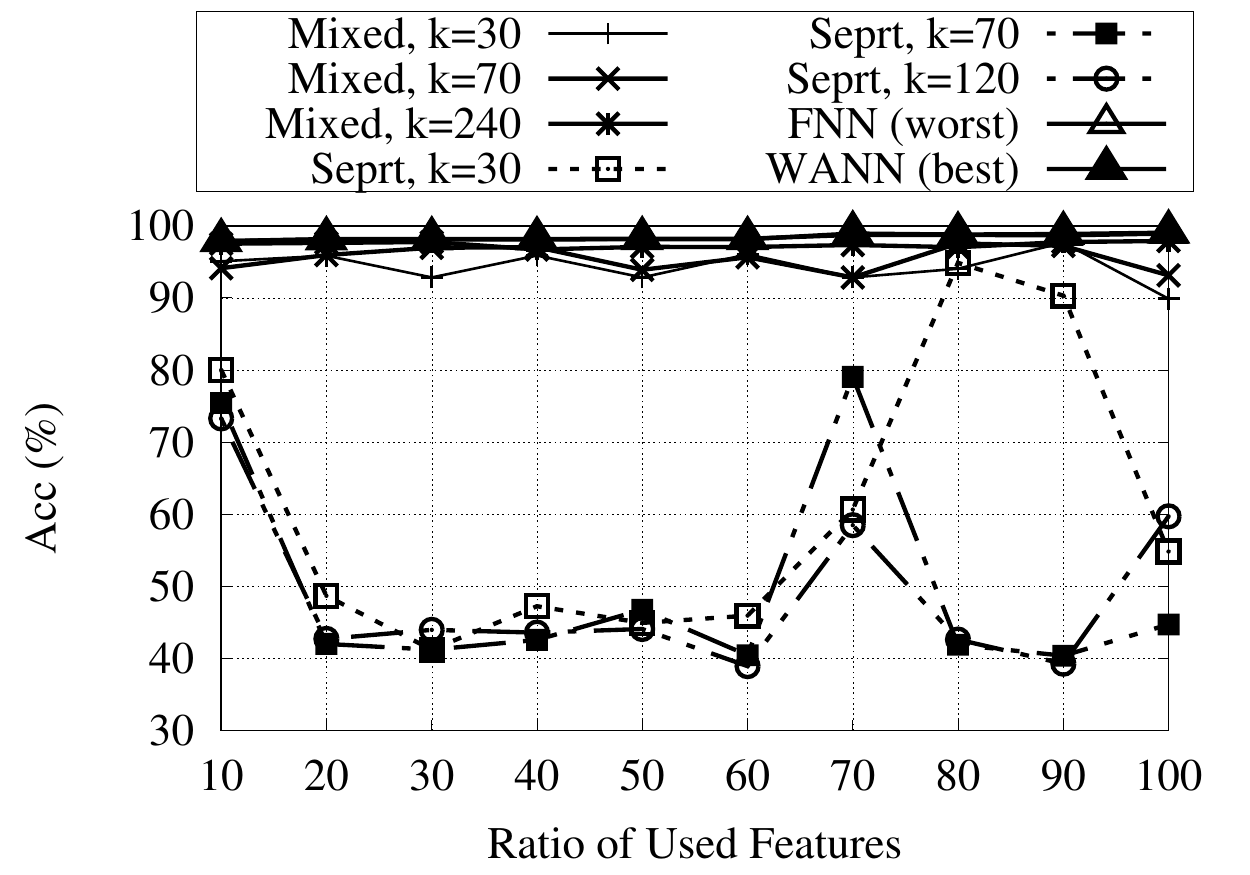}
		\caption{\small Genome-intent}
			\label{fig:fig4i}
 	\end{subfigure} 
 	\hfill
        \caption{\small Comparison accuracy value between \textit{Mixed} algorithm against \textit{Separated} (or \textit{Seprt}) algorithm reported in~\cite{Xiong2018} and our best (FNN) and worst (WANN) algorithms for API, intent and permission features for various datasets.\vspace{-10px}}
			\label{fig:fig4}
		\end{figure*}

\subsubsection{Fixed value for k in KNN-based algorithm}\label{results-of-conferance-paper}

In the reference~\cite{Xiong2018}, the authors propose two different methods based on the KNN algorithm. Hence, they consider fixed values for $k$ for two different methods called \textit{Mixed} and \textit{Separated} ones. In Fig.~\ref{fig:fig4}, we test them on different $k$ methods (i.e., $k=\{30, 40, 120, 240\}$) for the Drebin, Contagio, and Genome datasets. From this figure, we can draw \textit{three} conclusions. Firstly, we understand that our methods (i.e., FNN algorithm as the worst method and WANN as the best method) have the highest accuracy rate compared to the methods presented in~\cite{Xiong2018}: Mixed and Separated algorithms. Secondly, as we present in the three sets of comparison plots (Figs.~\ref{fig:fig4a}-\ref{fig:fig4c}, Figs.~\ref{fig:fig4d}-\ref{fig:fig4f}, and Figs.~\ref{fig:fig4g}-\ref{fig:fig4i}) the Mixed algorithm provides higher accuracy rather than the Separated algorithm (i.e., for each of four selected samples). This means that the average rate of accuracy (i.e., detecting malware from API, intent, and permission features) for the Separated method applied to all type of datasets is less than 90\% and for the Mixed algorithm less than 98\%. Finally, if we increase the $k$ value in KNN algorithm, the accuracy of both Separated and Mixed methods increase. The solutions presented in~\cite{Xiong2018} depending on the optimal $k$ value. Hence, they require optimization algorithms or trial and error methods to find the appropriate $k$ value to approach current accuracy ratios which raise the complexity of their methods. While in the methods proposed in this paper, we try to find the nearest neighbors regardless of the value of $k$ and compare the proposed methods with the modification in the number of features.

\subsubsection{Comparing methods based of precision, recall and f1-Score}\label{R-P}

\begin{figure*}[!htb]
    \centering
	\begin{subfigure}{0.32\textwidth}
 		\centering 
 		\includegraphics[width=1.05\linewidth]{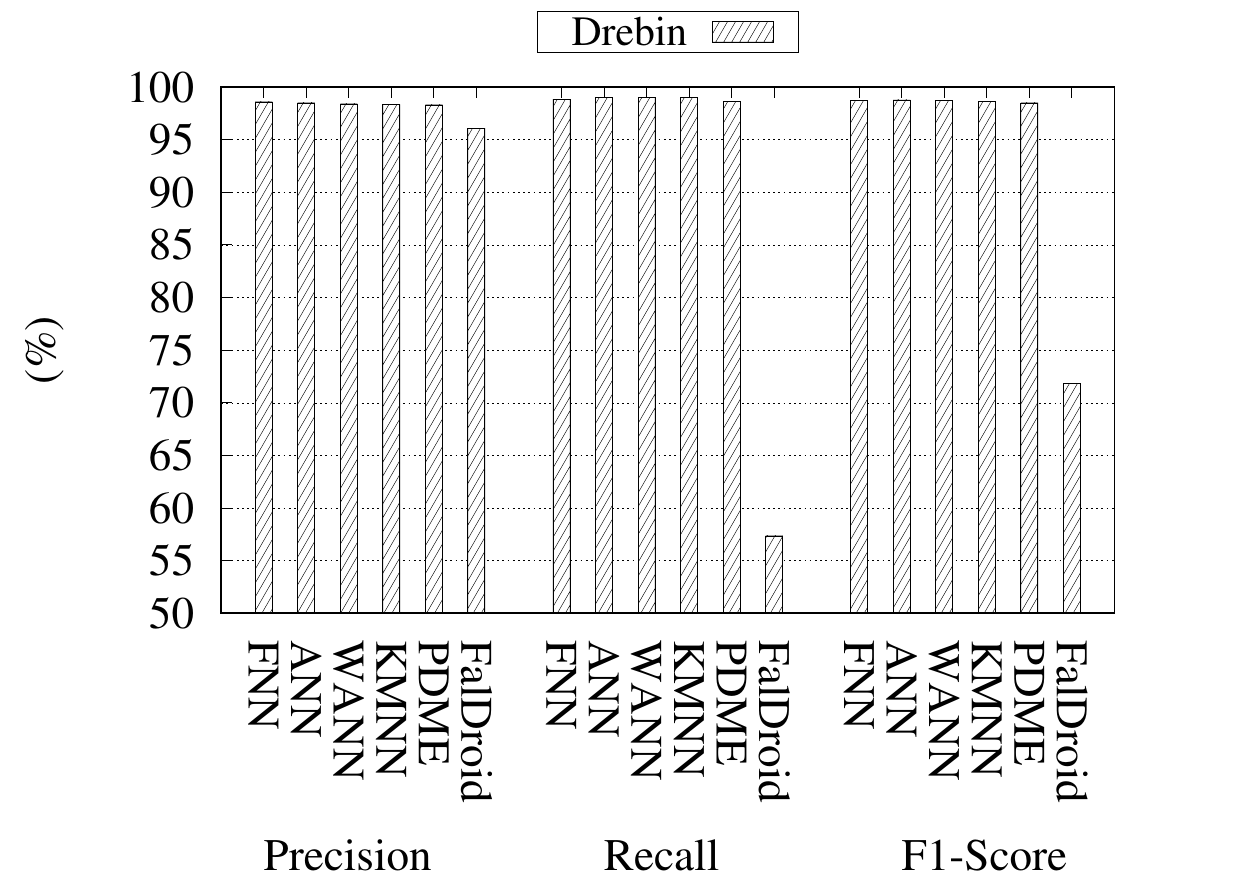}
			\caption{\small Drebin-API}
			\label{fig:fig10a}
 	\end{subfigure} 
   \begin{subfigure}{0.32\textwidth}
 		\centering
 		\includegraphics[width=1.05\linewidth]{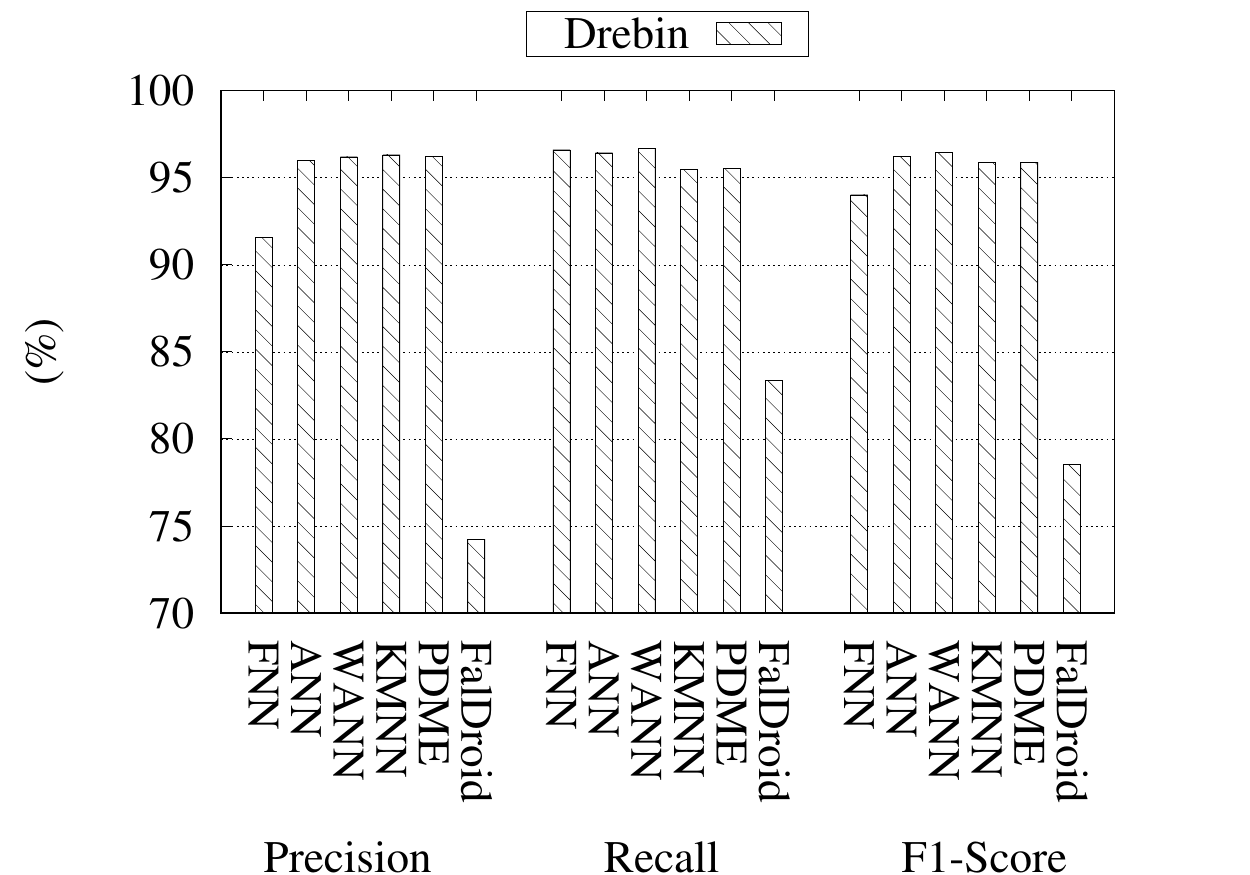}
			\caption{\small Drebin-permission}
			\label{fig:fig10b}
 	\end{subfigure} 
    \begin{subfigure}{0.32\textwidth}
 		\centering 
 		\includegraphics[width=1.05\linewidth]{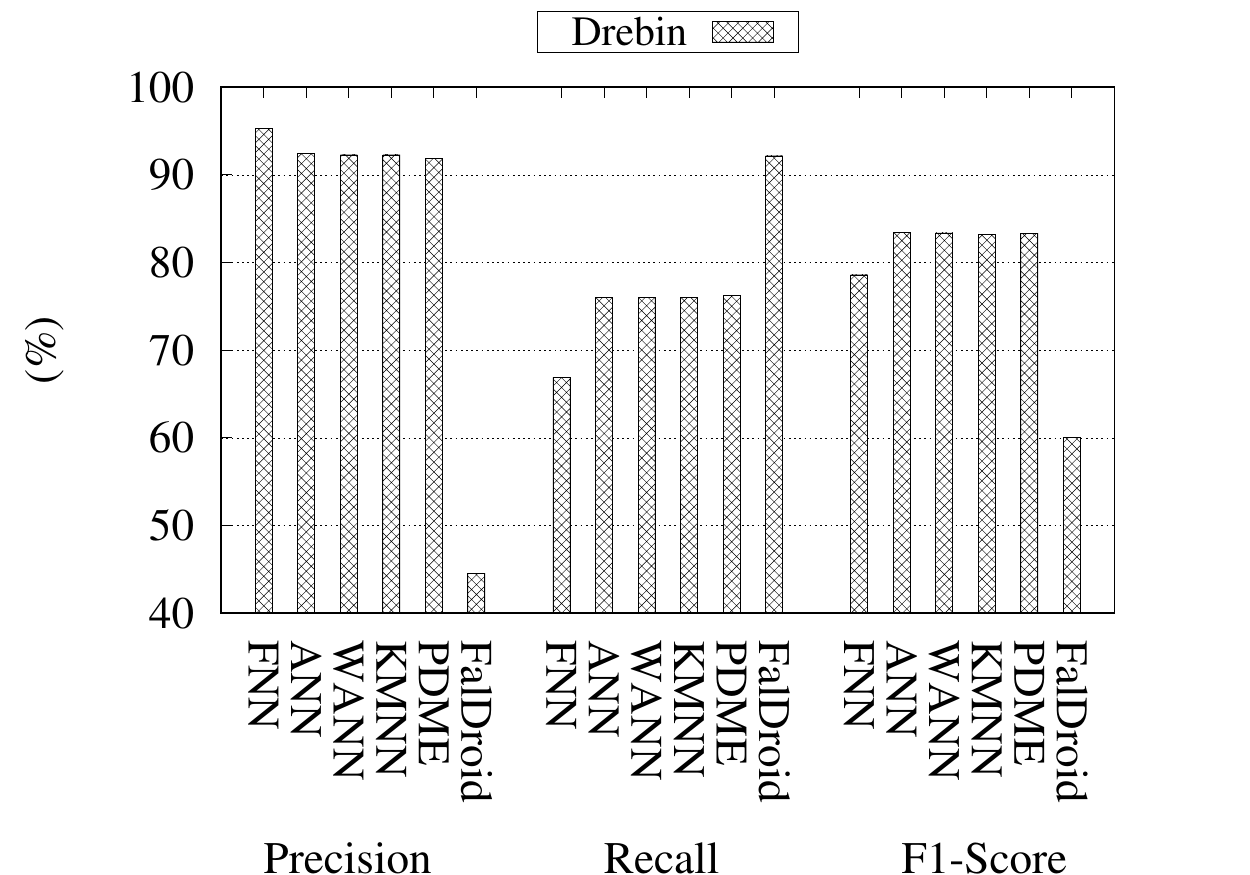}
		\caption{\small Drebin-intent}
			\label{fig:fig10c}
 	\end{subfigure}
 		\hfill
 	\begin{subfigure}{0.32\textwidth}
 		\centering
 		\includegraphics[width=1.05\linewidth]{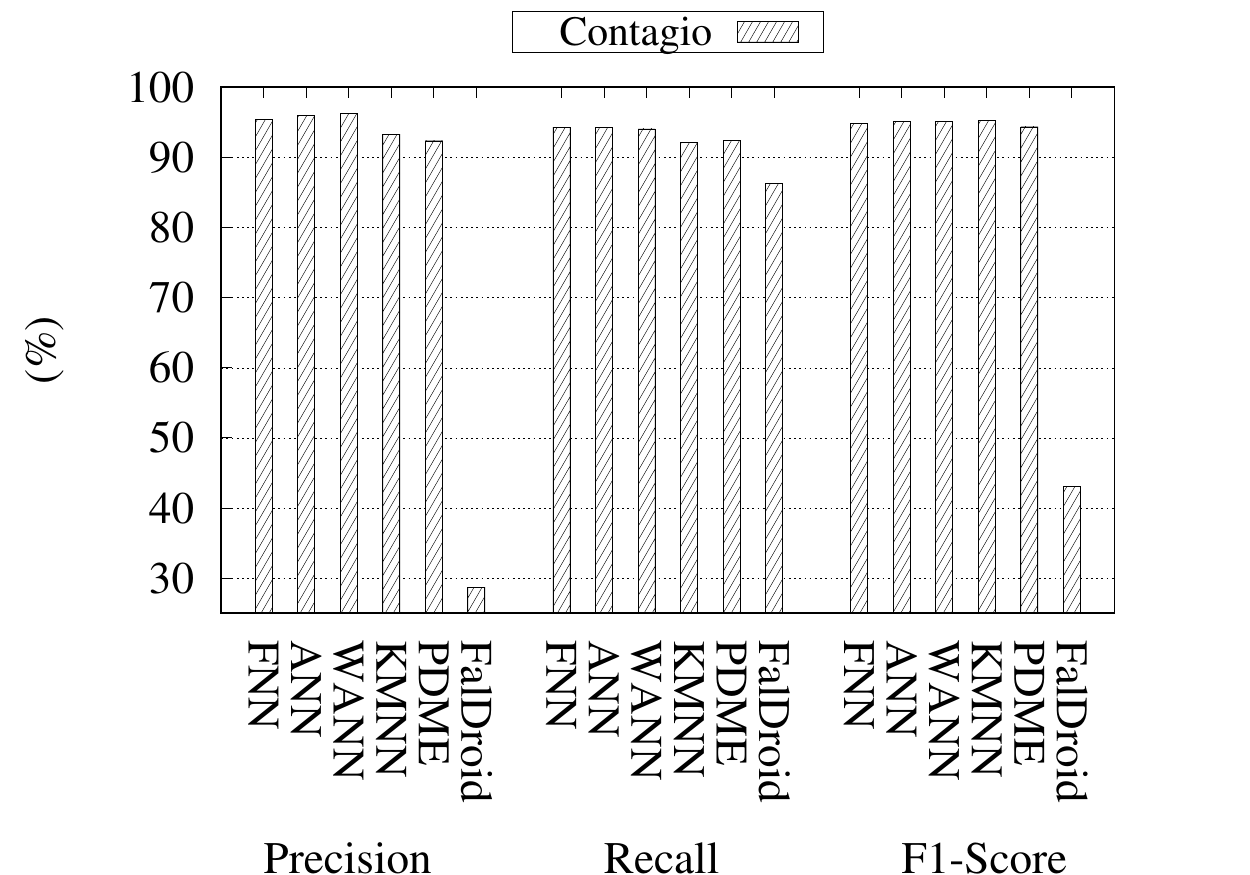}
			\caption{\small Contagio-API}
			\label{fig:fig10d}
 	\end{subfigure}  
    \begin{subfigure}{0.32\textwidth}
 		\centering 
 		\includegraphics[width=1.05\linewidth]{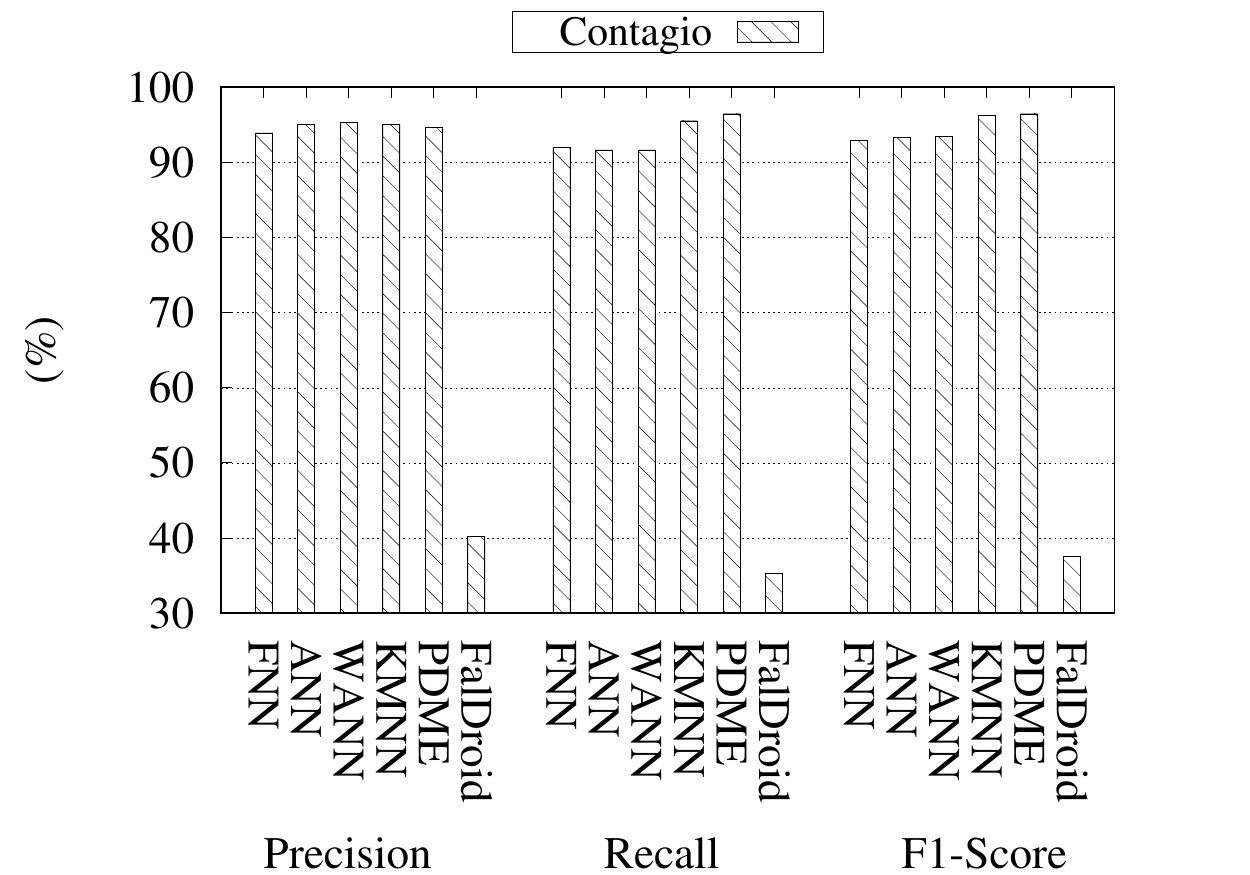}
			\caption{\small Contagio-permission}
			\label{fig:fig10e}
 	\end{subfigure}
   \begin{subfigure}{0.32\textwidth}
 		\centering
 		\includegraphics[width=1.05\linewidth]{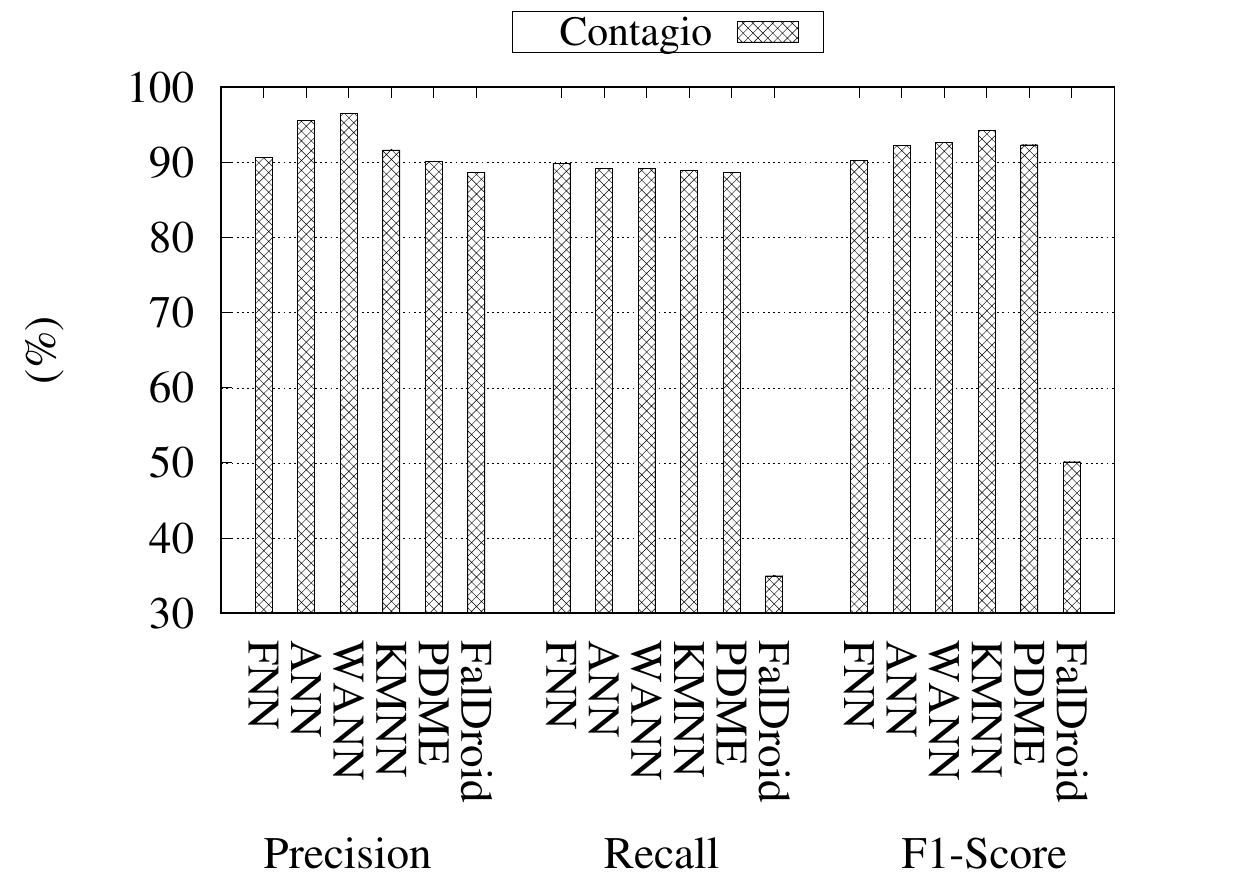}
		\caption{\small Contagio-intent}
			\label{fig:fig10f}
 	\end{subfigure} 
 	\begin{subfigure}{0.32\textwidth}
 		\centering
 		\includegraphics[width=1.05\linewidth]{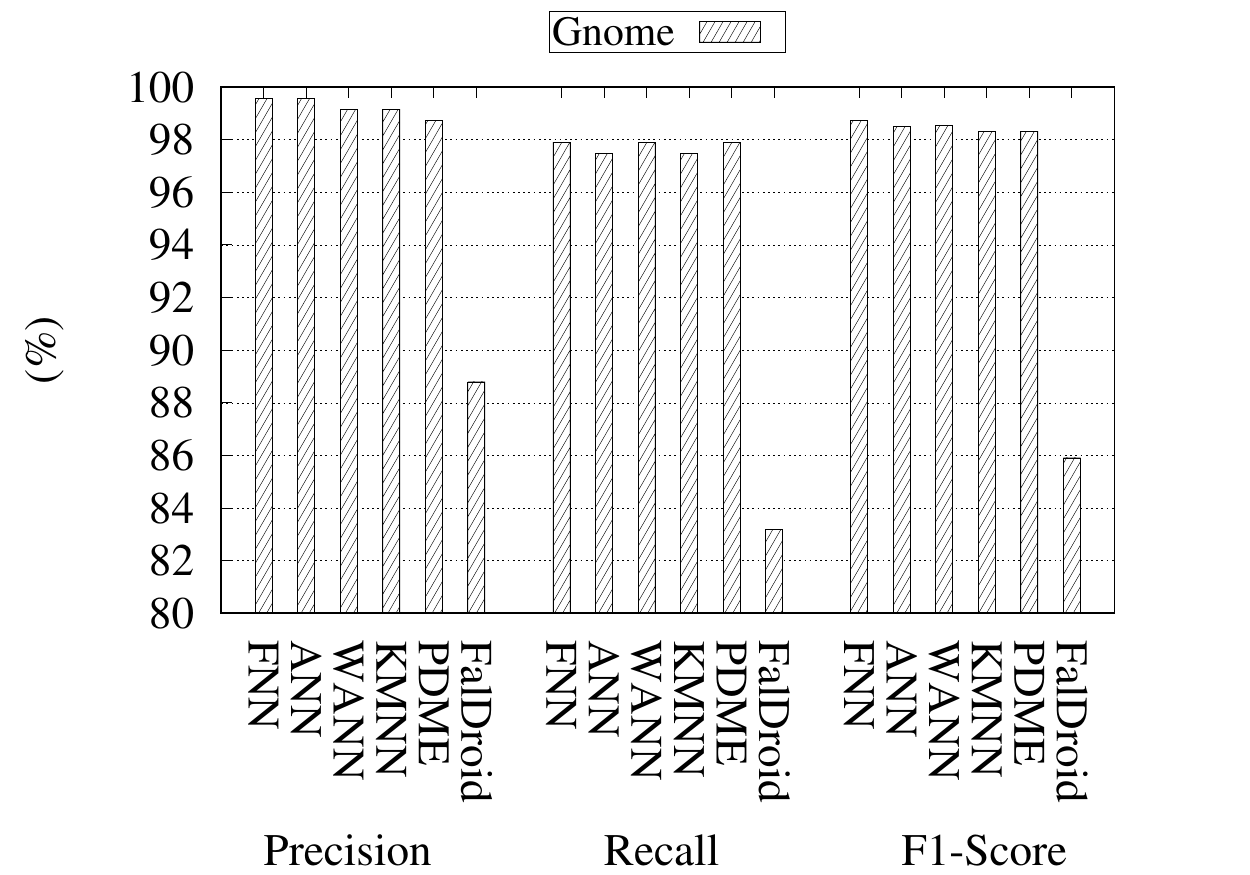}
			\caption{\small Genome-API}
			\label{fig:fig10g}
 	\end{subfigure}  
    \begin{subfigure}{0.32\textwidth}
 		\centering 
 		\includegraphics[width=1.05\linewidth]{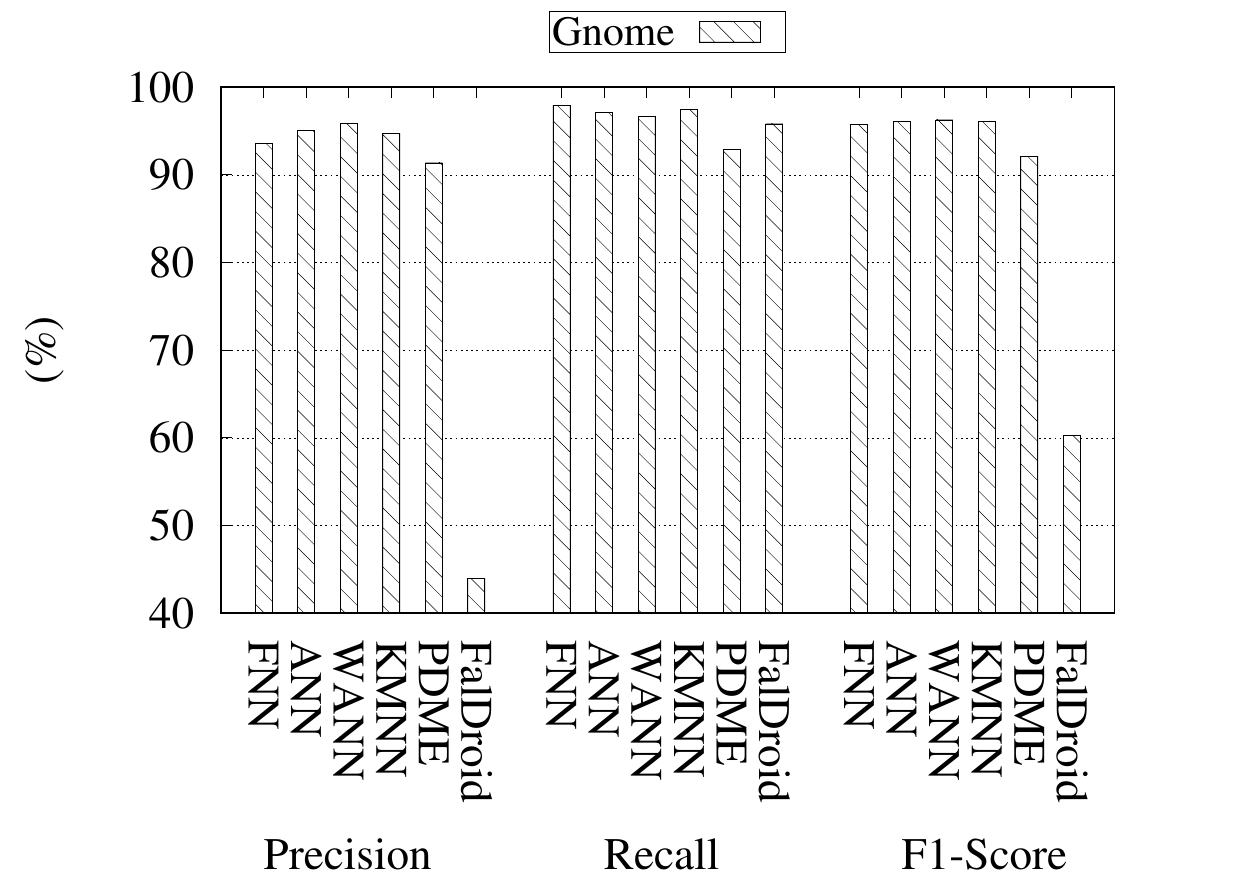}
			\caption{\small Genome-permission}
			\label{fig:fig10h}
 	\end{subfigure}
   \begin{subfigure}{0.32\textwidth}
 		\centering
 		\includegraphics[width=1.05\linewidth]{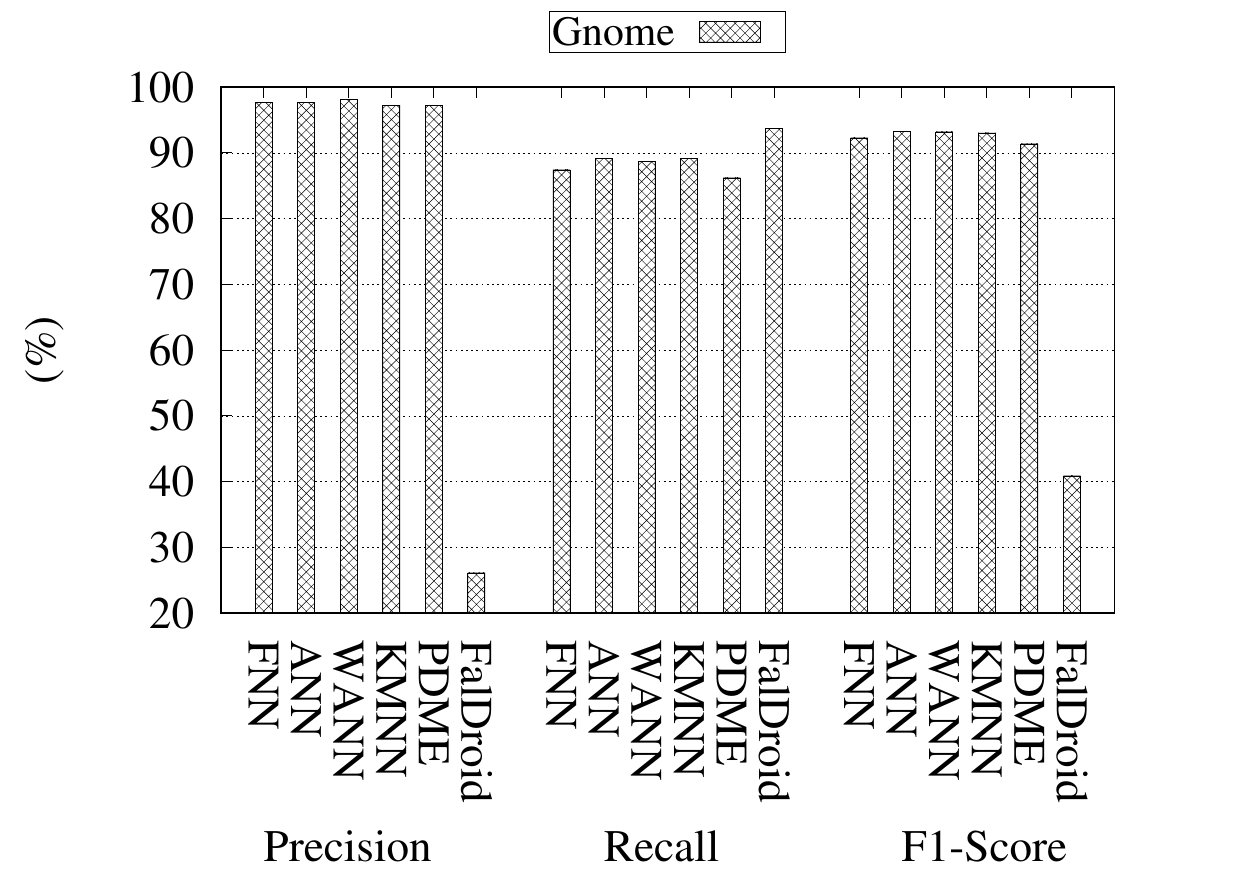}
		\caption{\small Genome-intent}
			\label{fig:fig10i}
 	\end{subfigure} 
 	\hfill
        \caption{\small Comparison between algorithms with reference to precision, recall, and f1-Score for API, intent and permission features in various datasets.\vspace{-10px}}
			\label{fig:fig10}
		\end{figure*}

In Fig.~\ref{fig:fig10}, we provide the precision and recall values for the different algorithms. Both recall (sensitivity) and precision (specificity) measures to use to determine generated errors. The recall is a measure that could show the rate of total detected malware. That is, the proportion of those correctly identified is the sum of all malware (i.e., those that are correctly identified by the malware plus those that are incorrectly detected by benign). Our goal in this section is to design a model with high recall that is more appropriate to identify malware. In this way, the set of  Figs.~\ref{fig:fig10a}-~\ref{fig:fig10c}, Figs.~\ref{fig:fig10d}-~\ref{fig:fig10f}, and Figs.~\ref{fig:fig10g}-~\ref{fig:fig10i} show the aforementioned values for the permission, API and intent data for the Drebin, Contagio, and Genome datasets, respectively. The precision measure shows the same concept for benign samples. It means, how many benign samples can we detect from all benign samples. Precision model is the proportion of samples that are not malware to the total benign samples (i.e., those that are detected as benign and those that are incorrectly considered as malware). With these explanations, the recall and precision metrics use instead of the accuracy metric and have a wider application in machine learning.
In most cases, these two metrics do not improve together. Sometimes we compute the precision of the model with more precise algorithms, that is, the ones we announce malware is most likely malware. Examples that are incorrectly classified as malware are very few, which means the precision of our algorithm is very high. But we may not consider the particular aspect of the data, and the total number of malware samples is much more than our declared examples, that is, we have a very low recall. On the other hand, it is possible to simplify our detection algorithm to increase the number of detected malware. In this case, the number of our incorrectly classified samples is increased, the precision value of the algorithm is low, and the recall value is high. On the other hand, if we can find a combination of both recall and precision values to measure the classification algorithms, the focus on that measure will be more appropriate than the simultaneous review of recall and precision. For example, we use the average of these two metrics as a new benchmark and try to raise the average of these two metrics. For this purpose, the mean harmonic is the two recall and precision criteria, which is called the \textit{f1-Score}. In this criterion, if the two values of recall and precision are small or even zero, the result will be small or zero.

Hence, in Fig.~\ref{fig:fig10}, as we can see, the FalDroid method has fewer values for precision and recall compared to other methods. The high precision measurements in the FNN algorithm on API features from the Drebin dataset shows that this algorithm is able to identify the more benign samples compared to other methods correctly. On the opposite side, the ANN, WANN, and KMNN algorithms have higher recall values. It indicates that these methods accurately detect malware samples more than other methods, and the higher accuracy of these algorithms are confirmed the result. Concerning the permission features of the WANN algorithm, it has a higher recall value and can detect more malware samples. While on these features, the KMNN algorithm has a higher precision value and therefore can detect more benign samples. The third group of features is intents. In these features, PDME~\cite{Radkani2017} and ANN algorithms have the highest recall value and can better identify malware samples. While in this feature group, the FNN algorithm has a higher precision value and can detect more benign samples correctly.

\subsubsection{Comparing methods based of different f1-Score values}\label{R-P-F1}

\begin{figure*}[!htb]
    \centering
	\begin{subfigure}{0.32\textwidth}
 		\centering 
 		\includegraphics[width=1.05\linewidth]{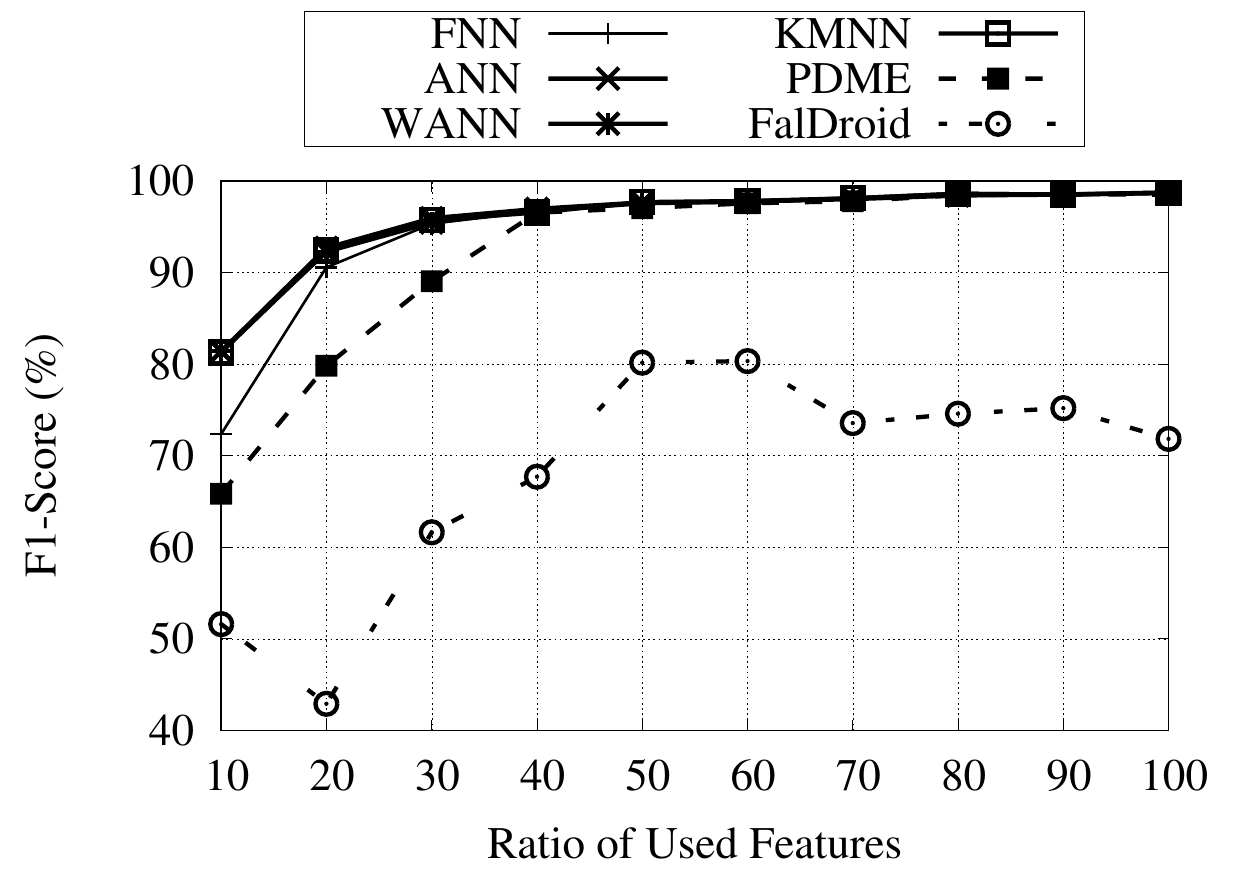}
			\caption{\small Drebin-API}
			\label{fig:fig11a}
 	\end{subfigure} 
   \begin{subfigure}{0.32\textwidth}
 		\centering
 		\includegraphics[width=1.05\linewidth]{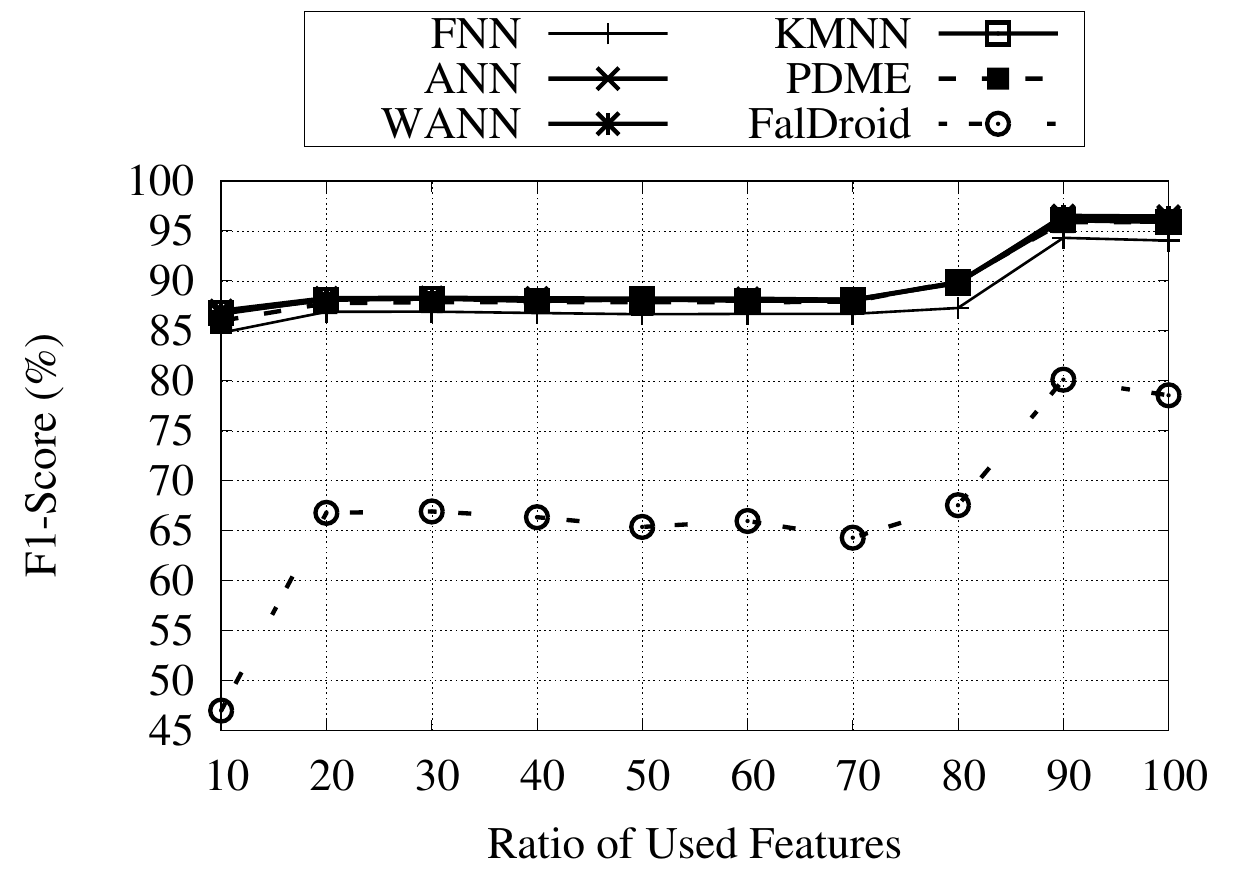}
			\caption{\small Drebin-permission}
			\label{fig:fig11b}
 	\end{subfigure} 
    \begin{subfigure}{0.32\textwidth}
 		\centering 
 		\includegraphics[width=1.05\linewidth]{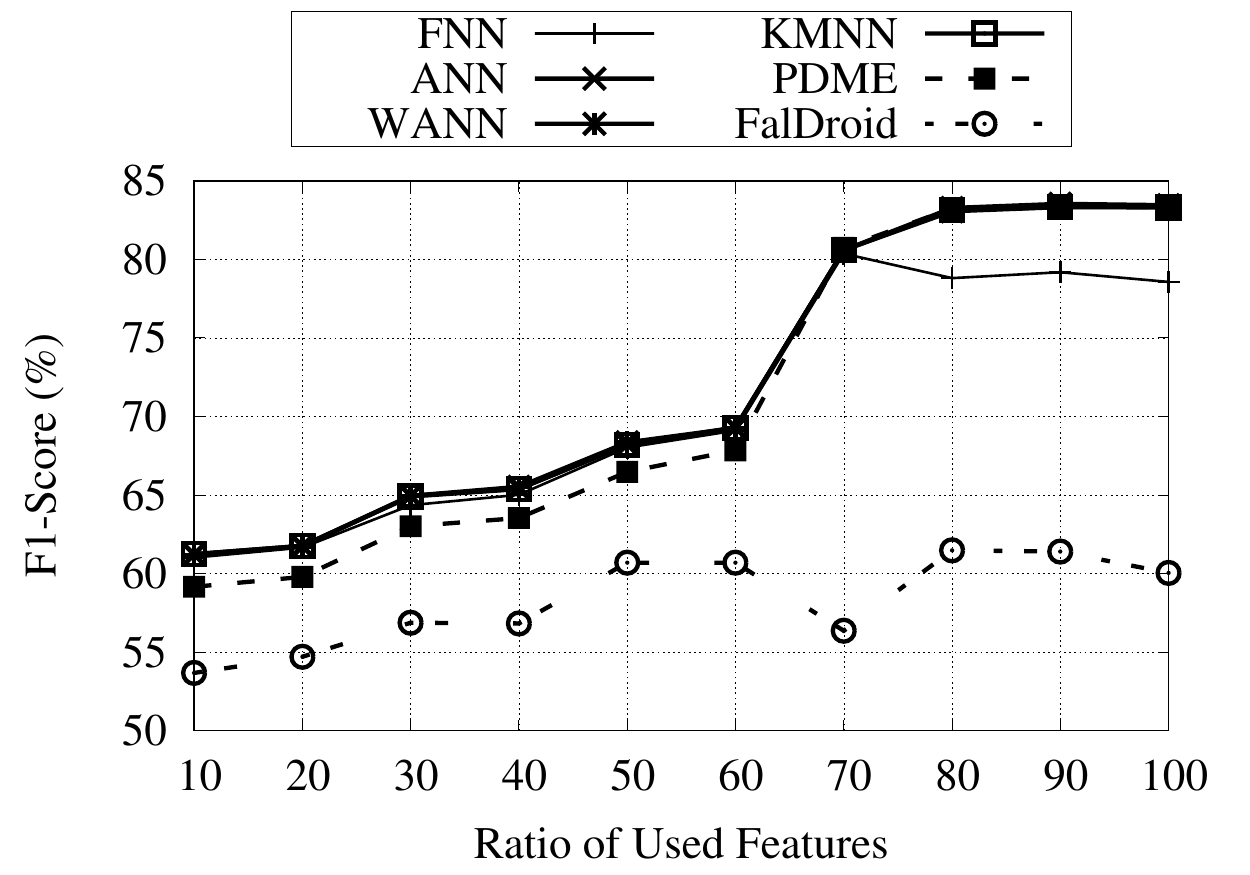}
		\caption{\small Drebin-intent}
			\label{fig:fig11c}
 	\end{subfigure}
 		\hfill
 	\begin{subfigure}{0.32\textwidth}
 		\centering
 		\includegraphics[width=1.05\linewidth]{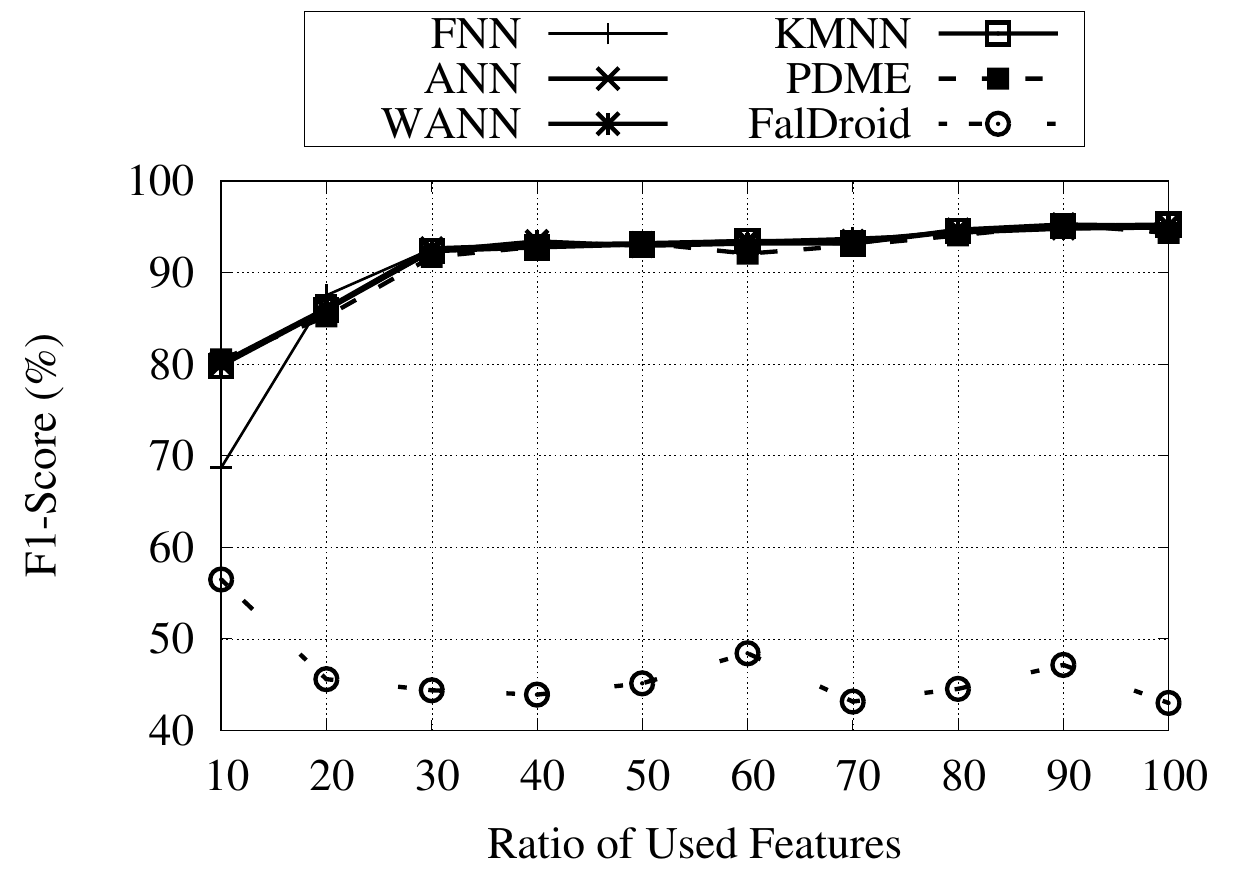}
			\caption{\small Contagio-API}
			\label{fig:fig11d}
 	\end{subfigure}  
    \begin{subfigure}{0.32\textwidth}
 		\centering 
 		\includegraphics[width=1.05\linewidth]{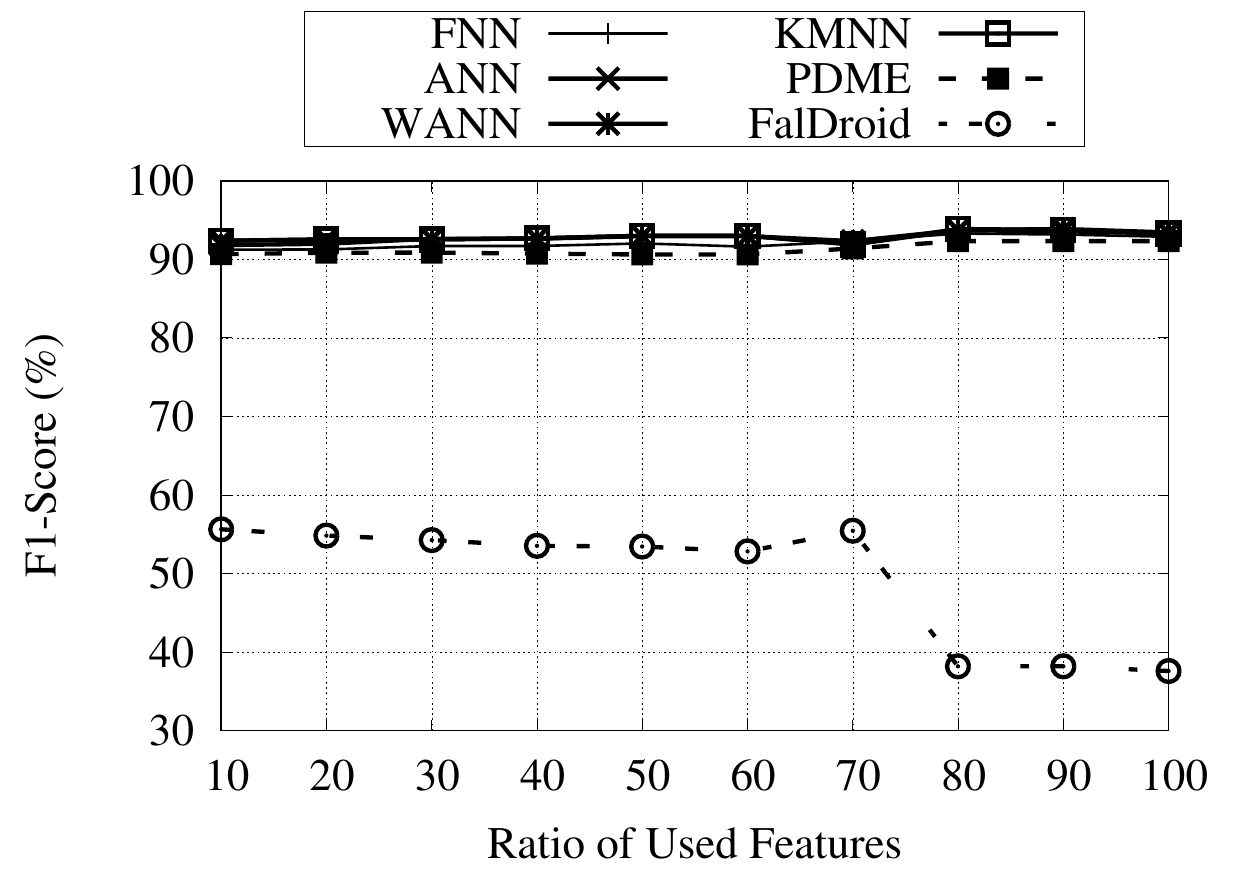}
			\caption{\small Contagio-permission}
			\label{fig:fig11e}
 	\end{subfigure}
   \begin{subfigure}{0.32\textwidth}
 		\centering
 		\includegraphics[width=1.05\linewidth]{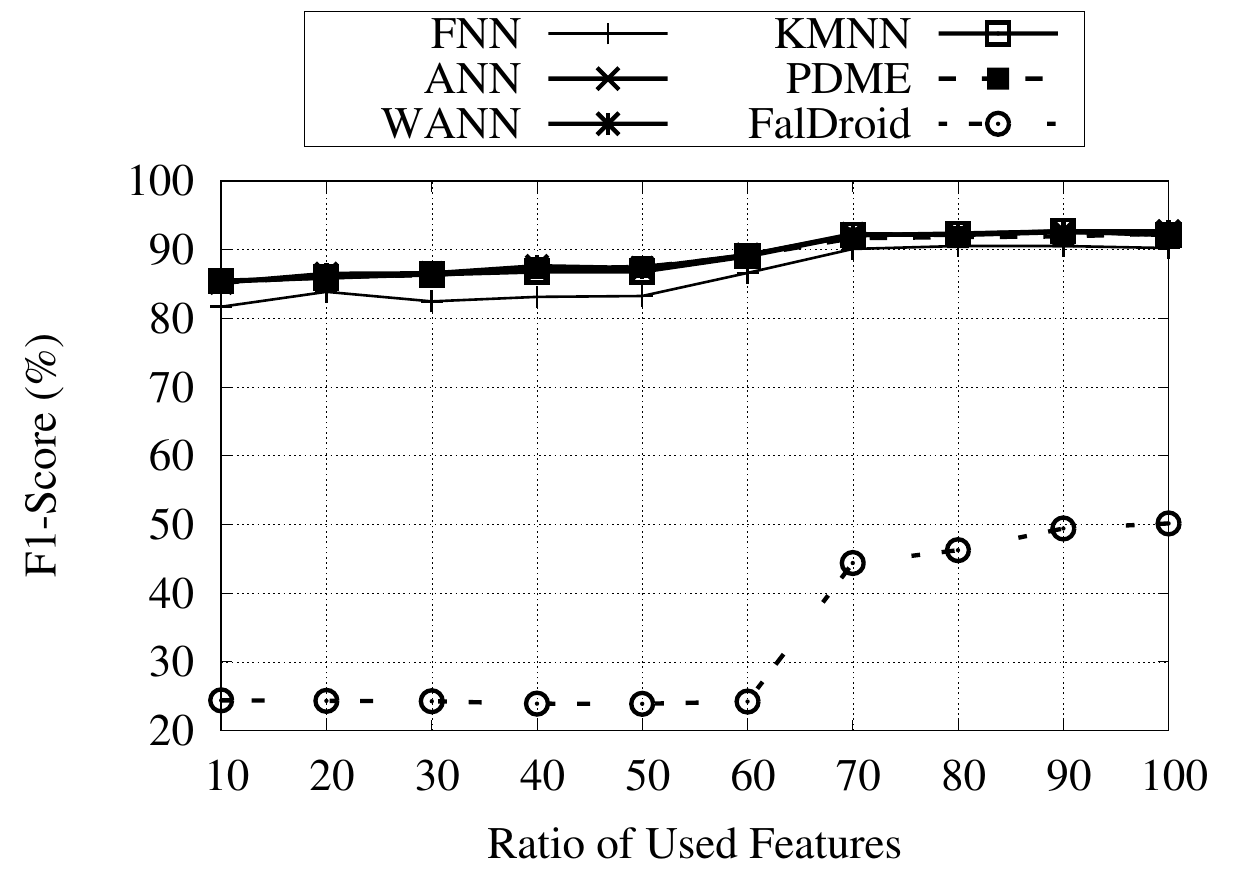}
		\caption{\small Contagio-intent}
			\label{fig:fig11f}
 	\end{subfigure} 
 	\begin{subfigure}{0.32\textwidth}
 		\centering
 		\includegraphics[width=1.05\linewidth]{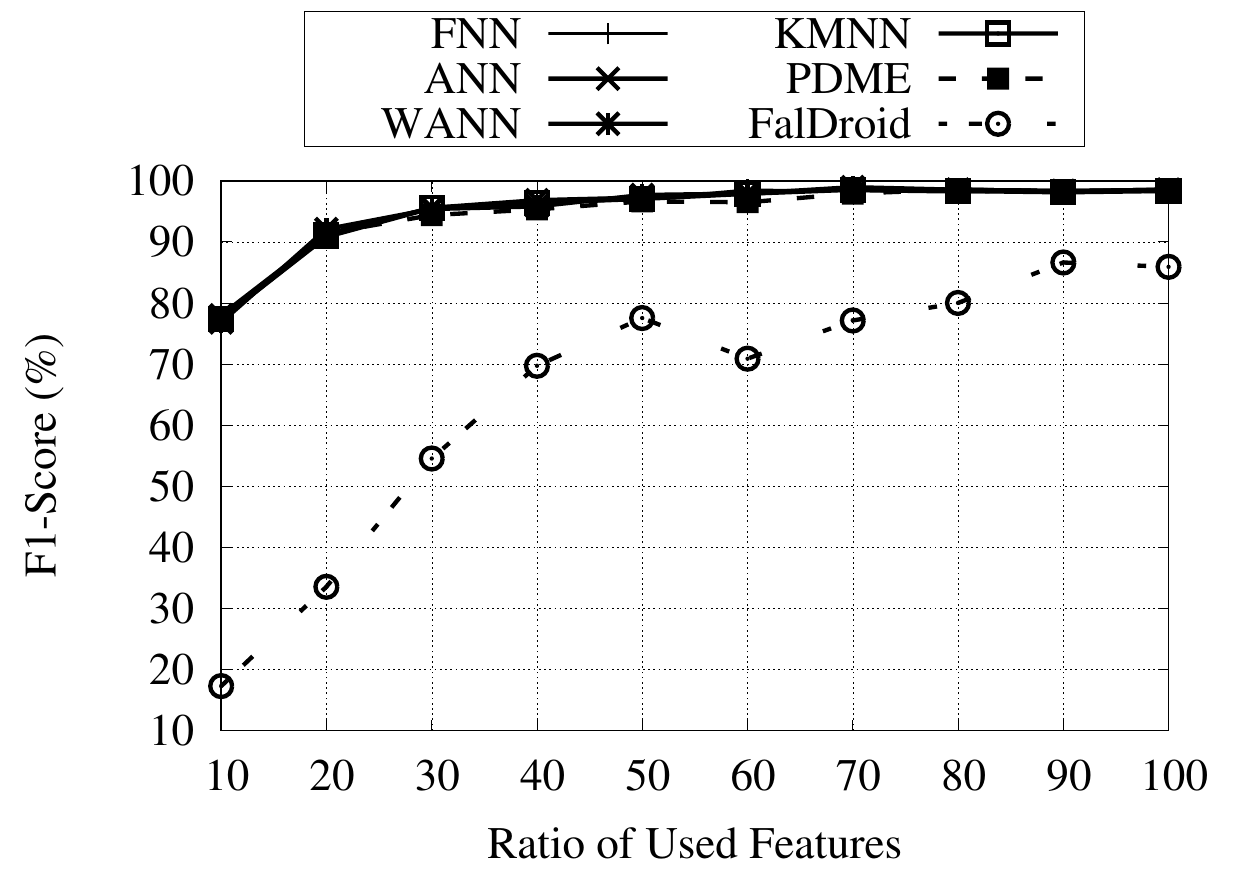}
			\caption{\small Genome-API}
			\label{fig:fig11g}
 	\end{subfigure}  
    \begin{subfigure}{0.32\textwidth}
 		\centering 
 		\includegraphics[width=1.05\linewidth]{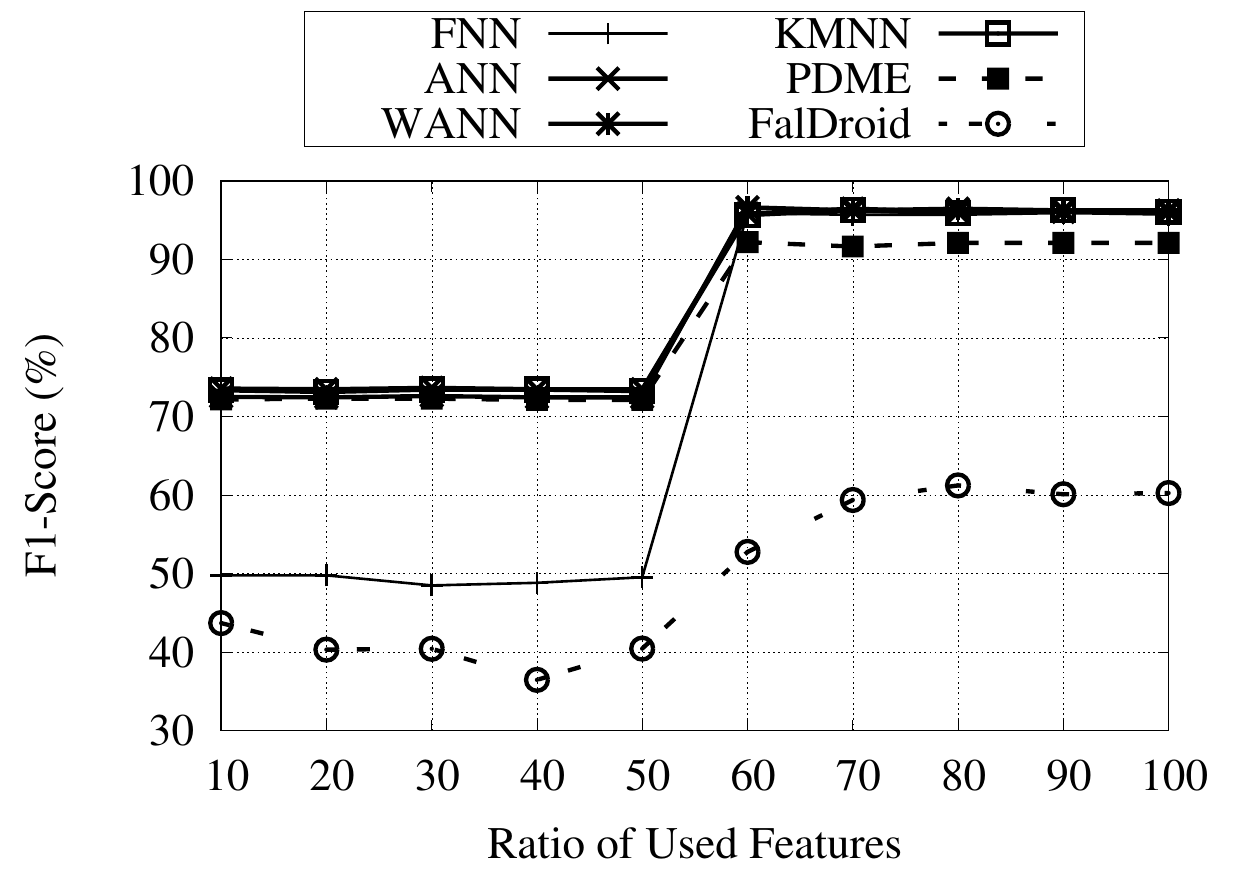}
			\caption{\small Genome-permission}
			\label{fig:fig11h}
 	\end{subfigure}
   \begin{subfigure}{0.32\textwidth}
 		\centering
 		\includegraphics[width=1.05\linewidth]{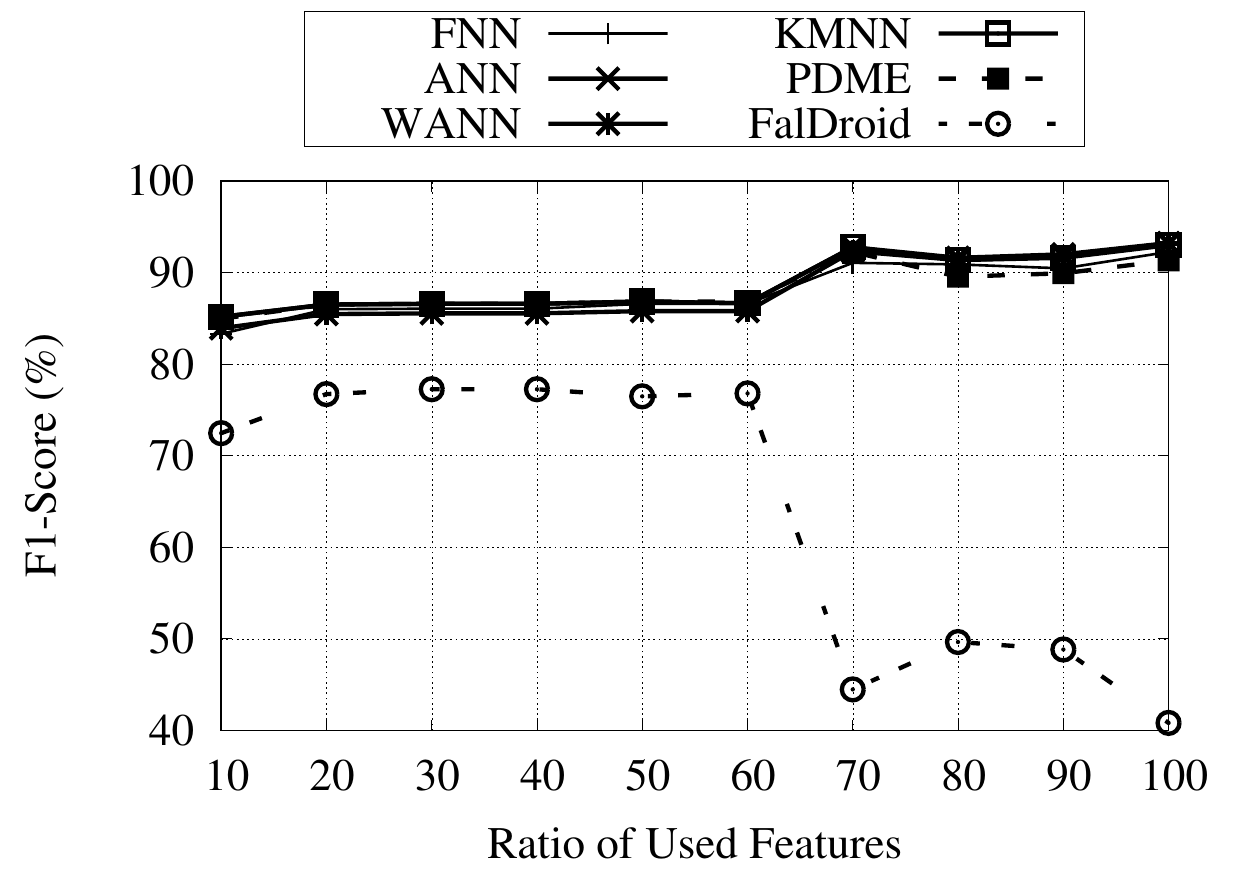}
		\caption{\small Genome-intent}
			\label{fig:fig11i}
 	\end{subfigure} 
 	\hfill
        \caption{\small Comparison f1-score value between algorithms for API, intent and permission features for various datasets.\vspace{-10px}}
			\label{fig:fig11}
		\end{figure*}

In this part, we aim to present the different f1-Score for different feature types in different datasets. To do so, we rely our results on the equation~\eqref{eq:eq182} which shows the f1-Score formula. In this equation, we use two criteria: recall and precision. These criteria can be between zero and one. f1-Score is calculated based on multiple of these two criteria values. Thus, the final result tends to be smaller than each of these criteria values. If both of them are large numbers (approach to one), the final result will be near to one. With this explanation, the higher value for the f1-Score means that the algorithm could detect more malware and benign. In Fig.~\ref{fig:fig11}, we present the f1-Score for the API, permission, and intent features of different algorithm for different datasets. To be precise, the f1-Score for API and intent features of the Drebin dataset has the highest value and especially this rate is higher for the ANN algorithm. After that, the f1-Score for the FNN and WANN algorithms are the second and third biggest f1-Score which can present more malware and benign detection. Interestingly, in the three types of API, permission, and intent features the three ANN, WANN and KMNN algorithms have a higher f1-Score and this value for the PDME method~\cite{Radkani2017} is placed in the next rank. FalDroid algorithm always has the lowest f1-Score. In Fig.~\ref{fig:fig11}, we notice that the f1-Score value increases with increasing number of features. The best results of the three datasets, are from the Contagio dataset. By examining Fig.~\ref{fig:fig11}, it can be concluded that f1-Score value for the API features has the highest rate.

\subsubsection{Comparing methods based on Accuracy, FPR and AUC}\label{ACCFPRAUC}

In this section, we compare algorithms based on Accuracy, FPR and, AUC metrics with both PDME~\cite{Radkani2017} and FalDroid~\cite{FanMing2018} algorithms which we present them in Tables~\ref{tab5}-\ref{tab7}. To be precise, in Table~\ref{tab5} for Drebin dataset, for each algorithm, increasing the number of features increases the accuracy and AUC values, and decreases the FPR values. Specifically, focusing on FNN algorithm by considering all API features and 10\% API features, the value of Acc and AUC increase about 15\% and 21\%, respectively while the FPR value decreases exponentially approximately 95\% and approaches to 0.004. Focusing on ANN algorithm, by considering all API features, the AUC value is 98.96\%, FPR value is 0.004, and Acc value is 99.33\% which is about 0.02\% above  algorithm. The WANN and KMNN algorithms also achieve FPR value about 0.004 for both methods; their AUC values are 98.90\% and 98.84\%, and their accuracy values are 99.31\% and 99.28\%, respectively. Focusing on the state-of-the art method like PDME algorithm~\cite{Radkani2017}, the highest accuracy is 99.21\%, which is obtained for 90\% of the API features with the FPR value of 0.005 and the AUC 98.96\%, while the FalDroid algorithm~\cite{FanMing2018} has the best accuracy by considering 60\% of the API features, which is about 90.89\% with AUC value 96.17\% and with FPR of 0.099, and this method has the worst results (in all metrics presented in Table~\ref{tab5}) than the proposed methods. By considering the results of the API features, the accuracy of FNN, ANN, WANN, KMNN, and PDME~\cite{Radkani2017} methods are more than 99\% with 80\% of the features, and the FalDroid method~\cite{FanMing2018} has no accuracy of more than 90.89\%. As a result, focusing on Tables~\ref{tab5}-\ref{tab7}, we can understand \textit{five} conclusions. First, by examining the permissions and intent features of the Drebin dataset in Table~\ref{tab5} and the API, permission, and intent features from the two Contagio and Genome datasets reported in Tables~\ref{tab6} and ~\ref{tab7}, we realize that the described ML metric results for all algorithms are roughly the same. Second, focusing on our proposed algorithms, interestingly, the highest accuracy is usually achieved with the use of 70\% or 80\% or 90\% of the features, and we do not have to choose all the features. Third, if the highest accuracy of different algorithms is ranked with respect to intent, permission, API features and their average ratings, the WANN algorithm has the highest accuracy value compared to others and the ANN, KMNN algorithms are in the next rank, and the PDME algorithm proposed in~\cite{Radkani2017} is in third place. The FalDroid algorithm~\cite{FanMing2018} has the least accuracy in all cases. Fourth, in all of the proposed methods, the accuracy of the API features are highest. Finally, the presented results of the Genome dataset are better than the Drebin and Contagio datasets.
\begin{table*}[!htb]
  \centering
  \caption{\small Accuracy, FPR, AUC values for API, permission, intent data of the Drebin dataset (Acc= accuracy; FPR= false positive rate; AUC= area under cover; FR= feature length; MAA= maximum available accuracy).\vspace{-10px}}
\label{tab5}
  \scriptsize{
\setlength\tabcolsep{0.5pt} 
  \begin{tabular}{|c|c|c|c|c||c|c|c||c|c|c||c|c|c||c|c|c||c|c|c||c|c|c|}
  \hline
  \rowcolor{LightCyan} 
  \multicolumn{23}{|c|}{\textbf{Drebin Dataset}}\\\hline\hline
\rowcolor{yellow}
   &&\multicolumn{3}{c|}{\textbf{FNN}}&\multicolumn{3}{c|}{\textbf{ANN}}&\multicolumn{3}{c|}{\textbf{WANN}}&\multicolumn{3}{c|}{\textbf{KMNN}}&\multicolumn{3}{c|}{\textbf{PDME~\cite{Radkani2017}}}&\multicolumn{3}{c|}{\textbf{\textbf{FalDroid~\cite{FanMing2018}}}}&\multicolumn{3}{c|}{\textbf{MAA}}\\\hline
   \rowcolor{maroon!10}
    &\textbf{FR}&\textbf{Acc}&\textbf{FPR}&\textbf{AUC}&\textbf{Acc}&\textbf{FPR}&\textbf{AUC}&\textbf{Acc}&\textbf{FPR}&\textbf{AUC}&\textbf{Acc}&\textbf{FPR}&\textbf{AUC}&\textbf{Acc}&\textbf{FPR}&\textbf{AUC}&\textbf{Acc}&\textbf{FPR}&\textbf{AUC}&\textbf{Acc}&\textbf{FPR}&\textbf{AUC}\\
 \hline
\cellcolor[HTML]{FAEC0D}&\cellcolor[HTML]{E8E8AB}10\%&84.76&11.81&78.93&90.36&5.74&88.79&90.36&5.74&88.79&90.31&5.71&88.82&84.97&4.09&89.36&76.74&12.46&72.53&99.48&0.29&99.45\\\cline{2-23}
\cellcolor[HTML]{FAEC0D}&\cellcolor[HTML]{E8E8AB}20\%&95.01&3.18&93.95&96.09&2.92&94.55&95.85&3.21&94.03&95.94&2.95&94.47&90.53&2.27&94.77&78.94&3.38&86.39&99.21&0.45&99.14\\\cline{2-23}
\cellcolor[HTML]{FAEC0D}&\cellcolor[HTML]{E8E8AB}30\%&97.47&2.24&95.86&97.81&1.98&96.34&97.54&2.17&95.98&97.69&2.08&96.16&94.54&1.56&96.77&84.83&1.17&96.13&99.45&0.42&99.21\\\cline{2-23}
\cellcolor[HTML]{FAEC0D}&\cellcolor[HTML]{E8E8AB}40\%&98.26&1.75&96.77&98.35&1.62&97.00&98.23&1.79&96.71&98.14&1.79&96.70&98.14&1.43&97.32&86.52&1.49&95.65&99.33&0.55&98.96\\\cline{2-23}
\cellcolor[HTML]{FAEC0D}&\cellcolor[HTML]{E8E8AB}50\%&98.71&1.14&97.88&98.74&1.04&98.05&98.69&1.10&97.94&98.76&0.97&98.17&98.43&0.84&98.39&90.79&1.75&95.88&99.50&0.42&99.21\\\cline{2-23}
\cellcolor[HTML]{FAEC0D}&\cellcolor[HTML]{E8E8AB}60\%&98.71&1.30&97.59&98.78&1.23&97.71&98.78&1.17&97.83&98.81&1.23&97.71&98.69&1.04&98.05&90.89&1.62&96.17&99.50&0.42&99.21\\\cline{2-23}
\cellcolor[HTML]{FAEC0D}&\cellcolor[HTML]{E8E8AB}70\%&98.95&1.14&97.89&98.97&1.10&97.95&98.93&1.14&97.89&98.97&1.14&97.89&98.81&1.01&98.12&88.50&1.33&96.45&99.59&0.32&99.39\\\cline{2-23}
\cellcolor[HTML]{FAEC0D}\multirow{-5}{*}{\begin{sideways}{\textbf{API}}\end{sideways}}&\cellcolor[HTML]{E8E8AB}80\%&99.21&0.75&98.60&99.24&0.65&98.78&99.14&0.78&98.54&99.24&0.65&98.78&99.17&0.58&98.90&89.00&0.84&97.73&99.62&0.19&99.63\\\cline{2-23}
\cellcolor[HTML]{FAEC0D}&\cellcolor[HTML]{E8E8AB}90\%&99.26&0.62&98.84&99.19&0.68&98.72&99.19&0.68&98.72&99.19&0.68&98.72&99.21&0.55&98.96&89.24&0.75&98.00&99.55&0.26&99.51\\\cline{2-23}
\cellcolor[HTML]{FAEC0D}&\cellcolor[HTML]{E8E8AB}100\%&99.31&0.52&99.02&99.33&0.55&98.96&99.31&0.58&98.90&99.28&0.62&98.84&99.19&0.62&98.84&88.07&0.84&97.62&99.57&0.29&99.45\\
\hline\hline
\cellcolor[HTML]{FAEC0D}&\cellcolor[HTML]{E8E8AB}10\%&91.58 &6.36&	88.58&	93.27&	2.69&	94.65&	93.13&	2.76&	94.52&	93.13&3.02&	94.06&	92.75&	3.08&	93.88&	80.24&	1.71&	92.94&99.40&0.59&	98.92\\\cline{2-23}
\cellcolor[HTML]{FAEC0D}&\cellcolor[HTML]{E8E8AB}20\%&92.72&	5.64&	89.90&	93.89&	2.46&	95.15&	93.84&2.53&	95.03&	93.80&	2.56&	94.97&	93.63&	2.59&	94.88&	85.33&	2.95&	92.17&99.31&	0.72&98.69\\\cline{2-23}
\cellcolor[HTML]{FAEC0D}&\cellcolor[HTML]{E8E8AB}30\%&92.75&	5.58&	90.00&	93.92&	2.40&	95.27&	93.87&	2.46&	95.15&	93.87&	2.46&	95.15&	93.70&	2.49&	95.07&	85.35&	3.02&	92.04&99.36&	0.66&	98.81\\\cline{2-23}
\cellcolor[HTML]{FAEC0D}&\cellcolor[HTML]{E8E8AB}40\%&92.65&	5.71&	89.79&	93.89&	2.43&	95.21&	93.84&	2.49&	95.09&	93.75&	2.53&	95.02&	93.70&	2.49&	95.07&	85.11&	3.15&	91.67&99.33&	0.69&	98.75\\\cline{2-23}
\cellcolor[HTML]{FAEC0D}&\cellcolor[HTML]{E8E8AB}50\%&92.58&	5.81&	89.63&	93.87&	2.46&	95.15&	93.82&	2.49&	95.09&	93.82&	2.53&	95.03&	93.70&	2.49&	95.07&	84.73&	3.31&	91.18&99.36&	0.69&	98.75\\\cline{2-23}
\cellcolor[HTML]{FAEC0D}&\cellcolor[HTML]{E8E8AB}60\%&92.60&	5.77&	89.69&	93.87&	2.46&	95.15&	93.82&	2.49&	95.09&	93.75&	2.59&	94.90&	93.68&	2.53&	95.01&	84.75&	3.77&	90.28&99.36&	0.69&	98.75\\\cline{2-23}
\cellcolor[HTML]{FAEC0D}&\cellcolor[HTML]{E8E8AB}70\%&92.60&	5.77&	89.69&	93.80&	2.56&	94.97&	93.82&	2.49&	95.09&	93.75&	2.66&	94.79&	93.72&	2.49&	95.07&	83.73&	5.02&	87.52&99.36&	0.69&	98.75\\\cline{2-23}
\cellcolor[HTML]{FAEC0D}\multirow{-6}{*}{\begin{sideways}{\textbf{Permission}}\end{sideways}}&\cellcolor[HTML]{E8E8AB}80\%&92.91&	5.64&	89.95&	94.49&	3.64&	93.28&	94.44&	3.61&	93.33&	94.46&	3.67&	93.23&	94.56&	3.44&	93.61&	83.13&	9.84&	80.60&99.36&0.59&98.92\\\cline{2-23}
\cellcolor[HTML]{FAEC0D}&\cellcolor[HTML]{E8E8AB}90\%&96.80&	3.12&	94.48&	98.00&	1.41&	97.42&	98.09&	1.31&	97.59&	97.85&	1.28&	97.64&	97.73&	1.41&	97.11&	88.33&	10.83&	82.03&99.48&	0.56&	98.99\\\cline{2-23}
\cellcolor[HTML]{FAEC0D}&\cellcolor[HTML]{E8E8AB}100\%&96.64&	3.35&	94.10&	97.92&	1.51&	97.24&	98.04&	1.44&	97.36&	97.76&	1.38&	97.46&	97.76&	1.41&	97.40&	87.57&	10.86&	81.68&99.36&	0.72&	98.69\\
\hline\hline
\cellcolor[HTML]{FAEC0D}&\cellcolor[HTML]{E8E8AB}10\%&84.20&	1.21&	96.06&	84.35&	1.02&	96.67&	84.32&	1.05&	96.57&	84.35&	1.02&	96.67&	83.92&	0.62&	97.81&	63.21&	42.39&	49.24&99.88&	0.07&	99.88\\\cline{2-23}
\cellcolor[HTML]{FAEC0D}&\cellcolor[HTML]{E8E8AB}20\%&84.32&	1.38&	95.63&	84.51&	0.98&	96.80&	84.49&	1.02&	96.70&	84.51&	0.98&	96.80&	84.11&	0.59&	97.95&	64.71&	40.32&	50.88&99.81&	0.10&	99.82\\\cline{2-23}
\cellcolor[HTML]{FAEC0D}&\cellcolor[HTML]{E8E8AB}30\%&84.35&	3.44&	90.75&	85.47&	0.98&	96.98&	85.47&	0.98&	96.98&	85.45&	1.02&	96.89&	85.04&	0.59&	98.07&	65.86&	40.39&	51.49&99.81&	0.10&	99.82\\\cline{2-23}
\cellcolor[HTML]{FAEC0D}&\cellcolor[HTML]{E8E8AB}40\%&84.56&	3.44&	90.84&	85.61&	0.98&	97.01&	85.66&	0.95&	97.11&	85.59&	1.02&	96.91&	85.21&	0.56&	98.20&	65.81&	40.45&	51.44&99.83&	0.07&	99.88\\\cline{2-23}
\cellcolor[HTML]{FAEC0D}&\cellcolor[HTML]{E8E8AB}50\%&86.33&	1.28&	96.35&	86.47&	1.05&	96.98&	86.54&	0.98&	97.16&	86.47&	1.05&	96.98&	86.09&	0.59&	98.19&	71.10&	32.91&	57.67&99.83&	0.07&	99.88\\\cline{2-23}
\cellcolor[HTML]{FAEC0D}&\cellcolor[HTML]{E8E8AB}60\%&86.69&	1.28&	96.42&	86.83&	1.05&	97.03&	86.81&	1.08&	96.94&	86.81&	1.08&	96.94&	86.52&	0.59&	98.24&	71.10&	32.91&	57.67&99.83&	0.07&	99.88\\\cline{2-23}
\cellcolor[HTML]{FAEC0D}&\cellcolor[HTML]{E8E8AB}70\%&90.07&	4.07&	91.61&	90.36&	3.25&	93.09&	90.43&	2.95&	93.64&	90.43&	2.95&	93.64&	90.48&	2.99&	93.59&	59.68&	53.74&	43.12&99.43&	0.52&	99.04\\\cline{2-23}
\cellcolor[HTML]{FAEC0D}\multirow{-6}{*}{\begin{sideways}{\textbf{Intent}}\end{sideways}}&\cellcolor[HTML]{E8E8AB}80\%&90.07&	1.51&	96.44&	91.67&	2.43&	94.86&	91.55&	2.56&	94.59&	91.60&	2.49&	94.72&	91.65&	2.49&	94.73&	67.10&	43.83&	50.66&99.31&0.56&98.98\\\cline{2-23}
\cellcolor[HTML]{FAEC0D}&\cellcolor[HTML]{E8E8AB}90\%&90.26&	1.35&	96.82&	91.82&	2.30&	95.13&	91.70&	2.43&	94.86&	91.70&	2.43&	94.86&	91.74&	2.43&	94.87&	67.07&	43.77&	50.68&99.40&	0.43&	99.22\\\cline{2-23}
\cellcolor[HTML]{FAEC0D}&\cellcolor[HTML]{E8E8AB}100\%&90.05&	1.25&	97.01&	91.77&	2.33&	95.06&	91.72&	2.39&	94.92&	91.70&	2.40&	94.92&	91.67&	2.53&	94.68&	66.55&	43.04&	50.74&99.26&	0.39&	99.27\\
\hline
  \end{tabular}}
  \end{table*}

\begin{table*}[!htb]
  \centering
  \caption{\small Accuracy, FPR, AUC values for API, permission, intent data of the Contagio dataset (Acc= accuracy; FPR= false positive rate; AUC= area under cover; FR= feature length; MAA= maximum available accuracy).\vspace{-10px}}
\label{tab6}
  \scriptsize{
\setlength\tabcolsep{0.5pt} 
  \begin{tabular}{|c|c|c|c|c||c|c|c||c|c|c||c|c|c||c|c|c||c|c|c||c|c|c|}
  \hline
  \rowcolor{LightCyan} 
  \multicolumn{23}{|c|}{\textbf{Contagio Dataset}}\\\hline\hline
\rowcolor{yellow}
   &&\multicolumn{3}{c|}{\textbf{FNN}}&\multicolumn{3}{c|}{\textbf{ANN}}&\multicolumn{3}{c|}{\textbf{WANN}}&\multicolumn{3}{c|}{\textbf{KMNN}}&\multicolumn{3}{c|}{\textbf{PDME~\cite{Radkani2017}}}&\multicolumn{3}{c|}{\textbf{\textbf{FalDroid~\cite{FanMing2018}}}}&\multicolumn{3}{c|}{\textbf{MAA}}\\\hline
   \rowcolor{maroon!10}
    &\textbf{FR}&\textbf{Acc}&\textbf{FPR}&\textbf{AUC}&\textbf{Acc}&\textbf{FPR}&\textbf{AUC}&\textbf{Acc}&\textbf{FPR}&\textbf{AUC}&\textbf{Acc}&\textbf{FPR}&\textbf{AUC}&\textbf{Acc}&\textbf{FPR}&\textbf{AUC}&\textbf{Acc}&\textbf{FPR}&\textbf{AUC}&\textbf{Acc}&\textbf{FPR}&\textbf{AUC}\\
 \hline
 \cellcolor[HTML]{FAEC0D}&\cellcolor[HTML]{E8E8AB}10\%&93.35&	4.14&	81.04&	96.34&	0.91&	94.54&	96.31&	0.95&	94.37&	96.28&	0.95&	94.35&	96.43&	0.82&	95.08&	90.48&	4.58&	77.94&99.62&	0.07&	99.67\\\cline{2-23}
\cellcolor[HTML]{FAEC0D}&\cellcolor[HTML]{E8E8AB}20\%&97.36&	1.83&	91.54&	97.42&	0.36&	97.88&	97.39&	0.42&	97.52&	97.39&	0.36&	97.87&	97.22&	0.62&	96.44&	89.22&	9.47&	50.14&99.38&	0.23&	98.86\\\cline{2-23}
\cellcolor[HTML]{FAEC0D}&\cellcolor[HTML]{E8E8AB}30\%&98.48&	0.85&	95.85&	98.56&	0.23&	98.77&	98.51&	0.29&	98.43&	98.51&	0.26&	98.60&	98.39&	0.29&	98.41&	91.85&	7.30&	62.25&99.33&	0.10&	99.50\\\cline{2-23}
\cellcolor[HTML]{FAEC0D}&\cellcolor[HTML]{E8E8AB}40\%&98.56&	0.69&	96.59&	98.62&	0.33&	98.28&	98.71&	0.20&	98.96&	98.56&	0.39&	97.95&	98.59&	0.29&	98.44&	92.15&	7.43&	61.33&99.38&	0.13&	99.34\\\cline{2-23}
\cellcolor[HTML]{FAEC0D}&\cellcolor[HTML]{E8E8AB}50\%&98.56&	0.69&	96.59&	98.65&	0.36&	98.13&	98.59&	0.42&	97.80&	98.62&	0.39&	97.96&	98.65&	0.42&	97.81&	92.53&	7.40&	61.34&99.30&	0.23&	98.85\\\cline{2-23}
\cellcolor[HTML]{FAEC0D}&\cellcolor[HTML]{E8E8AB}60\%&98.68&	0.49&	97.51&	98.71&	0.26&	98.62&	98.65&	0.33&	98.29&	98.71&	0.26&	98.62&	98.42&	0.55&	97.14&	92.76&	7.10&	63.07&99.36&	0.13&	99.34\\\cline{2-23}
\cellcolor[HTML]{FAEC0D}&\cellcolor[HTML]{E8E8AB}70\%&98.74&	0.52&	97.37&	98.71&	0.26&	98.62&	98.65&	0.33&	98.29&	98.65&	0.33&	98.29&	98.59&	0.49&	97.49&	92.44&	7.59&	60.25&99.36&	0.16&	99.18\\\cline{2-23}
\cellcolor[HTML]{FAEC0D}\multirow{-5}{*}{\begin{sideways}{\textbf{API}}\end{sideways}}&\cellcolor[HTML]{E8E8AB}80\%&98.83&	0.62&	96.95&	98.89&	0.52&	97.41&	98.92&	0.52&	97.41&	98.89&	0.52&	97.41&	98.80&	0.33&	97.54&	92.56&	7.21&	60.88&99.33&0.07&99.50\\\cline{2-23}
\cellcolor[HTML]{FAEC0D}&\cellcolor[HTML]{E8E8AB}90\%&98.92&	0.55&	97.27&	98.95&	0.49&	97.57&	99.03&	0.42&	97.89&	99.00&	0.42&	97.88&	99.03&	0.33&	98.35&	92.65&	7.21&	62.44&99.38&	0.07&	99.67\\\cline{2-23}
\cellcolor[HTML]{FAEC0D}&\cellcolor[HTML]{E8E8AB}100\%&98.95&	0.52&	97.42&	99.00&	0.46&	97.73&	99.00&	0.42&	97.88&	99.03&	0.42&	97.89&	98.86&	0.39&	98.01&	92.24&	7.55&	60.55&99.33&	0.10&	99.50\\
\hline\hline
\cellcolor[HTML]{FAEC0D}&\cellcolor[HTML]{E8E8AB}10\%&98.33&	0.65&	96.51&	98.57&	0.39&	97.86&	98.48&	0.32&	98.18&	98.57&	0.39&	97.86&	98.24&	0.58&	96.80&	92.48&	5.42&	71.46&99.77&	0.13&	99.33\\\cline{2-23}
\cellcolor[HTML]{FAEC0D}&\cellcolor[HTML]{E8E8AB}20\%&98.33&	0.68&	96.36&	98.60&	0.39&	97.86&	98.51&	0.32&	98.18&	98.60&	0.39&	97.86&	98.27&	0.58&	96.81&	92.19&	5.41&	71.62&99.77&	0.16&	99.17\\\cline{2-23}
\cellcolor[HTML]{FAEC0D}&\cellcolor[HTML]{E8E8AB}30\%&98.42&	0.62&	96.70&	98.60&	0.39&	97.86&	98.62&	0.29&	98.37&	98.60&	0.39&	97.86&	98.27&	0.62&	96.65&	91.95&	5.40&	71.78&99.80&	0.13&	99.34\\\cline{2-23}
\cellcolor[HTML]{FAEC0D}&\cellcolor[HTML]{E8E8AB}40\%&98.42&	0.62&	96.70&	98.62&	0.36&	98.04&	98.62&	0.29&	98.37&	98.62&	0.36&	98.04&	98.24&	0.62&	96.48&	91.66&	5.39&	71.93&99.80&	0.13&	99.34\\\cline{2-23}
\cellcolor[HTML]{FAEC0D}&\cellcolor[HTML]{E8E8AB}50\%&98.48&	0.62&	96.71&	98.68&	0.36&	98.05&	98.68&	0.29&	98.38&	98.68&	0.36&	98.05&	98.22&	0.65&	96.16&	91.63&	5.39&	71.93&99.77&	0.16&	99.17\\\cline{2-23}
\cellcolor[HTML]{FAEC0D}&\cellcolor[HTML]{E8E8AB}60\%&98.39&	0.71&	96.23&	98.68&	0.36&	98.05&	98.68&	0.29&	98.38&	98.68&	0.36&	98.05&	98.22&	0.71&	96.16&	91.43&	5.40&	71.92&99.80&	0.16&	99.17\\\cline{2-23}
\cellcolor[HTML]{FAEC0D}&\cellcolor[HTML]{E8E8AB}70\%&98.54&	0.52&	97.20&	98.48&	0.42&	97.67&	98.57&	0.32&	98.19&	98.48&	0.42&	97.67&	98.39&	0.45&	97.49&	92.33&	5.37&	71.79&99.65&	0.26&	98.68\\\cline{2-23}
\cellcolor[HTML]{FAEC0D}\multirow{-6}{*}{\begin{sideways}{\textbf{Permission}}\end{sideways}}&\cellcolor[HTML]{E8E8AB}80\%&98.71&	0.55&	97.09&	98.80&	0.42&	97.75&	98.80&	0.42&	97.75&	98.83&	0.39&	97.91&	98.54&	0.55&	97.04&	87.30&	6.53&	66.86&99.50&0.23&98.83\\\cline{2-23}
\cellcolor[HTML]{FAEC0D}&\cellcolor[HTML]{E8E8AB}90\%&98.68&	0.58&	96.93&	98.77&	0.45&	97.58&	98.83&	0.39&	97.91&	98.80&	0.45&	97.59&	98.54&	0.55&	97.04&	87.30&	6.53&	66.86&99.50&	0.23&	98.83\\\cline{2-23}
\cellcolor[HTML]{FAEC0D}&\cellcolor[HTML]{E8E8AB}100\%&98.62&	0.65&	96.61&	98.71&	0.52&	97.25&	98.74&	0.49&	97.41&	98.71&	0.52&	97.25&	98.54&	0.55&	97.04&	86.98&	6.55&	66.84&99.44&	0.29&	98.50\\
\hline\hline
\cellcolor[HTML]{FAEC0D}&\cellcolor[HTML]{E8E8AB}10\%&96.64&	1.23&	92.92&	97.31&	0.84&	95.14&	97.28&	0.88&	94.97&	97.34&	0.81&	95.31&	97.28&	0.88&	94.97&	45.41&	2.49&	93.95&99.91&	0.03&	99.83\\\cline{2-23}
\cellcolor[HTML]{FAEC0D}&\cellcolor[HTML]{E8E8AB}20\%&97.05&	0.97&	94.38&	97.43&	0.75&	95.68&	97.54&	0.62&	96.38&	97.43&	0.75&	95.68&	97.45&	0.71&	95.85&	45.23&	2.51&	93.94&99.88&	0.06&	99.67\\\cline{2-23}
\cellcolor[HTML]{FAEC0D}&\cellcolor[HTML]{E8E8AB}30\%&96.90&	0.71&	95.58&	97.54&	0.68&	96.05&	97.54&	0.68&	96.05&	97.51&	0.68&	96.04&	97.48&	0.75&	95.70&	45.23&	2.58&	93.75&99.85&	0.10&	99.50\\\cline{2-23}
\cellcolor[HTML]{FAEC0D}&\cellcolor[HTML]{E8E8AB}40\%&96.99&	0.75&	95.46&	97.60&	0.75&	95.76&	97.72&	0.65&	96.29&	97.57&	0.75&	95.74&	97.66&	0.71&	95.94&	45.06&	2.96&	92.81&99.85&	0.10&	99.50\\\cline{2-23}
\cellcolor[HTML]{FAEC0D}&\cellcolor[HTML]{E8E8AB}50\%&97.02&	0.71&	95.64&	97.63&	0.71&	95.93&	97.66&	0.71&	95.94&	97.57&	0.75&	95.74&	97.69&	0.71&	96.11&	44.97&	2.97&	92.81&99.91&	0.03&	99.83\\\cline{2-23}
\cellcolor[HTML]{FAEC0D}&\cellcolor[HTML]{E8E8AB}60\%&97.57&	0.58&	96.56&	98.04&	0.32&	98.09&	98.04&	0.36&	97.91&	98.01&	0.32&	98.08&	98.01&	0.39&	97.73&	45.47&	2.71&	93.39&99.91&	0.03&	99.83\\\cline{2-23}
\cellcolor[HTML]{FAEC0D}&\cellcolor[HTML]{E8E8AB}70\%&98.10&	0.88&	95.37&	98.54&	0.45&	97.52&	98.51&	0.45&	97.52&	98.51&	0.45&	97.52&	98.42&	0.55&	97.01&	78.29&	1.53&	93.68&99.77&	0.10&	99.50\\\cline{2-23}
\cellcolor[HTML]{FAEC0D}\multirow{-6}{*}{\begin{sideways}{\textbf{Intent}}\end{sideways}}&\cellcolor[HTML]{E8E8AB}80\%&98.19&	0.78&	95.85&	98.51&	0.49&	97.35&	98.57&	0.39&	97.85&	98.54&	0.45&	97.52&	98.45&	0.52&	97.18&	79.93&	1.54&	93.53&99.74&0.13&99.33\\\cline{2-23}
\cellcolor[HTML]{FAEC0D}&\cellcolor[HTML]{E8E8AB}90\%&98.16&	0.97&	94.97&	98.60&	0.39&	97.86&	98.62&	0.32&	98.20&	98.62&	0.32&	98.20&	98.48&	0.45&	97.51&	82.33&	1.49&	93.55&99.68&	0.13&	99.33\\\cline{2-23}
\cellcolor[HTML]{FAEC0D}&\cellcolor[HTML]{E8E8AB}100\%&98.10&	1.00&	94.80&	98.54&	0.45&	97.52&	98.62&	0.36&	98.03&	98.51&	0.45&	97.52&	98.57&	0.36&	98.02&	82.83&	1.48&	93.56&99.68&	0.13&	99.33\\
\hline
  \end{tabular}}
  \end{table*}

  \begin{table*}[!htb]
  \centering
  \caption{\small Accuracy, FPR, AUC values for API, permission, intent data of the Genome dataset (Acc= accuracy; FPR= false positive rate; AUC= area under cover; FR= feature length; MAA= maximum available accuracy).\vspace{-10px}}
\label{tab7}
  \scriptsize{
\setlength\tabcolsep{0.5pt} 
  \begin{tabular}{|c|c|c|c|c||c|c|c||c|c|c||c|c|c||c|c|c||c|c|c||c|c|c|}
  \hline
  \rowcolor{LightCyan} 
  \multicolumn{23}{|c|}{\textbf{Genome Dataset}}\\\hline\hline
\rowcolor{yellow}
   &&\multicolumn{3}{c|}{\textbf{FNN}}&\multicolumn{3}{c|}{\textbf{ANN}}&\multicolumn{3}{c|}{\textbf{WANN}}&\multicolumn{3}{c|}{\textbf{KMNN}}&\multicolumn{3}{c|}{\textbf{PDME~\cite{Radkani2017}}}&\multicolumn{3}{c|}{\textbf{\textbf{FalDroid~\cite{FanMing2018}}}}&\multicolumn{3}{c|}{\textbf{MAA}}\\\hline
   \rowcolor{maroon!10}
    &\textbf{FR}&\textbf{Acc}&\textbf{FPR}&\textbf{AUC}&\textbf{Acc}&\textbf{FPR}&\textbf{AUC}&\textbf{Acc}&\textbf{FPR}&\textbf{AUC}&\textbf{Acc}&\textbf{FPR}&\textbf{AUC}&\textbf{Acc}&\textbf{FPR}&\textbf{AUC}&\textbf{Acc}&\textbf{FPR}&\textbf{AUC}&\textbf{Acc}&\textbf{FPR}&\textbf{AUC}\\
 \hline
 \cellcolor[HTML]{FAEC0D}&\cellcolor[HTML]{E8E8AB}10\%&96.37&	2.84&	83.56&	96.58&	2.52&	84.81&	96.34&	2.78&	83.68&	96.46&	2.65&	84.24&	96.49&	2.62&	84.38&	55.73&	4.70&	32.50&99.43&	0.35&	97.54\\\cline{2-23}
\cellcolor[HTML]{FAEC0D}&\cellcolor[HTML]{E8E8AB}20\%&98.65&	0.81&	94.45&	98.74&	0.74&	94.88&	98.86&	0.68&	95.32&	98.71&	0.78&	94.67&	98.80&	0.45&	96.68&	86.19&	4.24&	57.31&99.28&	0.42&	97.08\\\cline{2-23}
\cellcolor[HTML]{FAEC0D}&\cellcolor[HTML]{E8E8AB}30\%&99.34&	0.36&	97.51&	99.37&	0.32&	97.73&	99.34&	0.36&	97.51&	99.37&	0.32&	97.73&	99.19&	0.48&	96.65&	90.19&	1.47&	65.77&99.43&	0.32&	97.75\\\cline{2-23}
\cellcolor[HTML]{FAEC0D}&\cellcolor[HTML]{E8E8AB}40\%&99.46&	0.23&	98.39&	99.55&	0.16&	98.85&	99.40&	0.29&	97.95&	99.49&	0.19&	98.62&	99.34&	0.26&	98.15&	95.41&	2.02&	81.49&99.67&	0.12&	99.08\\\cline{2-23}
\cellcolor[HTML]{FAEC0D}&\cellcolor[HTML]{E8E8AB}50\%&99.58&	0.23&	98.85&	99.61&	0.16&	99.08&	99.67&	0.29&	98.86&	99.58&	0.19&	99.07&	99.52&	0.23&	98.40&	96.85&	1.84&	88.74&99.79&	0.06&	99.54\\\cline{2-23}
\cellcolor[HTML]{FAEC0D}&\cellcolor[HTML]{E8E8AB}60\%&99.79&	0.06&	99.54&	99.70&	0.13&	99.09&	99.70&	0.16&	98.87&	99.70&	0.10&	99.31&	99.49&	0.36&	97.56&	96.25&	2.74&	89.16&99.82&	0.06&	99.54\\\cline{2-23}
\cellcolor[HTML]{FAEC0D}&\cellcolor[HTML]{E8E8AB}70\%&99.76&	0.10&	99.32&	99.82&	0.06&	99.54&	99.85&	0.06&	99.55&	99.79&	0.06&	99.54&	99.70&	0.19&	98.65&	96.94&	2.11&	90.76&99.85&	0.06&	99.55\\\cline{2-23}
\cellcolor[HTML]{FAEC0D}\multirow{-5}{*}{\begin{sideways}{\textbf{API}}\end{sideways}}&\cellcolor[HTML]{E8E8AB}80\%&99.73&	0.13&	99.09&	99.79&	0.10&	99.32&	99.79&	0.10&	99.32&	99.76&	0.10&	99.32&	99.79&	0.16&	98.88&	97.21&	1.67&	90.31&99.82&	0.06&	99.54\\\cline{2-23}
\cellcolor[HTML]{FAEC0D}&\cellcolor[HTML]{E8E8AB}90\%&99.76&	0.06&	99.54&	99.76&	0.06&	99.54&	99.76&	0.06&	99.54&	99.73&	0.06&	99.54&	99.76&	0.10&	99.32&	98.11&	1.10&	93.31&99.82&	0.03&	99.77\\\cline{2-23}
\cellcolor[HTML]{FAEC0D}&\cellcolor[HTML]{E8E8AB}100\%&99.82&	0.03&	99.77&	99.79&	0.03&	99.77&	99.79&	0.06&	99.54&	99.76&	0.06&	99.54&	99.76&	0.10&	99.32&	98.05&	1.29&	93.99&99.85&	0.03&	99.77\\
\hline\hline
\cellcolor[HTML]{FAEC0D}&\cellcolor[HTML]{E8E8AB}10\%&90.85&	7.05&	66.94&	96.85&	0.42&	95.70&	96.79&	0.32&	96.53&	96.79&	0.52&	94.83&	96.76&	0.32&	96.51&	93.04&	2.71&	74.50&99.73&	0.19&	98.66\\\cline{2-23}
\cellcolor[HTML]{FAEC0D}&\cellcolor[HTML]{E8E8AB}20\%&90.85&	7.05&	66.94&	96.79&	0.55&	94.57&	96.73&	0.45&	95.32&	96.73&	0.61&	94.00&	96.73&	0.42&	95.60&	90.85&	5.49&	66.12&99.70&	0.23&	98.44\\\cline{2-23}
\cellcolor[HTML]{FAEC0D}&\cellcolor[HTML]{E8E8AB}30\%&90.37&	7.56&	65.83&	96.82&	0.52&	94.86&	96.76&	0.42&	95.62&	96.76&	0.61&	94.04&	96.73&	0.42&	95.60&	89.83&	6.98&	63.88&99.73&	0.19&	98.66\\\cline{2-23}
\cellcolor[HTML]{FAEC0D}&\cellcolor[HTML]{E8E8AB}40\%&90.37&	7.63&	65.85&	96.79&	0.55&	94.57&	96.73&	0.45&	95.32&	96.76&	0.61&	94.04&	96.70&	0.45&	95.29&	89.02&	7.53&	61.77&99.73&	0.19&	98.66\\\cline{2-23}
\cellcolor[HTML]{FAEC0D}&\cellcolor[HTML]{E8E8AB}50\%&90.46&	7.63&	66.08&	96.79&	0.55&	94.57&	96.73&	0.45&	95.32&	96.73&	0.65&	93.76&	96.70&	0.45&	95.29&	88.60&	8.76&	61.75&99.73&	0.19&	98.66\\\cline{2-23}
\cellcolor[HTML]{FAEC0D}&\cellcolor[HTML]{E8E8AB}60\%&99.40&	0.42&	97.13&	99.40&	0.39&	97.33&	99.52&	0.23&	98.40&	99.37&	0.42&	97.12&	98.89&	0.55&	96.11&	87.91&	12.61&	61.99&99.79&	0.10&	99.32\\\cline{2-23}
\cellcolor[HTML]{FAEC0D}&\cellcolor[HTML]{E8E8AB}70\%&99.37&	0.48&	96.72&	99.49&	0.32&	97.76&	99.46&	0.32&	97.76&	99.46&	0.36&	97.55&	98.80&	0.68&	95.29&	91.00&	9.08&	67.36&99.76&	0.16&	98.88\\\cline{2-23}
\cellcolor[HTML]{FAEC0D}\multirow{-6}{*}{\begin{sideways}{\textbf{Permission}}\end{sideways}}&\cellcolor[HTML]{E8E8AB}80\%&99.37&	0.48&	96.72&	99.43&	0.39&	97.34&	99.49&	0.29&	97.97&	99.40&	0.42&	97.13&	98.86&	0.68&	95.32&	91.33&	9.02&	67.98&99.73&	0.19&	98.66\\\cline{2-23}
\cellcolor[HTML]{FAEC0D}&\cellcolor[HTML]{E8E8AB}90\%&99.40&	0.48&	96.73&	99.43&	0.39&	97.34&	99.46&	0.32&	97.76&	99.46&	0.39&	97.35&	98.86&	0.68&	95.32&	90.91&	9.47&	67.15&99.73&	0.19&	98.66\\\cline{2-23}
\cellcolor[HTML]{FAEC0D}&\cellcolor[HTML]{E8E8AB}100\%&99.37&	0.52&	96.53&	99.43&	0.39&	97.34&	99.46&	0.32&	97.76&	99.43&	0.42&	97.14&	98.86&	0.68&	95.32&	90.97&	9.41&	67.26&99.73&	0.19&	98.66\\
\hline\hline
\cellcolor[HTML]{FAEC0D}&\cellcolor[HTML]{E8E8AB}10\%&97.87&	0.32&	97.16&	98.02&	0.55&	95.60&	97.87&	0.58&	95.28&	98.02&	0.55&	95.60&	97.99&	0.58&	95.36&	96.01&	2.26&	84.58&99.91&	0.00&	100.00\\\cline{2-23}
\cellcolor[HTML]{FAEC0D}&\cellcolor[HTML]{E8E8AB}20\%&98.17&	0.32&	97.30&	98.23&	0.32&	97.33&	98.11&	0.32&	97.27&	98.23&	0.32&	97.33&	98.23&	0.32&	97.33&	96.76&	1.55&	88.60&99.94&	0.00&	100.00\\\cline{2-23}
\cellcolor[HTML]{FAEC0D}&\cellcolor[HTML]{E8E8AB}30\%&98.17&	0.36&	97.06&	98.23&	0.36&	97.09&	98.11&	0.36&	97.03&	98.23&	0.36&	97.09&	98.23&	0.36&	97.09&	96.82&	1.55&	88.70&99.94&	0.00&	100.00\\\cline{2-23}
\cellcolor[HTML]{FAEC0D}&\cellcolor[HTML]{E8E8AB}40\%&98.17&	0.36&	97.06&	98.23&	0.36&	97.09&	98.11&	0.36&	97.03&	98.23&	0.36&	97.09&	98.23&	0.36&	97.09&	96.82&	1.55&	88.70&99.94&	0.00&	100.00\\\cline{2-23}
\cellcolor[HTML]{FAEC0D}&\cellcolor[HTML]{E8E8AB}50\%&98.23&	0.36&	97.09&	98.26&	0.36&	97.10&	98.14&	0.36&	97.04&	98.26&	0.36&	97.10&	98.26&	0.36&	97.10&	96.64&	1.81&	87.33&99.94&	0.00&	100.00\\\cline{2-23}
\cellcolor[HTML]{FAEC0D}&\cellcolor[HTML]{E8E8AB}60\%&98.23&	0.36&	97.09&	98.23&	0.39&	96.85&	98.14&	0.36&	97.04&	98.23&	0.39&	96.85&	98.26&	0.36&	97.10&	96.70&	1.75&	87.69&99.94&	0.00&	100.00\\\cline{2-23}
\cellcolor[HTML]{FAEC0D}&\cellcolor[HTML]{E8E8AB}70\%&98.74&	0.55&	96.03&	99.01&	0.19&	98.52&	98.95&	0.23&	98.27&	99.01&	0.19&	98.52&	98.95&	0.16&	98.75&	83.37&	17.39&	55.91&99.94&	0.06&	99.55\\\cline{2-23}
\cellcolor[HTML]{FAEC0D}\multirow{-6}{*}{\begin{sideways}{\textbf{Intent}}\end{sideways}}&\cellcolor[HTML]{E8E8AB}80\%&98.74&	0.42&	96.86&	98.83&	0.42&	96.90&	98.80&	0.42&	96.89&	98.80&	0.42&	96.89&	98.56&	0.48&	96.35&	86.49&	14.03&	59.91&99.91&	0.03&	99.77\\\cline{2-23}
\cellcolor[HTML]{FAEC0D}&\cellcolor[HTML]{E8E8AB}90\%&98.68&	0.45&	96.62&	98.89&	0.36&	97.36&	98.86&	0.36&	97.34&	98.83&	0.42&	96.90&	98.62&	0.42&	96.81&	85.98&	14.61&	59.21&99.91&	0.03&	99.77\\\cline{2-23}
\cellcolor[HTML]{FAEC0D}&\cellcolor[HTML]{E8E8AB}100\%&98.95&	0.16&	98.75&	99.07&	0.16&	98.77&	99.07&	0.13&	99.01&	99.04&	0.19&	98.53&	98.83&	0.19&	98.48&	80.61&	20.39&	52.86&99.91&	0.03&	99.77\\
\hline
  \end{tabular}}
  \end{table*}
Some other remarks for Table~\ref{tab5} are in order. We understand that increasing the number of features increases the accuracy and AUC, and decreases the FPR. Increasing the amount of AUC means that algorithms with higher precision can separate malware and benign samples. It is also observed that the four proposed methods and the PDME algorithm~\cite{Radkani2017} with accuracy more than 99\% have a high AUC. This process has also been repeated for other features, namely permission and intent, as well as Contagio and Genome datasets. Hence, in best case, accuracy and AUC rates are not 100\% and the FPR value is not even zero.

\subsubsection{Comparing methods based on ROC}\label{ROC}

\begin{figure*}[!htb]
    \begin{subfigure}{0.32\textwidth}
 		\centering 		\includegraphics[width=1.05\linewidth]{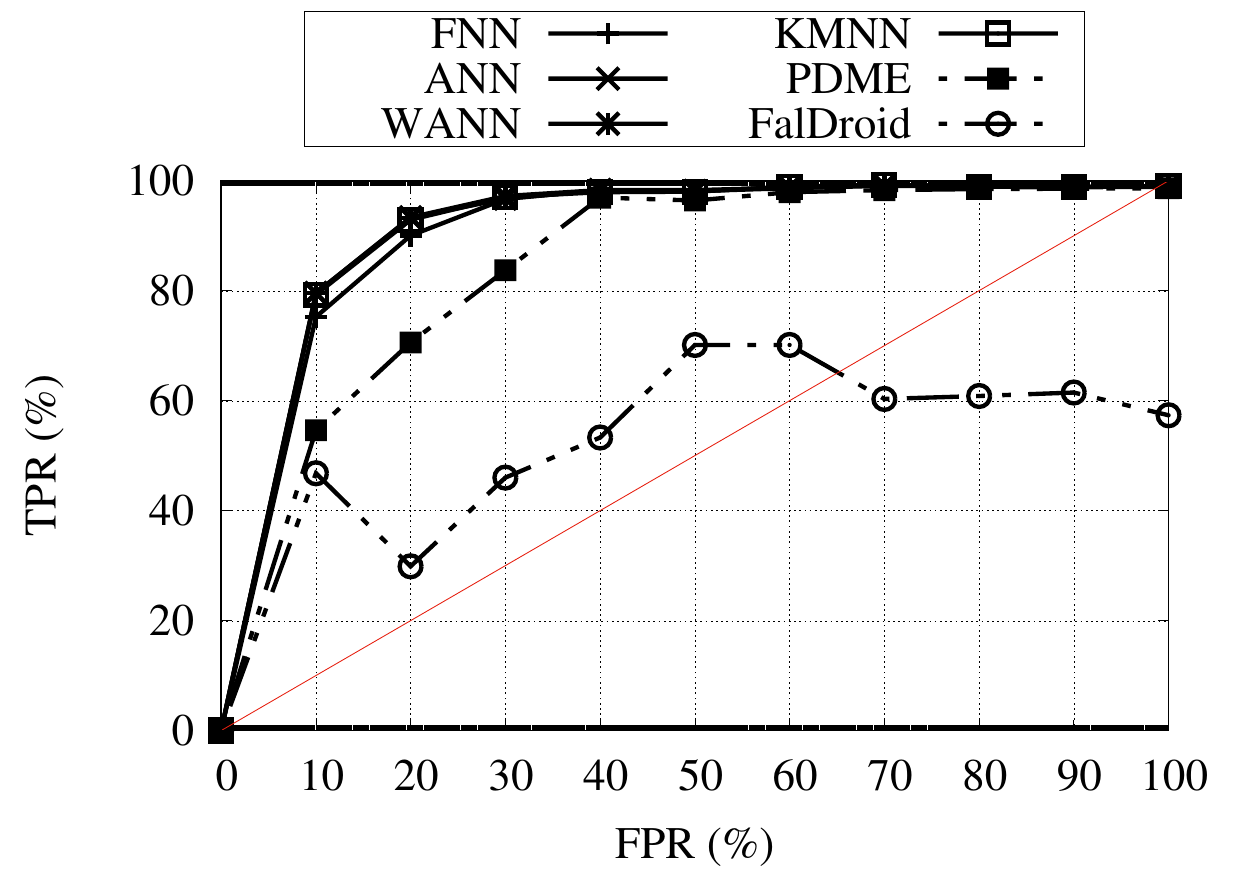}
 		\caption{\small Drebin-API\vspace{-5px}}
			\label{fig:fig33a}
 	\end{subfigure} 
   \begin{subfigure}{0.32\textwidth}
 		\centering 		\includegraphics[width=1.05\linewidth]{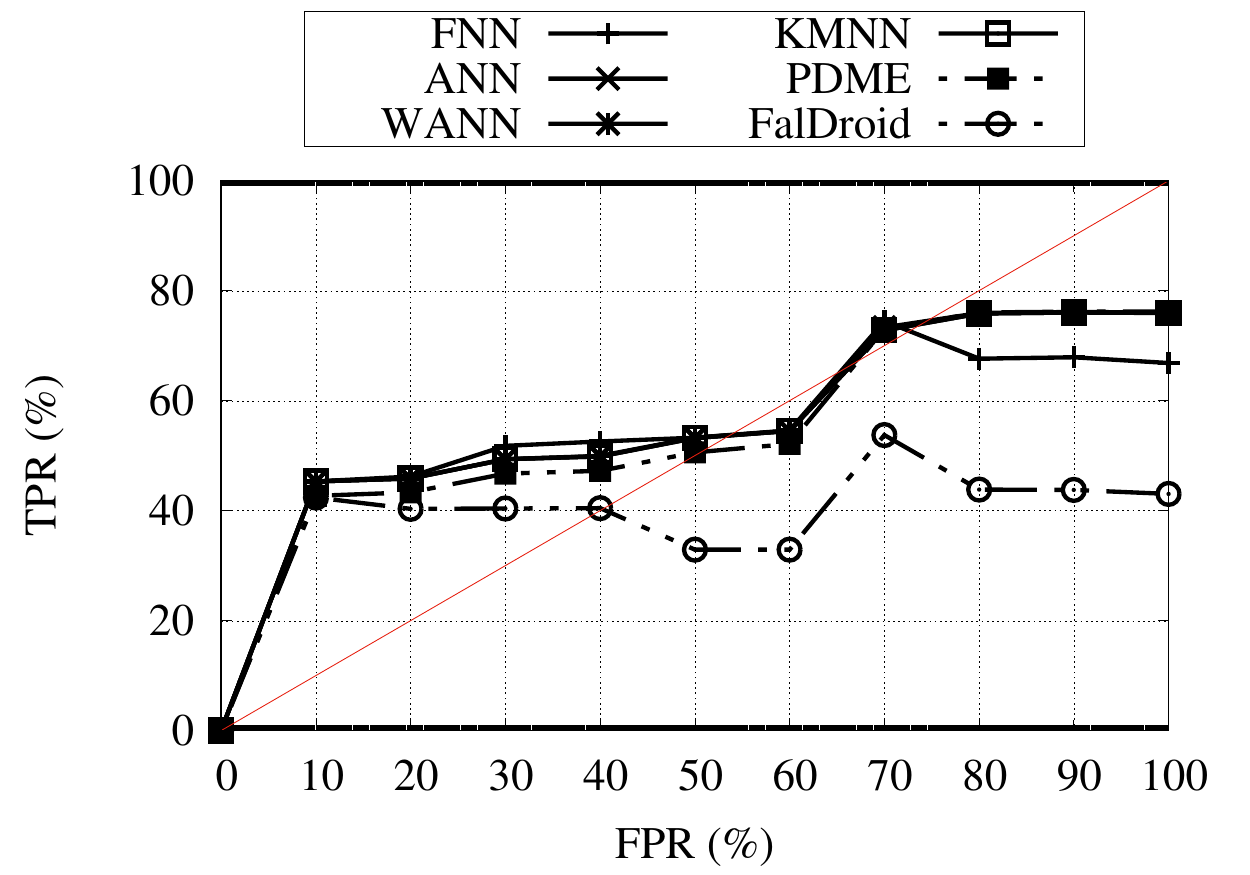}
			\caption{\small Drebin-intent\vspace{-5px}}
			\label{fig:fig33b}
 	\end{subfigure} 
    \begin{subfigure}{0.32\textwidth}
 		\centering 		\includegraphics[width=1.05\linewidth]{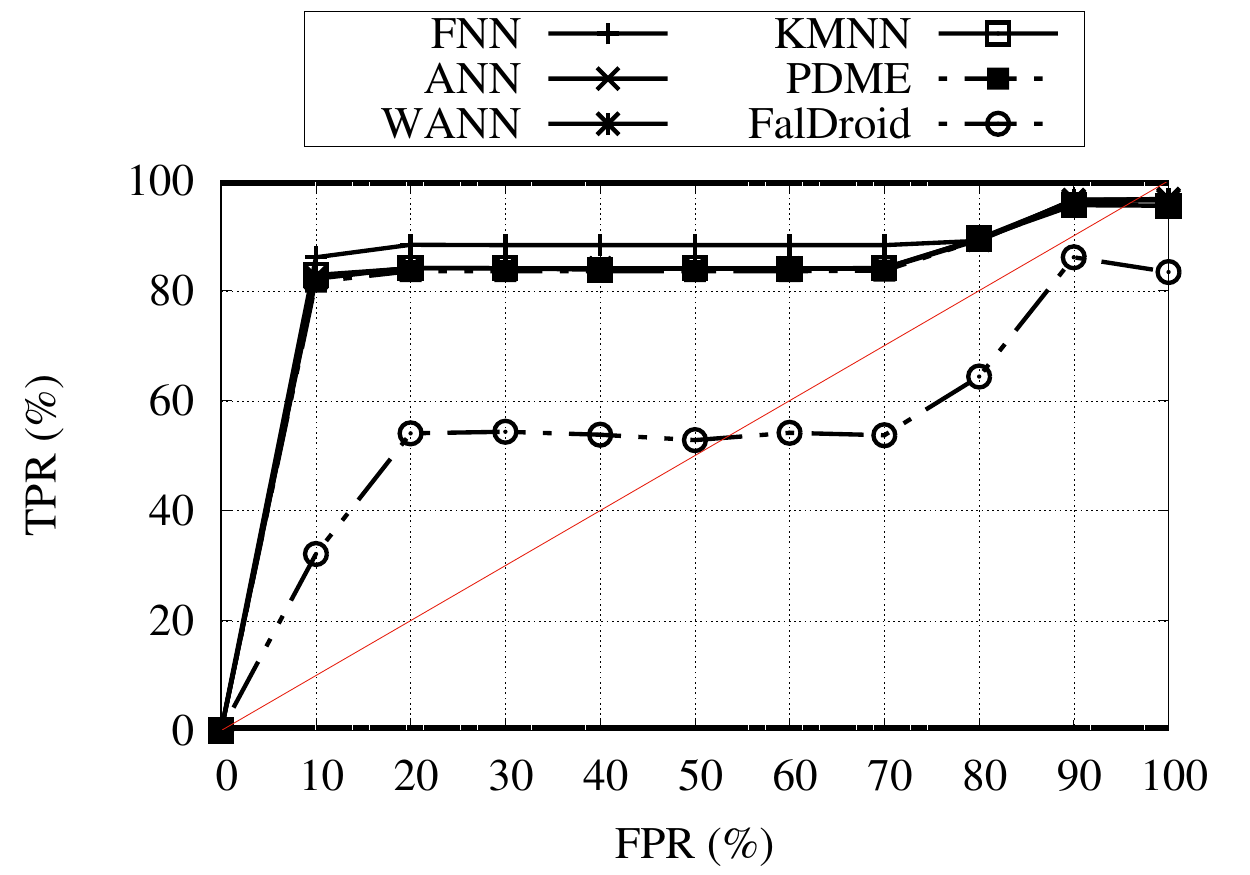}
		\caption{\small Drebin-permission\vspace{-5px}}
			\label{fig:fig33c}
 	\end{subfigure}
        \caption{\small Comparison of algorithms in Drebin dataset with reference to ROC. (TPR= true positive rate; FPR= false positive rate).\vspace{-10px}}
			\label{fig:fig33}
		\end{figure*}

\begin{figure*}[!htb]
    \begin{subfigure}{0.32\textwidth}
 		\centering 		\includegraphics[width=1.05\linewidth]{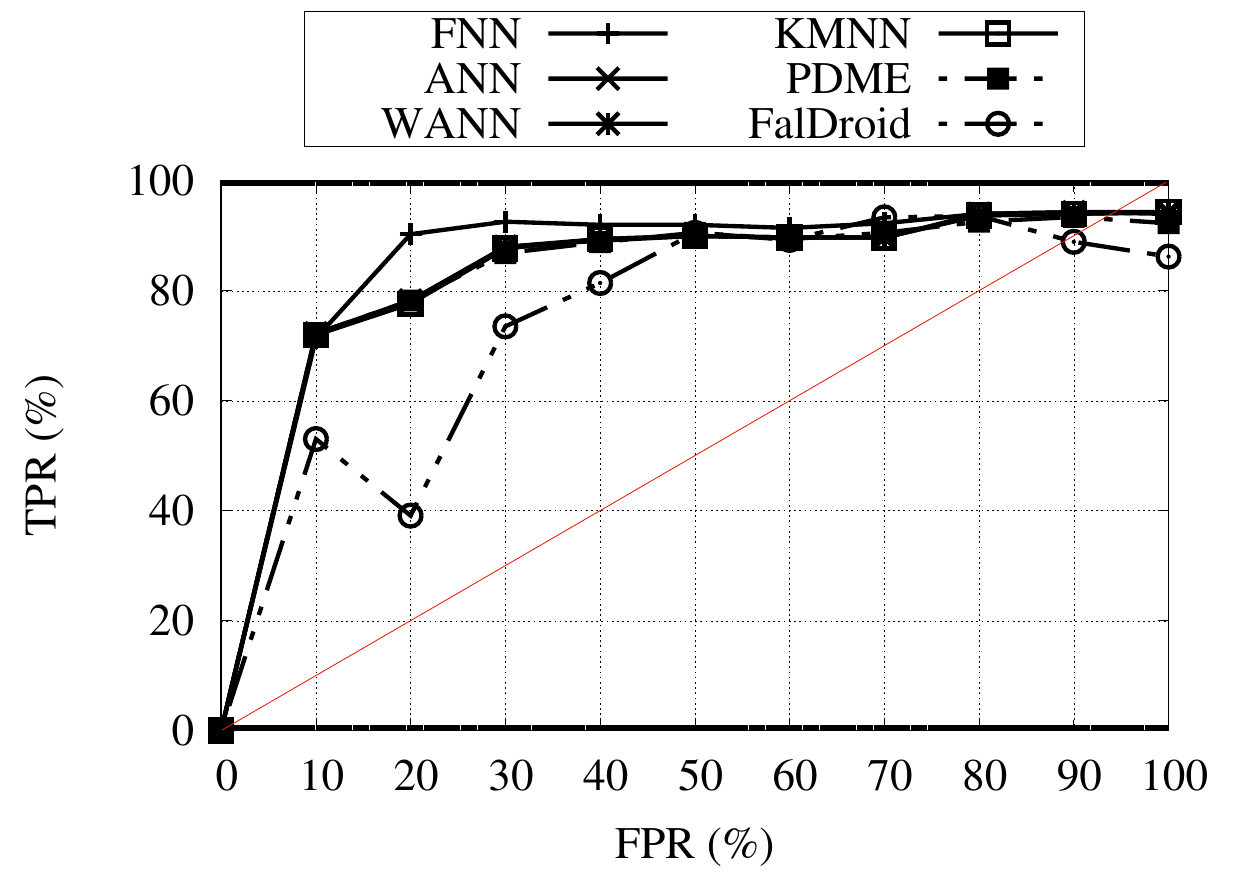}
 		\caption{\small Contagio-API\vspace{-5px}}
			\label{fig:fig331a}
 	\end{subfigure} 
   \begin{subfigure}{0.32\textwidth}
 		\centering 		\includegraphics[width=1.05\linewidth]{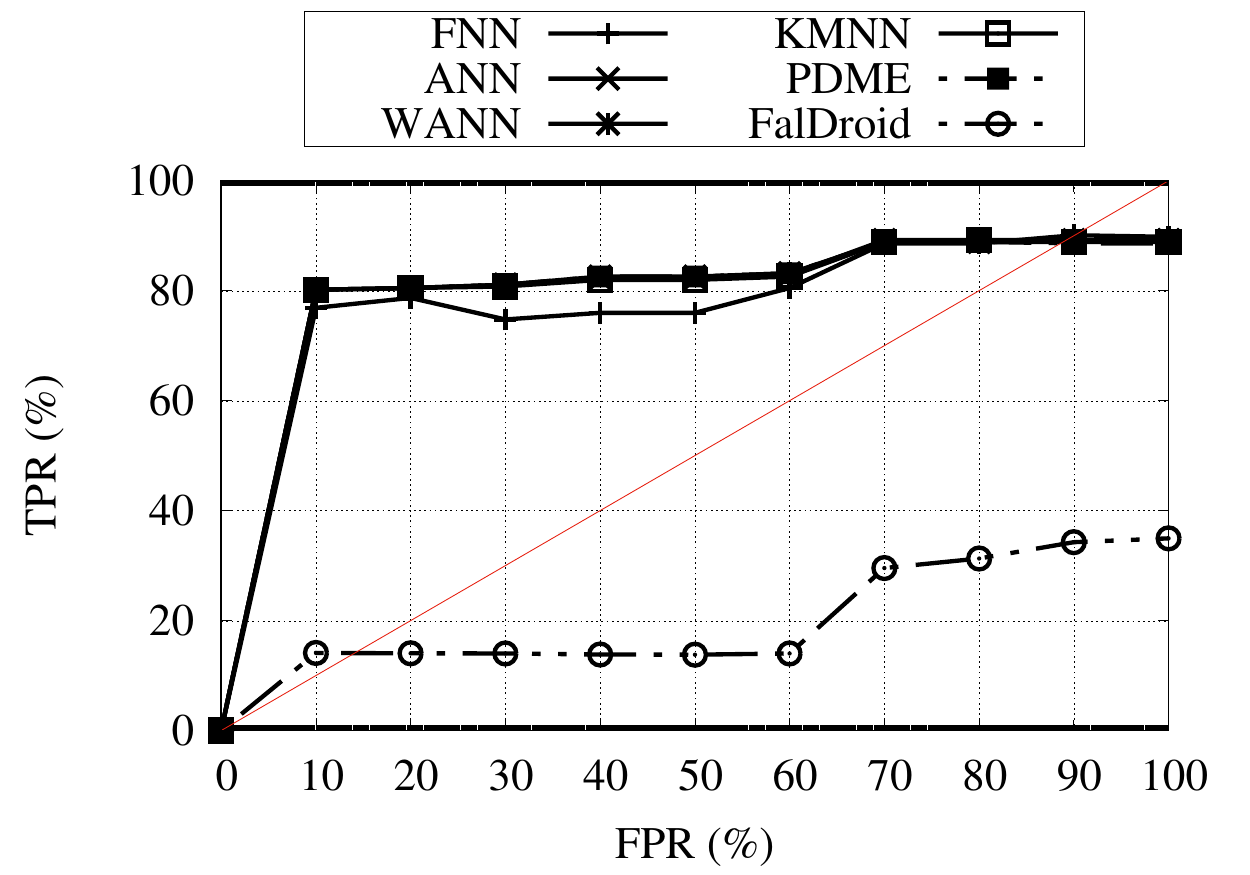}
			\caption{\small Contagio-intent\vspace{-5px}}
			\label{fig:fig331b}
 	\end{subfigure} 
    \begin{subfigure}{0.32\textwidth}
 		\centering 		\includegraphics[width=1.05\linewidth]{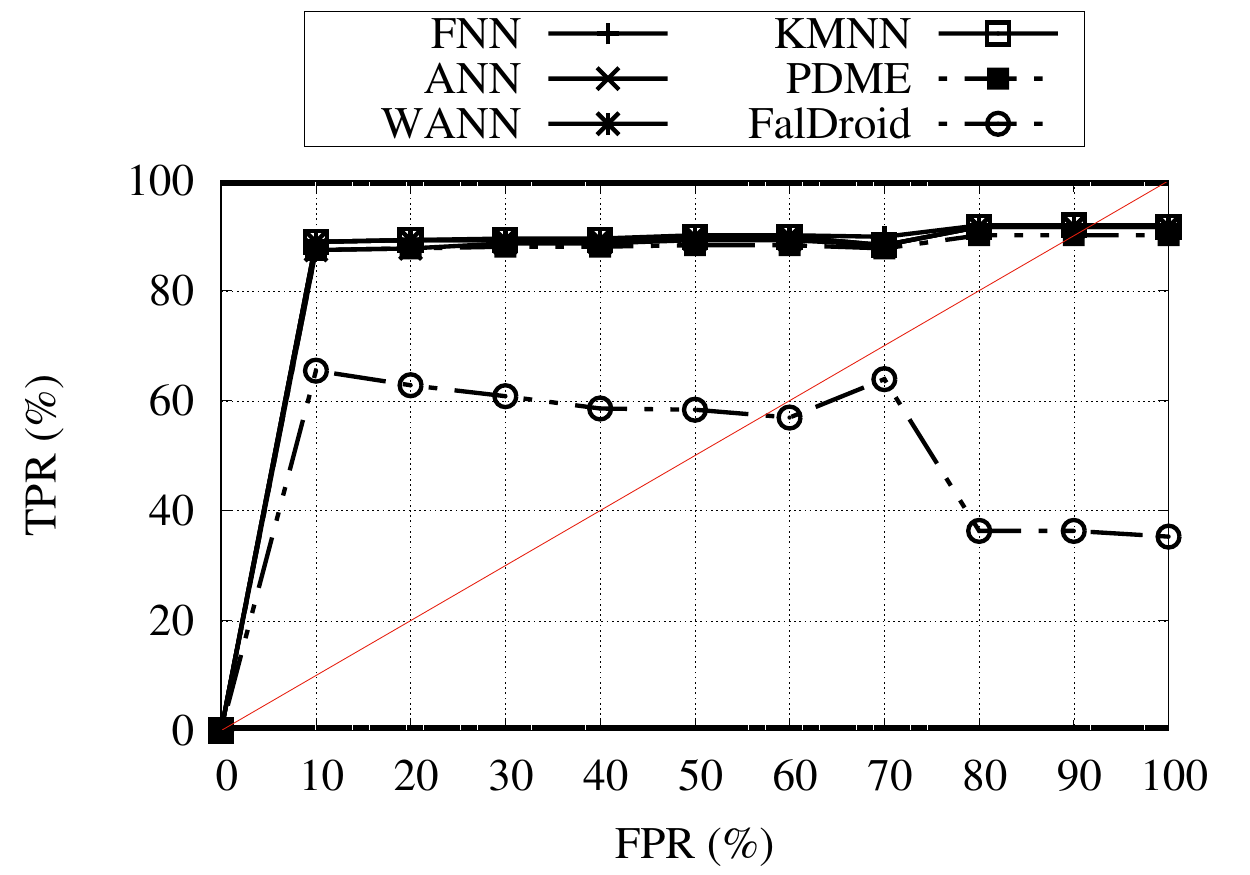}
		\caption{\small Contagio-permission\vspace{-5px}}
			\label{fig:fig331c}
 	\end{subfigure}
        \caption{\small Comparison of algorithms in Contagio dataset with reference to ROC. (TPR= true positive rate; FPR= false positive rate).\vspace{-10px}}
			\label{fig:fig331}
		\end{figure*}

\begin{figure*}[!htb]
    \begin{subfigure}{0.32\textwidth}
 		\centering 		\includegraphics[width=1.05\linewidth]{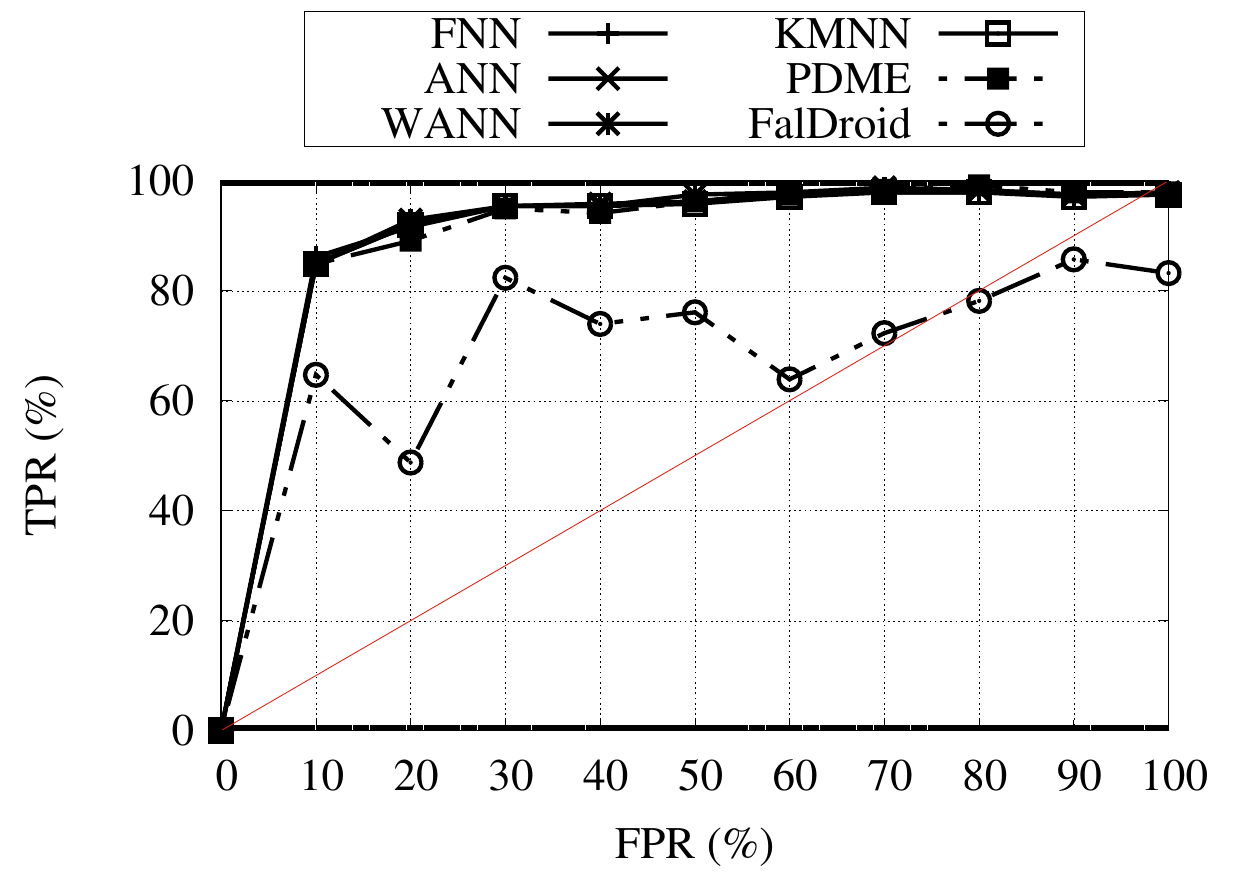}
 		\caption{\small Genome-API\vspace{-5px}}
			\label{fig:fig332a}
 	\end{subfigure} 
   \begin{subfigure}{0.32\textwidth}
 		\centering 		\includegraphics[width=1.05\linewidth]{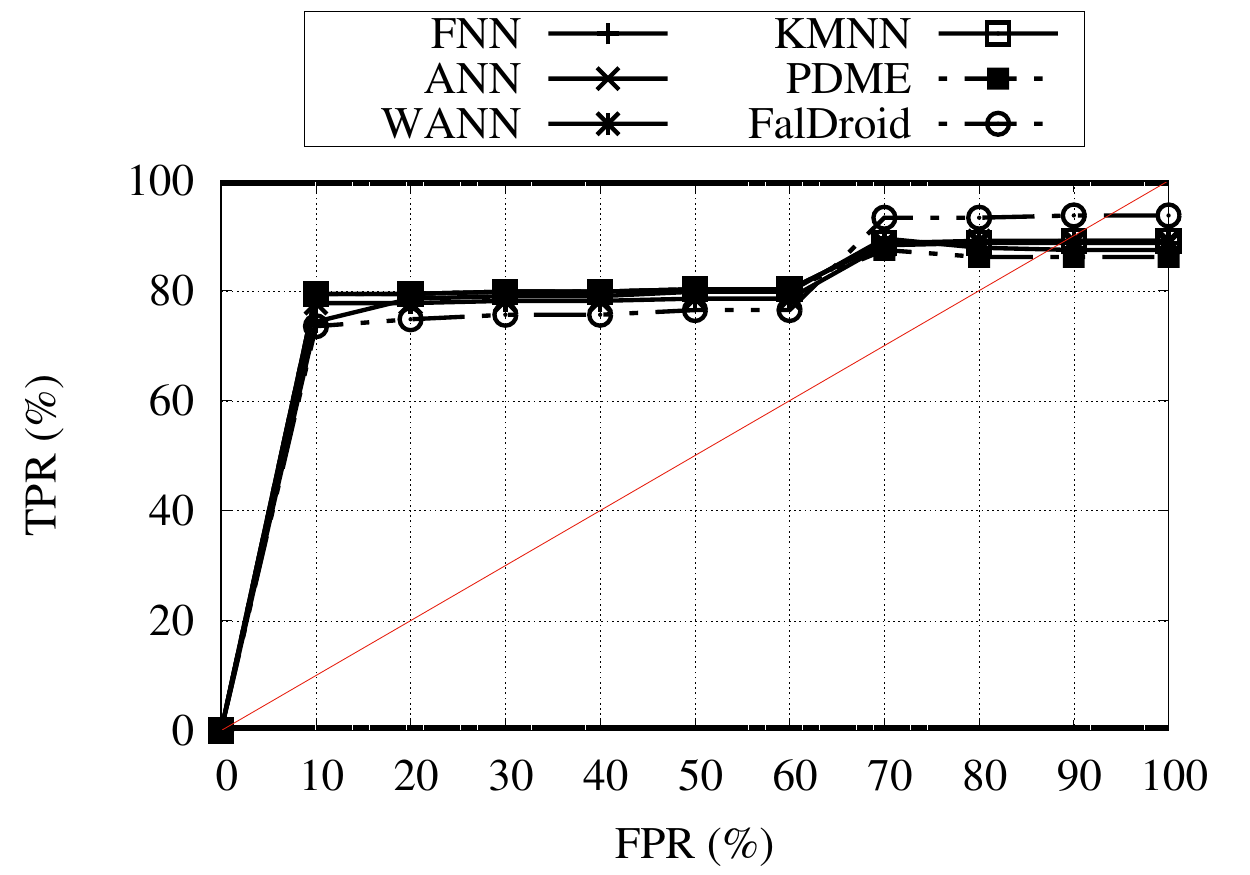}
			\caption{\small Genome-intent\vspace{-5px}}
			\label{fig:fig332b}
 	\end{subfigure} 
    \begin{subfigure}{0.32\textwidth}
 		\centering 		\includegraphics[width=1.05\linewidth]{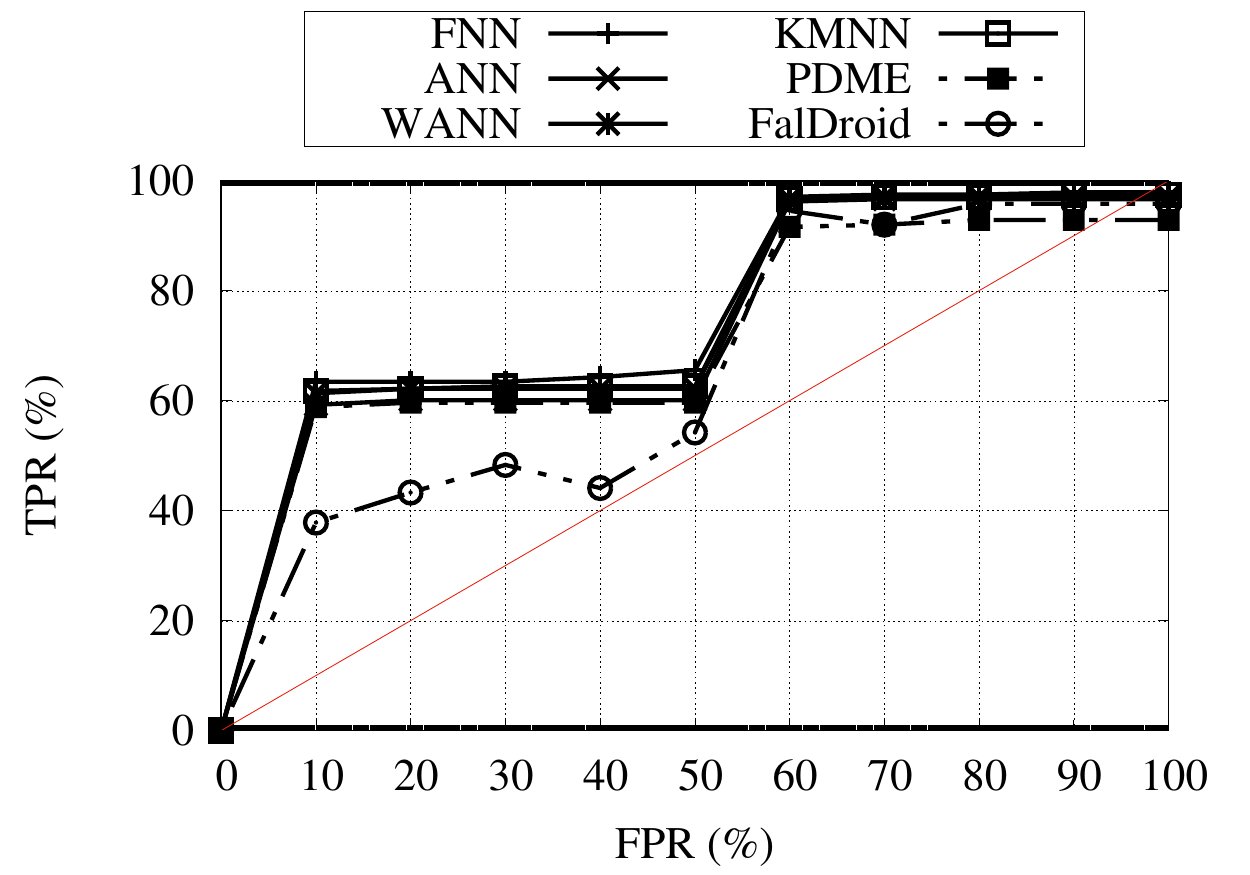}
		\caption{\small Genome-permission\vspace{-5px}}
			\label{fig:fig332c}
 	\end{subfigure}
        \caption{\small Comparison of algorithms in Genome dataset with reference to ROC. (TPR= true positive rate; FPR= false positive rate).\vspace{-10px}}
			\label{fig:fig332}
		\end{figure*}

In this section, we compare algorithms based on false positive rate (FPR) and true positive rate (TPR) metrics with both PDME~\cite{Radkani2017} and FalDroid~\cite{FanMing2018} algorithms which we present them in Figs.~\ref{fig:fig33}-\ref{fig:fig332}. Given that FPR and TPR are both numbers between zero and one, the area under the ROC obtained on the basis of these two measures, in the ideal case, represents the number one and in the random mode, the number is 0.5, and in most cases, the number between them will be closer to each other, indicating the greater accuracy of the proposed model in detecting malware samples. This area, which is shown by the ROC measure, is another yardstick for measuring the performance of a model. The TPR with a value close to one (i.e., 100\%) means that the model is more precise, and the fact that it is close to zero means the poor performance of the model in specifying the samples. 


Fig.~\ref{fig:fig33} plots the TPR of different algorithms over their FPR values. We can see that for Drebin with API features, over around 90\% of our designed learning model can correctly find malware samples even just below 20\% of FPR and the WANN and KMNN show the best rate among other methods. Focusing on the Drebin permission features, with over 90\% TPR rate and lower FPR around 20\% we can have the best performance of our ML model for FNN approach and this rate follow the same for other datasets. In all methods, FalDroid~\cite{FanMing2018} algorithm does not have acceptable performance for high Low FPR and high TPR. 
\subsubsection{  {Comparing methods based on various ML algorithms}}\label{comparewithML}
  {In this study, we also investigate the application of popular ML algorithms like SVM, Decision Tree, RF, and neural network. Using a random forest feature selection algorithm, we selected 300 important features. Then we implemented the above algorithms for 10\%, 20\%, 30\%, 40\%, 50\%, 60\%, 70\%, 80\%, 90\%, 100\% of the selected features. Examination of the results, according to Table~\ref{tabRFSVMDTNN} shows that the proposed methods in this paper have higher accuracy than these algorithms.}

\begin{table*}[!htb]
  \centering
  \caption{\small   {Accuracy and FPR values for API, permission, intent data of the Drebin, Contagio and Genome datasets (Acc= accuracy; FPR= false positive rate; FR= feature length; RF= Random Forest; SVM=Support Vector Machine; DT=Decision Tree; NN=Neural Network}).\vspace{-10px}}
\label{tabRFSVMDTNN}
  \scriptsize{
\setlength\tabcolsep{0.5pt} 
  \begin{tabular}{|c|c|c|c||c|c||c|c||c|c||c|c||c|c||c|c||c|c||c|c||c|c||c|c||c|c|}
  \hline
  \rowcolor{LightCyan} 
  \multicolumn{2}{|c|}{\textbf{features}}&\multicolumn{8}{|c|}{\textbf{Drebin Dataset}}&\multicolumn{8}{|c|}{\textbf{Contagio Dataset}}&\multicolumn{8}{|c|}{\textbf{Genome Dataset}}\\\hline\hline
\rowcolor{yellow}
   &&\multicolumn{2}{c|}{\textbf{RF}}&\multicolumn{2}{c|}{\textbf{SVM}}&\multicolumn{2}{c|}{\textbf{DT}}&\multicolumn{2}{c|}{\textbf{NN}}&\multicolumn{2}{c|}{\textbf{RF}}&\multicolumn{2}{c|}{\textbf{SVM}}&\multicolumn{2}{c|}{\textbf{DT}}&\multicolumn{2}{c|}{\textbf{NN}}&\multicolumn{2}{c|}{\textbf{RF}}&\multicolumn{2}{c|}{\textbf{SVM}}&\multicolumn{2}{c|}{\textbf{DT}}&\multicolumn{2}{c|}{\textbf{NN}}\\\hline
   \rowcolor{maroon!10}
    &\textbf{FR}&\textbf{Acc}&\textbf{FPR}&\textbf{Acc}&\textbf{FPR}&\textbf{Acc}&\textbf{FPR}&\textbf{Acc}&\textbf{FPR}&\textbf{Acc}&\textbf{FPR}&\textbf{Acc}&\textbf{FPR}&\textbf{Acc}&\textbf{FPR}&\textbf{Acc}&\textbf{FPR}&\textbf{Acc}&\textbf{FPR}&\textbf{Acc}&\textbf{FPR}&\textbf{Acc}&\textbf{FPR}&\textbf{Acc}&\textbf{FPR}\\
 \hline
\cellcolor[HTML]{FAEC0D}&\cellcolor[HTML]{E8E8AB}10\%&84.61&11.89&90.24&5.96&89.17&6.13&82.32&31.49&93.20&4.74&95.43&1.82&95.58&1.31&95.49&1.73&96.16&3.16&96.34&2.72&95.89&2.95&96.25&2.85\\\cline{2-26}
\cellcolor[HTML]{FAEC0D}&\cellcolor[HTML]{E8E8AB}20\%&94.87&3.25&95.97&3.14&94.66&3.57&87.95&23.80&97.22&2.44&96.51&1.27&96.66&0.79&96.60&1.14&98.44&1.13&98.50&0.94&98.41&0.84&98.50&0.97\\\cline{2-26}
\cellcolor[HTML]{FAEC0D}&\cellcolor[HTML]{E8E8AB}30\%&97.33&2.31&97.69&2.20&96.35&2.53&89.69&21.36&98.33&1.46&97.66&1.14&97.76&0.65&97.71&1.04&99.13&0.68&99.13&0.52&98.89&0.52&99.16&0.52\\\cline{2-26}
\cellcolor[HTML]{FAEC0D}&\cellcolor[HTML]{E8E8AB}40\%&98.12&1.82&98.23&1.85&97.04&2.13&90.15&20.54&98.42&1.30&97.71&1.24&97.98&0.56&97.77&1.17&99.25&0.55&99.31&0.36&98.95&0.45&99.28&0.39\\\cline{2-26}
\cellcolor[HTML]{FAEC0D}&\cellcolor[HTML]{E8E8AB}50\%&98.57&1.20&98.40&1.26&97.50&1.44&90.77&18.28&98.42&1.30&97.74&1.27&97.86&0.79&97.83&1.17&99.37&0.48&99.37&0.32&99.22&0.32&99.37&0.32\\\cline{2-26}
\cellcolor[HTML]{FAEC0D}&\cellcolor[HTML]{E8E8AB}60\%&98.57&1.36&98.59&1.46&97.59&1.51&90.82&19.00&98.54&1.10&97.80&1.17&97.92&0.69&97.92&1.04&99.58&0.39&99.46&0.32&99.25&0.32&99.43&0.29\\\cline{2-26}
\cellcolor[HTML]{FAEC0D}&\cellcolor[HTML]{E8E8AB}70\%&98.81&1.20&98.64&1.33&97.73&1.48&90.98&18.73&98.59&1.14&97.80&1.17&97.92&0.69&97.86&1.11&9.55&0.42&99.58&0.26&99.40&0.23&99.46&0.26\\\cline{2-26}
\cellcolor[HTML]{FAEC0D}\multirow{-5}{*}{\begin{sideways}{\textbf{API}}\end{sideways}}&\cellcolor[HTML]{E8E8AB}80\%&99.07&0.81&98.62&1.26&97.95&1.12&91.25&17.38&98.48&1.24&97.98&1.43&98.18&0.88&98.10&1.30&99.52&0.45&99.55&0.29&99.34&0.26&99.49&0.29\\\cline{2-26}
\cellcolor[HTML]{FAEC0D}&\cellcolor[HTML]{E8E8AB}90\%&99.12&0.68&98.66&1.46&98.01&1.02&91.20&17.47&98.59&1.17&98.04&1.40&98.30&0.79&98.21&1.21&99.55&0.39&99.52&0.26&99.31&0.23&99.52&0.26\\\cline{2-26}
\cellcolor[HTML]{FAEC0D}&\cellcolor[HTML]{E8E8AB}100\%&99.09&0.58&98.83&0.78&98.07&0.95&91.34&17.01&98.74&1.10&98.21&1.30&98.45&0.72&98.27&1.14&99.64&0.29&99.64&0.16&99.40&0.19&99.73&0.16\\
\hline\hline

\cellcolor[HTML]{FAEC0D}&\cellcolor[HTML]{E8E8AB}10\%&78.02 &18.87&	77.59&	17.64&	78.19&	16.77&	76.04&	18.27&	97.48&	2.17&96.46&	2.56&	97.45&	1.82&	96.28&	3.03&	86.16&	10.36&91.75&5.69&	92.05&3.77&91.90&4.54\\\cline{2-26}
\cellcolor[HTML]{FAEC0D}&\cellcolor[HTML]{E8E8AB}20\%&79.17&	18.23&	78.22&	17.44&	78.91&	16.57&	76.71&17.86&97.48&	2.20&	96.49&	2.56&	97.48&	1.82&	96.31&	3.03&	86.16&	10.36&91.69&	5.81&91.99&3.90&91.84&4.63\\\cline{2-26}
\cellcolor[HTML]{FAEC0D}&\cellcolor[HTML]{E8E8AB}30\%&79.19&18.17&	78.24&	17.38&	78.93&16.52&76.78&	17.77&	97.57&	2.14&	96.49&	2.56&	97.60&	1.79&	96.31&	3.03&	85.68&	10.87&91.72&	5.78&	92.02&3.87&91.87&4.63\\\cline{2-26}
\cellcolor[HTML]{FAEC0D}&\cellcolor[HTML]{E8E8AB}40\%&79.10&18.29&	78.22&	17.41&78.91&16.54&76.66&	17.83&	95.57&	2.14&96.52&2.52&97.60&	1.79&	96.34&3.00&	85.68&	10.93&91.69&5.81&91.99&3.90&91.87&4.63\\\cline{2-26}
\cellcolor[HTML]{FAEC0D}&\cellcolor[HTML]{E8E8AB}50\%&79.03&18.38&78.19&17.44&78.88&16.54&76.74&	17.83&97.63&2.14&96.58&2.52&97.66&1.79&	96.40&3.00&	85.77&	10.93&91.69&5.81&91.99&3.90&91.84&4.66\\\cline{2-26}
\cellcolor[HTML]{FAEC0D}&\cellcolor[HTML]{E8E8AB}60\%&79.05&18.35&78.19&17.44&78.88&16.54&76.66&	17.88&	97.54&2.23&	96.58&	2.52&	97.66&	1.79&96.40&	3.00&	94.72&	3.89&94.30&5.54&94.78&3.68&94.48&4.45\\\cline{2-26}
\cellcolor[HTML]{FAEC0D}&\cellcolor[HTML]{E8E8AB}70\%&79.05&18.35&	78.12&	17.52&	78.88&16.54&76.66&	17.94&97.69&	2.04&96.37&	2.59&	97.54&1.82&	96.20&3.07&	94.69&	3.95&94.39&	5.49&94.72&3.77&94.57&4.38\\\cline{2-26}
\cellcolor[HTML]{FAEC0D}\multirow{-6}{*}{\begin{sideways}{\textbf{Permission}}\end{sideways}}&\cellcolor[HTML]{E8E8AB}80\%&79.36&18.23&	78.81&18.48&79.50&17.51&77.38&	18.87&97.86&	2.07&96.58&	2.59&	97.69&	1.92&99.65&	3.03&	94.69&3.95&94.33&5.56&94.75&3.74&94.51&4.45\\\cline{2-26}
\cellcolor[HTML]{FAEC0D}&\cellcolor[HTML]{E8E8AB}90\%&83.25&16.00&	82.32&	16.51&	83.15&15.52&80.77&	16.70&	97.83&2.11&	96.61&	2.62&	97.78&	1.88&	96.52&	3.10&	94.72&	3.95&94.33&	5.57&94.72&3.77&94.57&4.41\\\cline{2-26}
\cellcolor[HTML]{FAEC0D}&\cellcolor[HTML]{E8E8AB}100\%&83.37&	15.92&	82.30&16.48&	83.20&15.91&	80.86&	16.64&	97.92&	2.07&	96.66&	2.59&	97.69&	1.98&	96.52&	3.10&	94.81&	3.88&94.42&	5.54&	94.72&3.83&94.63&4.44\\
\hline\hline
\cellcolor[HTML]{FAEC0D}&\cellcolor[HTML]{E8E8AB}10\%&85.09&3.47&	83.20&2.17&	83.23&2.92&	81.65&	2.85&	96.78&1.45&	97.31&	1.59&	97.22&1.49&	97.48&1.26&	95.53&	2.90&96.55&	2.58&	96.10&2.65&94.99&3.51\\\cline{2-26}
\cellcolor[HTML]{FAEC0D}&\cellcolor[HTML]{E8E8AB}20\%&85.21&	3.65&83.37&	2.13&	83.39&	2.89&81.82&	2.82&	97.19&	1.19&97.43&	1.49&97.48&1.23&97.57&	1.20&95.83&	2.90&96.76&2.36&96.34&2.40&95.20&3.29\\\cline{2-26}
\cellcolor[HTML]{FAEC0D}&\cellcolor[HTML]{E8E8AB}30\%&85.23&5.83&84.32&	2.13&84.37&2.85&	82.75&	2.85&	97.05&0.94&97.54&1.42&	97.48&1.29&97.66&1.13&95.83&	2.93&96.76&2.39&96.34&2.43&95.20&3.32\\\cline{2-26}
\cellcolor[HTML]{FAEC0D}&\cellcolor[HTML]{E8E8AB}40\%&85.45&5.83&84.47&2.13&84.56&2.82&82.89&	2.85&	97.13&0.97&97.60&	1.49&97.66&	1.26&97.72&	1.20&	95.83&	2.93&97.76&	2.39&	96.34&2.43&95.20&3.32\\\cline{2-26}
\cellcolor[HTML]{FAEC0D}&\cellcolor[HTML]{E8E8AB}50\%&87.21&3.54&85.33&2.20&84.45&2.85&	83.77&	2.89&	97.16&	0.94&97.63&1.46&97.60&1.32&97.72&1.20&95.89&	2.93&96.79&2.39&96.37&2.43&95.23&3.32\\\cline{2-26}
\cellcolor[HTML]{FAEC0D}&\cellcolor[HTML]{E8E8AB}60\%&87.57&	3.54&85.68&2.20&85.71&2.96&	84.11&	2.92&	97.72&0.81&	97.81&1.33&	97.98&	0.97&98.16&	0.78&	95.89&	2.93&96.76&	2.42&	96.37&2.43&995.20&3.35\\\cline{2-26}
\cellcolor[HTML]{FAEC0D}&\cellcolor[HTML]{E8E8AB}70\%&90.96&6.49&	89.21&	4.54&	89.33&	4.91&87.74&	4.93&	98.24&	1.10&	98.30&	1.26&98.45&	1.07&	98.65&	0.91&	96.40&3.12&97.54&	2.23&	97.18&2.30&95.75&3.40\\\cline{2-26}
\cellcolor[HTML]{FAEC0D}\multirow{-6}{*}{\begin{sideways}{\textbf{Intent}}\end{sideways}}&\cellcolor[HTML]{E8E8AB}80\%&90.96&	3.77&90.53&	3.67&90.46&	4.49&88.90&4.42&	98.33&1.00&	98.30&	1.23&98.54&	1.00&	98.68&	0.91&96.40&	3.00&97.36&2.46&97.03&2.49&95.74&3.38\\\cline{2-26}
\cellcolor[HTML]{FAEC0D}&\cellcolor[HTML]{E8E8AB}90\%&91.15&3.61&	90.67&	3.53&90.60&	4.35&89.05&	4.29&	98.30&1.19&	98.60&	1.13&	98.57&0.94&98.77&0.78&	96.34&	3.03&97.42&	2.39&	97.09&2.43&95.71&3.38\\\cline{2-26}
\cellcolor[HTML]{FAEC0D}&\cellcolor[HTML]{E8E8AB}100\%&91.48&	3.49&	90.93&	3.38&90.69&	4.49&88.93&	4.41&	98.45&	1.10&98.62&	1.17&98.68&	0.87&98.80&	0.78&96.34&	3.12&97.36&	2.49&97.09&2.49&95.74&3.38\\
\hline
  \end{tabular}}
  \end{table*}

\subsubsection{  {Comparing methods for without feature selection strategies}}\label{comparewithoutFeatureselection}
  {In this study, we aim to compare our proposed methods considering various popular ML algorithms, e.g., SVM, Decision Tree, RF, and neural network (NN). Using a RandomForestRegressor feature selection algorithm, we selected 300 important features. Then we implemented the above algorithms for 10\%, 20\%, 30\%, 40\%, 50\%, 60\%, 70\%, 80\%, 90\%, 100\% of the selected features. Examination of the results, according to Table~\ref{tab912} shows that the proposed methods in this paper have higher accuracy than these algorithms. To be precise, random forest (RF) is a meta estimator that fits a number of decision tree classifiers on various sub-samples of the dataset and uses averaging to improve the predictive accuracy and control over-fitting. We apply Neural Network (NN) as a multi-layer Perceptron classifier. This model optimizes the log-loss function using stochastic gradient descent. Additionally, we run the algorithm 200 epochs, and the learning rate is 0.01. It controls the step-size in updating the weights. The solver is 'adam' and hidden layer sizes=(5, 2). Focusing on SVM, we use the C-Support vector classification. The implementation is based on \textit{libsvm}. The fit time scales at least juridically with the number of samples and maybe impractical beyond tens of thousands of samples. Decision Trees are a non-parametric supervised learning method used for classification and regression. The goal is to \textit{create a model that predicts the value of a target variable by learning simple decision rules inferred from the data features}.}

\begin{table*}[!htb]
  \centering
  \caption{\small   {Accuracy and FPR values for API, permission, intent data of the Drebin, Contagio, and Genome datasets- Comparing Between ``Using 300 features'' and  "Using All Features" (Acc= accuracy; FPR= false positive rate).}\vspace{-10px}}
\label{tab912}
  \scriptsize{
\setlength\tabcolsep{0.5pt} 
  \begin{tabular}{|c|c|c|c||c|c||c|c||c|c||c|c||c|c||c|c||c|c||c|c|}
  \hline
  \rowcolor{LightCyan} 
  
  \multicolumn{2}{|c|}{\textbf{Algorithm}}&\multicolumn{6}{|c|}{\textbf{Drebin Dataset}}&\multicolumn{6}{|c|}{\textbf{Contagio Dataset}}&\multicolumn{6}{|c|}{\textbf{Genome Dataset}}\\\hline\hline
\rowcolor{yellow}
   &&\multicolumn{2}{c|}{\textbf{API}}&\multicolumn{2}{c|}{\textbf{Permission}}&\multicolumn{2}{c|}{\textbf{Intents}}&\multicolumn{2}{c|}{\textbf{API}}&\multicolumn{2}{c|}{\textbf{Permission}}&\multicolumn{2}{c|}{\textbf{Intents}}&\multicolumn{2}{c|}{\textbf{API}}&\multicolumn{2}{c|}{\textbf{Permission}}&\multicolumn{2}{c|}{\textbf{Intents}}\\\hline
   \rowcolor{maroon!10}
    &\textbf{FR}&\textbf{Acc}&\textbf{FPR}&\textbf{Acc}&\textbf{FPR}&\textbf{Acc}&\textbf{FPR}&\textbf{Acc}&\textbf{FPR}&\textbf{Acc}&\textbf{FPR}&\textbf{Acc}&\textbf{FPR}&\textbf{Acc}&\textbf{FPR}&\textbf{Acc}&\textbf{FPR}&\textbf{Acc}&\textbf{FPR}\\
 \hline
\cellcolor[HTML]{FAEC0D}&\cellcolor[HTML]{E8E8AB}FNN-300&99.31&0.52&96.64&3.35&90.05&1.25&98.95&0.52& 98.62&0.65&98.10&1.00&99.82&0.03&99.37&0.52&98.95&0.16\\\cline{2-20}
\cellcolor[HTML]{FAEC0D}&\cellcolor[HTML]{E8E8AB}FNN-All &99.36&0.52&96.83&3.25&90.29&1.18 &99.06&0.52& 98.77&0.62&98.77&0.62&99.88&0.01& 99.43&0.45&99.01&0.13\\\cline{2-20}
\hline\hline
\cellcolor[HTML]{FAEC0D}&\cellcolor[HTML]{E8E8AB}ANN-300&99.33&0.55&97.92&1.51&91.77&2.33&99.00&0.46&98.71&0.52&98.54&0.45&99.79&0.03&99.43&0.39&99.07&0.16\\\cline{2-19}
\cellcolor[HTML]{FAEC0D}&\cellcolor[HTML]{E8E8AB}ANN-All &99.48&0.55&98.02&1.48&91.79&2.23 &99.27&0.59& 99.00&0.42&99.09&0.29&99.88&0.03& 99.49&0.42&99.22&0.03\\\cline{2-20}
\hline\hline
\cellcolor[HTML]{FAEC0D}&\cellcolor[HTML]{E8E8AB}WANN-300&99.31&0.58&98.04&1.44&91.72&2.39&99.00&0.42&98.74&0.49&98.62&0.36&99.79&0.06&99.46&0.32&99.07&0.13\\\cline{2-19}
\cellcolor[HTML]{FAEC0D}&\cellcolor[HTML]{E8E8AB}WANN-All  &99.33&0.62&98.45&1.21&92.19&2.63 &99.06&0.42& 98.80&0.39&  99.06&0.23&99.82&0.06& 99.37&0.36&99.16&0.16\\\cline{2-19}
\hline\hline
\cellcolor[HTML]{FAEC0D}&\cellcolor[HTML]{E8E8AB}KMNN-300&99.28&0.62&97.76&1.38&91.70&2.40&99.03&0.42&98.71&0.52&98.51&0.45&99.76&0.06&99.43&0.42&99.04&0.19\\\cline{2-19}
\cellcolor[HTML]{FAEC0D}&\cellcolor[HTML]{E8E8AB}KMNN-All &99.26&0.65&97.88&1.54&91.91&2.36 &99.18&0.46& 98.83&0.52&98.89&0.36& 99.70&0.10& 99.43&0.29&99.13&0.39\\\cline{2-20}
\hline\hline
\cellcolor[HTML]{FAEC0D}&\cellcolor[HTML]{E8E8AB}PDME-300&99.19&0.62&97.76&1.41&91.67&2.53&98.86&0.39&98.54&0.55&98.57&0.36&99.76&0.1&98.86&0.68&98.83&0.19\\\cline{2-19}
\cellcolor[HTML]{FAEC0D}&\cellcolor[HTML]{E8E8AB}PDME-All&99.28&0.52&97.92&1.28&91.82&2.27 &998.92&0.59& 98.77&0.61&98.83&0.29&99.55&0.26& 98.89&0.55&98.89&0.35\\\cline{2-20}
\hline\hline
\cellcolor[HTML]{FAEC0D}&\cellcolor[HTML]{E8E8AB}FalDroid-300&88.07&0.84&87.57&10.86&66.55&43.04&92.24&7.55&86.98&6.55&82.83&1.48&98.05&1.29&90.97&9.41&80.61&20.39\\\cline{2-19}
\cellcolor[HTML]{FAEC0D}&\cellcolor[HTML]{E8E8AB}FalDroid-All &94.54&0.84&70.03&39.44&91.15&6.66 &93.58&6.34& 8.44&5.87&88.62&1.62&98.38&0.84& 91.81&8.60&85.11&15.60\\\cline{2-20}
\hline\hline
\cellcolor[HTML]{FAEC0D}&\cellcolor[HTML]{E8E8AB}MAA-300&99.57&0.29&99.36&0.72&99.26&0.39&99.33&0.10&99.44&0.29&93.56&0.13&99.85&0.03&99.73&0.19&99.91&0.03\\\cline{2-19}
\cellcolor[HTML]{FAEC0D}&\cellcolor[HTML]{E8E8AB}MAA-All&99.74&0.23&99.48&0.53&9.52&0.36 &99.41&0.10& 99.53&0.23&99.62&0.16&99.88&0.01& 99.85&0.13&99.91&0.03\\\cline{2-20}
\hline\hline
\cellcolor[HTML]{FAEC0D}&\cellcolor[HTML]{E8E8AB}RF-300&99.09&0.58&83.37&15.92&91.48&3.49&98.74&1.10&97.92&2.07&98.45&1.10&99.46&0.29&94.81&3.88&96.34&3.12\\\cline{2-19}
\cellcolor[HTML]{FAEC0D}&\cellcolor[HTML]{E8E8AB}RF-All&99.17&0.58&85.78&14.36&92.03&3.14 &98.92&0.94& 98.07&2.00&98.57&1.03&99.67&0.29& 95.10&3.67&96.82&2.87\\\cline{2-20}
\hline\hline
\cellcolor[HTML]{FAEC0D}&\cellcolor[HTML]{E8E8AB}SVM-300&98.83&0.78&82.20&16.48&90.93&3.382&98.21&1.30&96.66&2.59&98.62&1.17&99.64&0.16&94.42&5.54&97.36&2.49\\\cline{2-19}
\cellcolor[HTML]{FAEC0D}&\cellcolor[HTML]{E8E8AB}SVM-All&97.84&0.78&87.45&12.70&91.08&3.34 &98.30&1.27& 97.05&2.30&98.89&0.94&99.61&0.29& 94.57&5.25&97.45&2.25\\\cline{2-20}
\hline\hline
\cellcolor[HTML]{FAEC0D}&\cellcolor[HTML]{E8E8AB}DT-300&98.07&0.95&83.20&15.91&90.69&4.49&98.45&0.72&97.69&1.98&98.68&0.87&99.40&0.19&97.42&3.83&97.09&2.49\\\cline{2-19}
\cellcolor[HTML]{FAEC0D}&\cellcolor[HTML]{E8E8AB}DT-All&98.14&1.05&86.33&13.02&91.17&3.76 &98.39&0.85& 97.66&1.67&98.80&0.84&99.25&0.26& 94.96&3.83&97.12&2.52\\\cline{2-20}
\hline\hline
\cellcolor[HTML]{FAEC0D}&\cellcolor[HTML]{E8E8AB}NN-300&98.81&0.49&8.86&16.64&88.93&4.41&98.27&1.14&96.52&3.10&98.80&0.78&97.90&0.16&94.63&4.44&95.74&3.38\\
\cellcolor[HTML]{FAEC0D}&\cellcolor[HTML]{E8E8AB}NN-All&91.20&17.01&82.37&14.89&88.81&4.52 &98.36&0.98& 97.16&2.41&98.74&0.87&99.79&0.16& 94.72&4.34&95.68&3.80\\\cline{2-20}
\hline
 \end{tabular}}
  \end{table*}

\section{Discussion and Limitations}\label{discussion}

In the following, we explain the constraints on our similarity based detection algorithms and give some of our ongoing research directions. One of the limitations of our methods is that the Hamming distance measure calculates the distance between two programs based on the static features. The presented methods are similar to most of the other malware detection methods which can find two similar programs if they have similar features. For example, if we have two malware programs that have the same functionality with quite different features most of our methods are unable to find that these two programs are similar. 

Despite the effectiveness of the proposed methods, several issues are remaining to be solved. First, time complexity related to three of the proposed algorithms is $o(n\times m )$, and its rate for the KMNN algorithm is $o(n^2 \times m)$. However, a large amount of data in the smartphones that come with Android software rarely happens, but with a $o(n \times m)$ time complexity it needs a lot of time to run when the volume of data is enormous.  {Besides, as stated, static analysis usually separates and analyzes the various sources of the binary file without executing it. This analysis finds available features of malware at a decent speed. Dynamic analysis, also called \textit{behavioral analysis}, is performed by observing malware behavior while running on the host system. Compared to static analysis, dynamic analysis is more effective as there is no need to disassemble the infected file to analyze it. In addition, dynamic analysis can detect known and unknown malware.}  So, we can design the ML model and detect malicious apps faster using static features compared with dynamic features. However, we can not recognize some unknown behavior malware in many cases. Therefore, we need to enhance our method and use static and dynamic features together as hybrid features in which improve our accuracy in selecting the various type of malware based on similarity-based strategies.   { Hybrid analysis gathers information about malware from static analysis and dynamic analysis. By using hybrid analysis, security researchers gain the benefits of both analyses, static and dynamic. Therefore, increasing the ability to detect malicious programs correctly}. Third, to increase the efficiency of the similarity-based detection mechanism we can corporate deep learning, but it raises the time complexity of our methods. Hence, we should manage the convergence speed (i.e., detection speed) and accuracy rate which is another aspect of our future direction. Fourth, to provide robust detection algorithm, it is important to train our ML model using some adversarial examples and design new methods such as generative adversarial networks (GAN) to understand the simulated samples and detect compromised samples (i.e., malware samples in which change their behavior to fool the classifier). Finally, in this paper, we test our ML model on the three types of features (intent, permission, and API), some other features are existing in such datasets like URL-based Android services which are important for future ML model generation. Note Unlike two other used papers as a comparison,~\cite{Radkani2017} and ~\cite{Xiong2018}, which are accurately implemented, the FalDroid~\cite{FanMing2018} method originally is suggested for detecting malware families. In this paper, we re-code the ideas mentioned in the FalDroid paper to detect malware. It is the main reason for the inefficiency of FalDroid compared to other methods.

 \vspace{-5px}
\section{Conclusions and Future Work}\label{conclusion}
In this paper, we consider the Hamming distance as the benchmark for the similarity of samples and present four methods for detecting malware based on the nearest neighborhoods. To identify and analyze Android malware, and by using the binary features of Android applications, we provide machine learning-based methods to detect new malicious software with high precision and recall rates. Our approaches complement existing KNN-based solutions and validate our algorithms using three public datasets, namely the Drebin, Genome, and Contagio datasets, and applying API, intent and permission file types. Specifically, we demonstrate that the permission and intent-based method can classify malware from goodware inappropriate percent of cases. By considering API features of Drebin dataset, the accuracy of ANN, FNN, WANN, and KMNN methods are 99.31\%, 99.33\%, 99.31\%, 99.28\%, respectively, which are improved than PDME(99.19\%) and FalDroid(88.07\%). Similarly, with respect to the Permission features of the Drebin dataset, the accuracy of ANN, FNN WANN, and KMNN methods are 96.64\%, 97.92\%, 98.04\%, 97.76\%, respectively, while they are comparable with PDME with accuracy rate 97.76\% and have better accuracy than FalDroid with accuracy rate 87.57\%. Finally, about the Intents features of this dataset, the accuracy of ANN, FNN, WANN, and KMNN methods are 90.05\%, 91.77\%, 91.72\%, 91.70\%, respectively. In this type of features, the accuracy of PDME  is 91.67\% and FalDroid is 66.55\%. Similar enhancements have been made about the different type of features in the Contagio and Genome datasets. The results showed that there is a superiority of proposed methods to the newest researches. The results of the proposed methods showed that the WANN algorithm has the highest accuracy in terms of permissions and intents and the ANN algorithm is in the next rank. Also, with API features, KMNN and ANN algorithms can provide higher accuracy. By comparing the accuracy obtained in the proposed methods for the different features of the three used datasets, we always consider that the accuracy of these methods is higher than the PDME and FalDroid methods.
 
 For the future works, it is possible to define other (distance) similarity measures between programs that consider some other features of the program instead of just using features frequencies. One idea is to utilize the correlation between features. Thus, using several other features of two applications such as the correlation between them may be useful to detect the similarity between two malware programs. 

\section*{Acknowledgment}
Mohammad Shojafar is supported by Marie Curie Global Fellowship (MSCA-IF-GF) funded by European Commission agreement grant number MSCA-IF-GF-839255 and Mauro Conti is supported by a Marie Curie Fellowship funded by the European Commission (agreement PCIG11-GA-2012-321980).

\Urlmuskip=0mu plus 1mu\relax
\bibliographystyle{elsarticle-num}
\bibliography{references}

\begin{thebibliography}{10}
\bibitem{Demontis2017}
A.~Demontis, et~al., Yes, machine learning can be more secure! a case study on
  android malware detection, IEEE Transactions on Dependable and Secure
  Computing PP~(99).
\end{thebibliography}
\clearpage
\pagebreak
\newpage

\section*{Biographies}\label{sec:11}
\vspace{-10px}
\noindent\begin{minipage}{0.13\textwidth}
\includegraphics[width=1.15in,height=1.15in,clip,keepaspectratio]{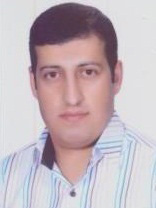} 
\end{minipage}%
\hfill%
\begin{minipage}{0.84\textwidth}
\textbf{Rahim~Taheri} received his B.Sc. degree of Computer engineering from Bahonar Technical College of Shiraz and M.Sc. degree of computer networks at the Shiraz University of Technology in 2007 and 2015, respectively. Now he is a Ph.D. candidate on Computer Networks at the Shiraz University of Technology. In February 2018, he joined to SPRITZ Security \& Privacy Research Group at the University of Padua as a visiting Ph.D. student. His main research interests include machine learning, data mining,  network securities and heuristic algorithms. His main research interests are in an adversarial machine and deep learning as a new trend in computer security.
\end{minipage}%

\hfill \break

\noindent\begin{minipage}{0.13\textwidth}
\includegraphics[width=1.15in,height=1.15in,clip,keepaspectratio]{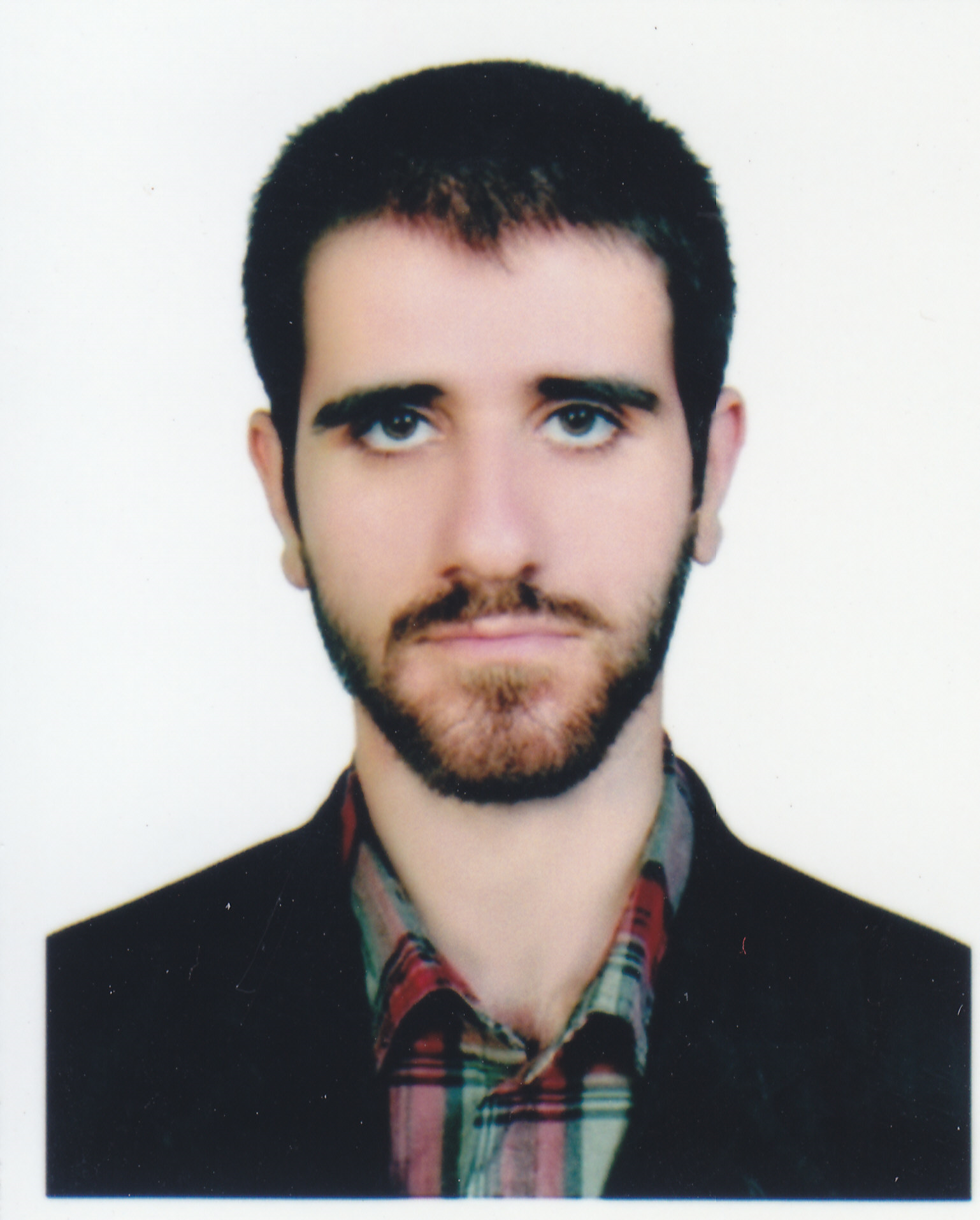} 
\end{minipage}%
\hfill%
\begin{minipage}{0.84\textwidth}
\textbf{Meysam~Ghahramani} graduated from BSc degree in mathematics and its applications, in 2014. He won the first rank at the ACM programming competitions of the university in 2013. He was admitted to the postgraduate in the field of cryptography. In 2016, he graduated with the first rank and received the award of a distinguished university student. Mr. Ghahramani is currently a Ph.D. student in the Department of Computer Engineering and Information Technology at the Shiraz University of Technology. His primary fields of interest are Post-Quantum Cryptography, Cryptographic Protocol Analysis, Applied Mathematics, and Information Security. 
 \end{minipage}%

\hfill \break

\noindent\begin{minipage}{0.13\textwidth}
\includegraphics[width=1in,height=1.15in,clip,keepaspectratio]{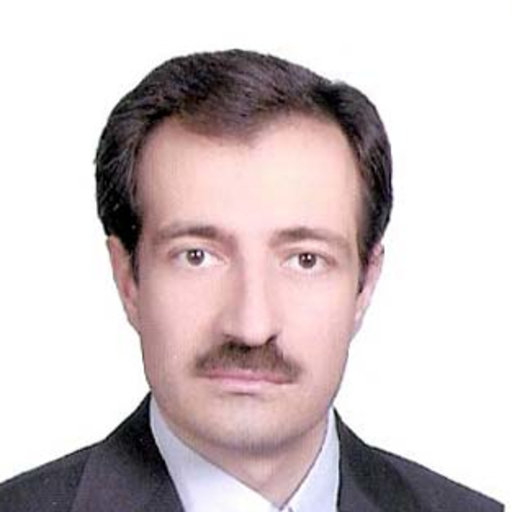}  
\end{minipage}%
\hfill
\begin{minipage}{0.84\textwidth}	
 \textbf{Reza~Javidan} received M.Sc. Degree in Computer Engineering (Machine Intelligence and Robotics) from Shiraz University in 1996. He received a Ph.D. degree in Computer Engineering (Artificial Intelligence) from Shiraz University in 2007. Dr. Javidan has many publications in international conferences and journals regarding Image Processing, Underwater Wireless Sensor Networks (UWSNs) and Software Defined Networks (SDNs). His major fields of interest are Network security, Underwater Wireless Sensor Networks (UWSNs), Software Defined Networks (SDNs), Internet of Things, artificial intelligence, image processing, and SONAR systems. Dr. Javidan is an associate professor in the Department of Computer Engineering and Information Technology at the Shiraz University of Technology.
\end{minipage}

\hfill\break 

\noindent\begin{minipage}{0.13\textwidth}
\includegraphics[width=0.9in,height=1.15in,clip,keepaspectratio]{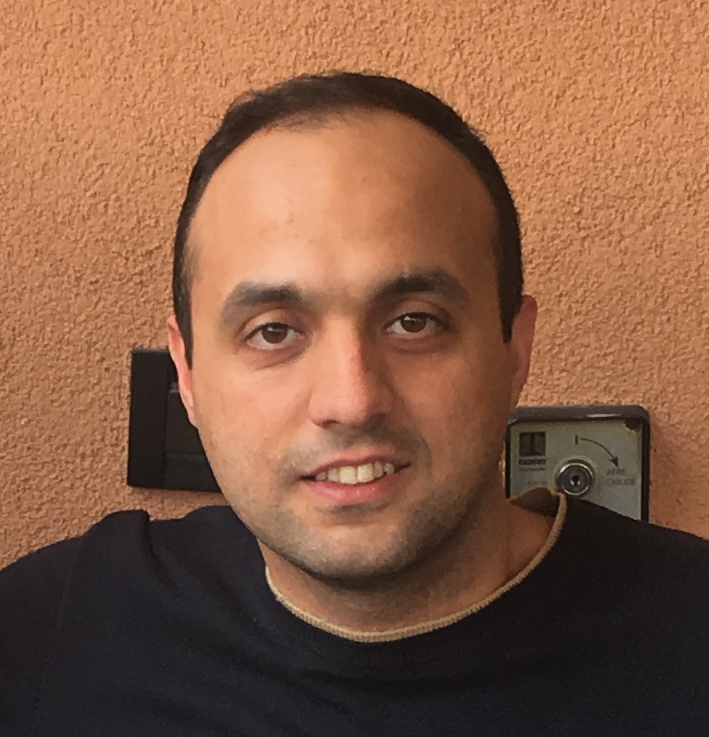}
\end{minipage}%
\hfill
\begin{minipage}{0.84\textwidth}		
\textbf{Mohammad Shojafar} is a Marie Curie Fellow, Intel Innovator, and Senior Researcher in SPRITZ Security and Privacy Research group at the University of Padua, Italy in since Jan. 2018. Also, he was CNIT Senior Researcher at the University of Rome Tor Vergata contributed on European H2020 ``SUPERFLUIDITY'' project. He received the Ph.D. degree from Sapienza University of Rome, Italy, in 2016 with an ``Excellent'' degree. His main research interest is in the area of Network and network security and privacy. In this area, he published more than 99 papers in top-most international peer-reviewed journals and conference, e.g., IEEE TCC, IEEE TNSM, IEEE TGCN, and IEEE ICC/GLOBECOM (h-index=25, 2.3k+ citations). He is an Associate Editor in IEEE Transactions on Consumer Electronics, IET Communication, Cluster Computing, and Ad Hoc \& Sensor Wireless Networks Journals. He is a Senior Member of the IEEE. For additional information: \url{http://mshojafar.com} 
\end{minipage}%

\hfill \break

\noindent\begin{minipage}{0.13\textwidth}
\includegraphics[width=1in,height=1.15in,clip,keepaspectratio]{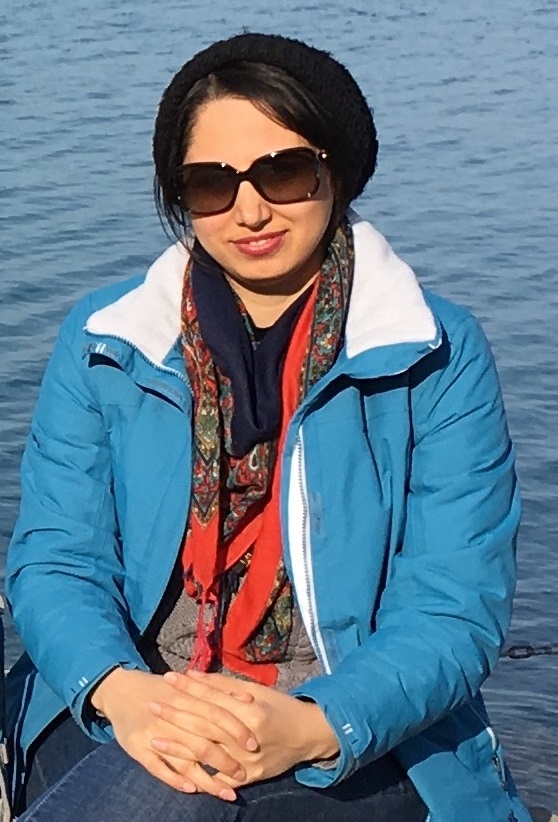}  
\end{minipage}%
\hfill%
\begin{minipage}{0.84\textwidth}	
 \textbf{Zahra~Pooranian} is currently a Postdoc in the SPRITZ Security and Privacy Research group at the University of Padua, Italy, since April 2017. She received her Ph.D. degree in Computer Science Sapienza University of Rome,  Italy, in February 2017. She is a (co)author of several peer-reviewed publications (h-index=15, citations=650+) in well-known conferences and journals. She is an Editor of KSSI transaction on internet and information systems and Future Internet. Her current research focuses on Machine Learning, Smart Grid, and Cloud/Fog Computing. She was a programmer in several companies in Iran from 2009-2014, respectively. She is a member of IEEE. For additional information: \url{https://www.math.unipd.it/~zahra/} 
\end{minipage}

\hfill \break

\noindent\begin{minipage}{0.13\textwidth}
\includegraphics[width=1.15in,height=1.15in,clip,keepaspectratio]{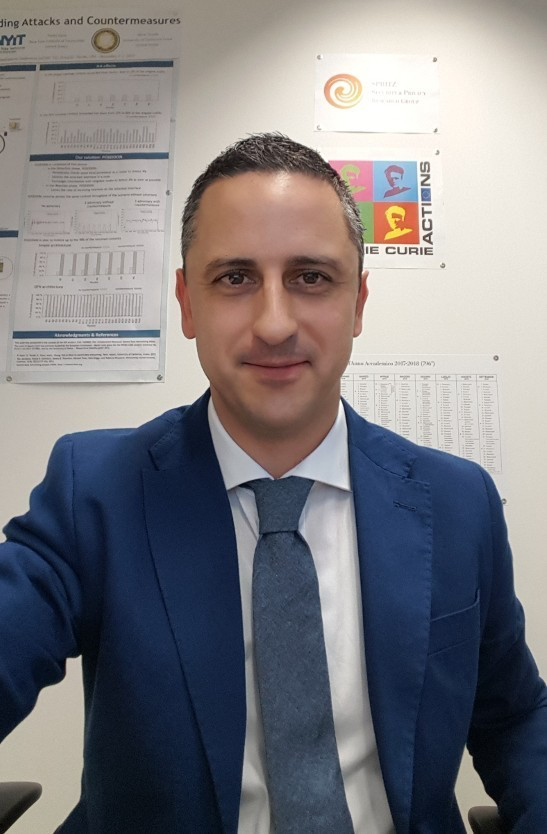}
\end{minipage}%
\hfill%
\begin{minipage}{0.84\textwidth}
\textbf{Mauro Conti} received his MSc and his PhD in Computer Science from Sapienza University of Rome, Italy, in 2005 and 2009. He has been Visiting Researcher at GMU (2008, 2016), UCLA (2010), UCI (2012, 2013, 2014), TU Darmstadt (2013), UF (2015), and FIU (2015, 2016). In 2015 he became Associate Professor, and Full Professor in 2018. He has been awarded with a Marie Curie Fellowship (2012) by the European Commission, and with a Fellowship by the German DAAD (2013). His main research interest is in the area of security and privacy. In this area, he published more than 300 papers in topmost international peer-reviewed journals and conference. He is Associate Editor for several journals, including IEEE Communications Surveys \& Tutorials, IEEE Transactions on Network and Service Management, and IEEE Transactions on Information Forensics and Security. He is Senior Member of the IEEE. For additional information: \url{ http://www.math.unipd.it/~conti/}
\end{minipage}

\hfill \break

\clearpage
\pagebreak
\newpage
\appendices

\color{black}
\section{Toy example of presented methods}\label{appendix1}
In this section, we aim to give a simple example how our proposed similarity-based algorithms adopt to detect Android malware in a binary dataset. 

\begin{thm}
Suppose $X$ is the sample that we want to predict its label. As an example, vector $X$ can be defined as following:\\

 $\quad \quad \quad X=\left[ \begin{matrix}
0&0&0&1&0&0&0&1&0&0\end{matrix}\right]$\\

\noindent Also, this vector can be written as follows. The numbers of this vector are the sample locations that have a value of 1. So, we have:\\ 
$ \quad \quad \quad X=\left[ \begin{matrix}
4&8\end{matrix}\right]$\\
\end{thm}

\begin{thm}
Suppose the training set namely $S$ is used as follows. Given the fact that this matrix is sparse, it is possible to write the matrix only by storing the features of the value of 1. Hence, we have:\\

\noindent$S=$
\resizebox{8.25cm}{!}{
\setlength\tabcolsep{0.5pt} 

\begin{blockarray}{cccccccccccc}
Sample&$f_1$ & $f_2$& $f_3$ & $f_4$& $f_5$& $f_6$& $f_7$& $f_8$& $f_9$& $f_{10}$ &Label\\
\begin{block}{c(cccccccccc)(c)}
$S_1$&0&1&0&0&0&0&1&1&0&0&0\\‎
$S_2$&0&0&0&1&0&0&0&0&1&0&0\\‎
$S_3$&1&1&0&0&0&0&0&0&0&0&1\\‎
$S_4$&0&0&1&0&0&0&0&1&0&0&1\\‎
$S_5$&0&1&0&0&0&0&0&0&1&0&1\\‎
$S_6$&0&0&0&0&1&0&0&0&0&1&0\\‎
$S_7$&1&0&0&0&0&0&0&1&0&0&1\\‎
$S_8$&0&0&0&1&0&0&1&0&0&0&1\\‎
$S_9$&0&0&1&0&0&1&0&0&0&0&0\\‎
$S_{10}$&1&0&0&1&0&0&0&0&1&0&0\\
\end{block}
\end{blockarray}
   $\xrightarrow{\text{Sparse form}}‎‎‎‎\left( \begin{matrix}
\{2,7,8 \}\\‎
\{4,9\}\\‎
\{1,2 \}\\‎
\{ 3,8\}\\‎
\{ 2,9\}\\‎
\{5,10 \}\\‎
\{1,8 \}\\‎
\{4,7 \}\\‎
\{3,6 \}\\‎
\{1,4,9 \}
\end{matrix}\right)$
\label{matrix}
}

Because, in all proposed algorithms, the distance of sample $X$ is used from all samples in training datasets. In Table~\ref{tab1}, we show the distance between each sample of the training set with the sample $X$, which is computed by the Hamming distance.
\end{thm}

\begin{table}[!h]
\centering
\footnotesize{
\setlength\tabcolsep{6pt} 
\caption{\small Distance of $X$ from each vector in sample dataset.}
\label{tab1}
\begin{tabular}{|c|c|c|c|c|c|c|c|c|c|}
\hline
$S_1$&$S_2$&$S_3$&$S_4$&$S_5$&$S_6$&$S_7$&$S_8$&$S_9$&$S_{10}$\\\hline
3&2&4&2&4&4&2&2&4&3\\\hline
\end{tabular}
}
\end{table}

Considering the presented definitions, in the following we examine our methods for the defined samples.\\

\noindent\textbf{Applying FNN Algorithm:} In Table~\ref{tab1}, the first nearest sample to $X$, which is selected by the FNN algorithm, is $S_2$. Since the label of sample $S_2$ is $0$, the value of $0$ is assigned to the sample $X$. \\

\noindent\textbf{Applying ANN Algorithm:}
Focusing on ANN algorithm, we select all similar samples. In this example, $S_2$, $S_4$, $S_7$, and $S_8$ have been selected according to Table~\ref{tab1} (i.e., see lower values; we select four vectors with value 2). By voting between labels of these samples, the value of $1$ is assigned to the sample $X$. \\

\noindent\textbf{Applying WANN Algorithm:} Focusing on WANN algorithm, we first count the number of features in the training samples to find the vector $w$ (see Table~\ref{tab2} which includes the weight of each feature). Now, we compute the weight of each sample. The weight of each sample is the total weight of the features of that sample, which is 1. Given that the weight of sample $X$ is equal to $6$ and as we can see from the Table~\ref{tab3}, samples $S_2$, $S_3$, $S_5$, and $S_7$ are similar to $X$ and by voting between them the label of sample $X$ is will be $1$.\\
\begin{table}
\centering
\footnotesize{
\setlength\tabcolsep{3pt} 
\caption{\small Weight of each feature in dataset}
\label{tab2}
\begin{tabular}{|c|c|c|c|c|c|c|c|c|c|}
\hline
$W_{f_1}$&$W_{f_2}$&$W_{f_3}$&$W_{f_4}$&$W_{f_5}$&$W_{f_6}$&$W_{f_7}$&$W_{f_8}$&$W_{f_9}$&$W_{f_{10}}$\\
\hline
3&3&2&3&1&1&2&3&3&1\\
\hline
\end{tabular}
}
\end{table}

\begin{table}
\centering
\footnotesize{
\setlength\tabcolsep{2pt} 
\caption{\small Weight of each sample in sample dataset}
\label{tab3}
\begin{tabular}{|c|c|c|c|c|c|c|c|c|c|}
\hline
$W_{S_1}$&$W_{S_2}$&$W_{S_3}$&$W_{S_4}$&$W_{S_5}$&$W_{S_6}$&$W_{S_7}$&$W_{S_8}$&$W_{S_9}$&$W_{S_{10}}$\\\hline
8&6&6&5&6&2&6&5&3&9\\\hline
\end{tabular}
}
\end{table}

\noindent\textbf{Applying KMNN Algorithm:} Focusing on KMNN method, we first select the same sample $X$ as the ANN method and then select $S_2$, $S_4$, $S_7$ and $S_8$ samples. Now, we create two clusters by placing similar samples in the same cluster. The similarity measure will be the distance between samples in each cluster. For this purpose, we determine the matrix of the intervals between these samples namely $I$ as follows:

\begin{center}
$I=$
\normalsize{
\setlength\tabcolsep{0.5pt} 
\begin{blockarray}{ccccc}
&$S_2$ & $S_4$& $S_7$ & $S_8$  \\
\begin{block}{c(cccc)}
$S_2$&  0 & 4 & 4 & 2  \\
$S_4$&  4 & 0 & 2& 4  \\
$S_7$&  4 & 2 & 0 & 4  \\
$S_8$&  2 & 4 & 4 & 0  \\
\end{block}
\end{blockarray}
}
\end{center}

Each entry of a matrix $I$ represents the distance between the two samples, which is obtained by comparing peer to peer elements of corresponding vectors. Focusing on matrix $I$, the distance between the $S_2$ and $S_8$ samples is the smallest distance, so we can place them in a cluster. Similarly, the samples of $S_4$ and $S_7$ are near each other and we can place them in another cluster. Now, in each cluster, we select one of the samples which has a minimum distance from other samples as a cluster head (CH). In this example, since we have only two samples per cluster, we can consider each cluster sample as a CH. Hence, we define $S_2$ as the CH in the first cluster and $S_4$ as the CH in the second cluster. Then, we compute the total distance (i.e., $d$) of all the samples from two CHs as (See Table~\ref{tab:tab41}). In the last step, we should leave a $k$ percentage of the most distant samples and vote among the other samples. In the proposed method, we consider $k = 10$, but for more clarity in these examples, we define $k = 25$, and we do not consider just the last sample. After that, we vote among the rest of the samples. As a result, the result of the voting obtains the value of 1 for the label of the sample $X$.

\begin{table}[!h]
\centering
\caption{\small Sum of distances from CHs for the selected samples.}
\footnotesize{
\setlength\tabcolsep{6pt} 
\label{tab:tab41}
\begin{tabular}{|c|c|c|c|}
\hline
$\sum_{d_{S_2}}$&$\sum_{d_{S_4}}$&$\sum_{d_{S_7}}$&$\sum_{d_{S_8}}$\\
\hline
4&4&6&6\\
\hline
\end{tabular}
}
\end{table}

\end{document}